\documentclass[12pt,oneside,reqno,dvipsnames]{article} \usepackage[UKenglish]{babel} \usepackage[utf8]{inputenc} 

\usepackage[a4paper,top=30mm,bottom=20mm,left=27.5mm,right=27.5mm]{geometry} \usepackage{lscape} \usepackage[table,xcdraw]{xcolor}
\usepackage[normalem]{ulem} \usepackage{tocloft} \usepackage[skip=0ex, indent=1.5em]{parskip}

\usepackage{amsmath} \usepackage{mathtools} \usepackage{array} 

\usepackage{titlesec} \titleformat*{\section}{\bfseries} \titleformat*{\subsection}{\bfseries} \titleformat*{\subsubsection}{\normalsize\bfseries} \titleformat*{\paragraph}{\normalsize\bfseries} \titleformat{name=\section}{\normalfont\bfseries}{\thesection}{0.7em}{\hspace{0em}}{} \titleformat{name=\subsection}{\normalfont\bfseries}{\thesubsection}{0.7em}{\hspace{0em}}{} \titleformat{name=\subsubsection}{\normalfont\normalsize\bfseries}{\thesubsubsection}{0.7em}{\hspace{0em}}{} \titleformat{name=\paragraph}[runin]{\normalfont\normalsize\bfseries}{\theparagraph}{0.7em}{\hspace{0em}}{} 

\usepackage[hang,flushmargin,symbol]{footmisc} \renewcommand*{\thefootnote}{\fnsymbol{footnote}} 

\usepackage{amssymb} \usepackage{marvosym} \usepackage{ifsym} \usepackage{wasysym} \usepackage[gen]{eurosym} \usepackage{textcomp} \usepackage{MnSymbol} 

\usepackage[font={small,it}]{caption} \usepackage{subcaption} \usepackage[capposition=top]{floatrow} \usepackage{float} \usepackage{afterpage} \usepackage{flafter}

\usepackage{xcolor} 

\usepackage{tabularx} \newcolumntype{L}[1]{>{\raggedright\arraybackslash}p{#1}} \newcolumntype{C}[1]{>{\centering\arraybackslash}p{#1}} \newcolumntype{R}[1]{>{\raggedleft\arraybackslash}p{#1}} \usepackage{multirow} \usepackage{hhline} \usepackage{rotating} \usepackage[flushleft]{threeparttable} 

\usepackage{booktabs}
\usepackage{longtable}
\usepackage[referable]{threeparttablex}

\usepackage{diagbox} \usepackage{colortbl} \usepackage{arydshln} 

\usepackage{pgfplots} \pgfplotsset{compat=newest}
    \usepgfplotslibrary{statistics} \pgfkeys{/pgf/number format/.cd,1000 sep={}} \usepackage{tikz} \usetikzlibrary{decorations.pathreplacing, patterns, intersections}

\usepackage{enumitem} 

\usepackage{graphicx} \graphicspath{{Figures/}}
\usepackage{pdfpages} 

\usepackage{comment} 

\usepackage{bm} 

\usepackage[maxcitenames=2, mincitenames=1, maxbibnames=99, citestyle=authoryear, style=authoryear, isbn=false, url=false, doi=false, clearlang=true, mergedate=true, bibencoding=utf8, backend=bibtex]{biblatex}

\usepackage{csquotes} 

\makeatletter
    \pretocmd{\blx@head@bibintoc}{\phantomsection}{}{\ddt}
    \makeatother

    \AtBeginBibliography{
 }

    \renewbibmacro{in:}{\ifentrytype{article}{}{\printtext{\bibstring{in}\intitlepunct}}}

\DeclareNameAlias{sortname}{family-given} \DeclareNameAlias{default}{family-given}

\DeclareFieldFormat[article,inbook,incollection,inproceedings,patent,thesis,unpublished]{title}{#1\isdot}

\makeatletter
    \AtEveryBibitem{
        \global\undef\bbx@lasthash}
    \makeatother

\DeclareCiteCommand{\citeyear}
         {}
         {\bibhyperref{\printdate}}
         {\multicitedelim}
         {}
    \DeclareCiteCommand{\citeyearpar}
         {}
         {\mkbibparens{\bibhyperref{\printdate}}}
         {\multicitedelim}
         {}

\usepackage[hyperfootnotes=true]{hyperref} \hypersetup{colorlinks=true,linkcolor=blue,citecolor=blue,urlcolor=blue,linkbordercolor=cyan,citebordercolor=cyan,urlbordercolor=cyan}

\newif\ifrednotes
\newif\ifgreennotes
\newif\ifbluenotes
\definecolor{notetextred}{rgb}{0.9,0,0}
\definecolor{notetextgreen}{rgb}{0,0.9,0}
\definecolor{notetextblue}{rgb}{0,0,0.9}

\makeatletter
{\@bsphack\@esphack}
{\@bsphack\@esphack}
{\@bsphack\@esphack}
\makeatother

\definecolor{bblue}{HTML}{4F81BD}
\definecolor{rred}{HTML}{C0504D}
\definecolor{ggreen}{HTML}{9BBB59}

\newif\ifinsertedhighlight
\insertedhighlighttrue
\newif\ifdeletedhighlight
\deletedhighlightfalse
\definecolor{revisedblue}{rgb}{0,0,0.6}
\definecolor{revisedboldblue}{rgb}{0,0.0,0.6}
\definecolor{deletedred}{rgb}{0.7,0,0}
\DeclareRobustCommand{\insertedfmt}[1]{\textcolor{revisedboldblue}{\textbf{#1}}}
\DeclareRobustCommand{\deletedfmt}[1]{\textcolor{deletedred}{\sout{#1}}}

\DeclareRobustCommand{\insertion}[1]{\ifinsertedhighlight
    \begingroup
      \let\originalfootnote\footnote
      \renewcommand\footnote[2][]{\ifx\relax##1\relax
          \originalfootnote{\insertedfmt{##2}}\else
          \originalfootnote[##1]{\insertedfmt{##2}}\fi}\insertedfmt{#1}\endgroup
  \else #1\fi}
\makeatletter
\newcommand{\deletionfootnotes}{}
\DeclareRobustCommand{\deletion}[1]{\ifdeletedhighlight
    \begingroup
      \let\originalfootnotemark\footnotemark
      \renewcommand\footnote[2][]{\ifx\relax##1\relax
          \mbox{\originalfootnotemark}\g@addto@macro\deletionfootnotes{\footnotetext{\deletedfmt{##2}}}\else
          \mbox{\originalfootnotemark[##1]}\g@addto@macro\deletionfootnotes{\footnotetext[##1]{\deletedfmt{##2}}}\fi}\deletedfmt{#1}\endgroup
    \deletionfootnotes
    \renewcommand{\deletionfootnotes}{}\fi}
\makeatother

\let\DIForigfootnote\footnote
\newcommand{\DIFgobblefootnote}[2][]{\stepcounter{footnote}}
\makeatletter
\newcommand{\DIFdisablefootnote}{\ifx\protect\@typeset@protect\ifdeletedhighlight\else\let\footnote\DIFgobblefootnote\fi\fi}
\newcommand{\DIFenablefootnote}{\ifx\protect\@typeset@protect\ifdeletedhighlight\else\let\footnote\DIForigfootnote\fi\fi}
\makeatother

\DeclareRobustCommand{\revisedfloatmarked}[2]{\ifinsertedhighlight
    \begingroup
      \let\revisedorigcaption\caption
      \renewcommand\caption[2][]{\ifx\relax##1\relax
          \revisedorigcaption{\insertion{#1}\ ##2}\else
          \revisedorigcaption[##1]{\insertion{#1}\ ##2}\fi}#2\endgroup
  \else #2\fi}
\DeclareRobustCommand{\revisedfloat}[1]{\revisedfloatmarked{[New in revision]}{#1}}
\DeclareRobustCommand{\updatedfloat}[1]{\revisedfloatmarked{[Updated in revision]}{#1}}

\def\code#1{\texttt{#1}}

\usepackage{amsfonts}
\definecolor{Gray}{gray}{0.9}

\usepackage{multicol}		\usepackage{pbox}           \usepackage{verbatim}
\usepackage{pdflscape}


\rednotesfalse \greennotesfalse     \bluenotesfalse 

\insertedhighlightfalse \deletedhighlightfalse

\begin{document}

\begin{titlepage}
\begin{center}

\vspace*{-1.5cm}
\Large\textbf{Non-Take-Up of Unemployment Benefit II in Germany: A Longitudinal Perspective\\ Using Administrative Data}

\normalsize
\vspace*{-3cm}
\begin{center}
\begin{minipage}[t][5cm][b]{0.6\textwidth}
\begin{center}
Jürgen Wiemers \\[0.1cm]
\small Institute for Employment Research (IAB) \\[-0.1cm]
\small juergen.wiemers@iab.de
\end{center}
\end{minipage}

\vspace*{0.5cm}
\end{center}

\today \\

\begin{abstract} \footnotesize
\noindent

Extensive research demonstrates that many households eligible for means-tested benefits do not claim them, a phenomenon known as non-take-up. Theoretically, long-term factors can substantially impact the take-up decision. For example, households facing income instability may anticipate only temporary eligibility, thus increasing the likelihood of non-take-up. Moreover, past benefits experiences can have varied effects: familiarity with the claiming process may lower informational and transactional barriers, while negative experiences could increase perceived stigma.

Despite the potential relevance of longitudinal aspects, evidence on their influence remains limited. This study addresses this gap by incorporating long-term factors in the analysis of non-take-up behaviour relating to \emph{Unemployment Benefit II} (UB~II), Germany's basic means-tested welfare programme. 

Using data from the German \textit{Panel Study Labour Market and Social Security} from 2008 to 2020, linked with administrative data from Germany's Federal Employment Agency, this study reconstructs households' benefit receipt and income histories, even during non-survey periods. In addition, the use of administrative data mitigates bias from self-reported benefit receipt. Household eligibility for UB~II is simulated using GETTSIM, an open-source microsimulation model.

Findings indicate that long-term factors are significantly associated with the probability of claiming UB~II. Specifically, a longer history of benefit receipt is associated with a higher probability, whereas higher income potential and positive income shocks are associated with a lower probability of claiming. Accounting for long-term factors substantially changes the estimated associations of traditionally used correlates of non-take-up, indicating a potential misspecification in existing models that neglect them. The analysis is descriptive throughout, documenting conditional associations rather than causal effects.

\end{abstract}

\vspace*{0.5cm}
\small

\textbf{JEL Classification:} \\[-0.1cm]
C15 / H53 / I38 \\[0.5cm]

\textbf{Keywords:} \\[-0.1cm]
non-take-up / means-tested benefits / microsimulation

\end{center}
\end{titlepage}

\renewcommand*{\thefootnote}{\arabic{footnote}}
\setcounter{footnote}{0}
\newpage
\normalsize

\section{Introduction}
\label{sec:Introduction}

Interest in the non-take-up of social benefits has grown considerably in recent years. Non-take-up refers to instances where households are entitled to social benefits yet do not claim them. Empirical evidence suggests that non-take-up is substantial in many welfare states. The magnitude of non-take-up is often quantified by the non-take-up rate (NTR), defined as the ratio of eligible households not receiving benefits to the total number of eligible households. Quantifying the NTR empirically presents significant challenges, as the entitlement of households to a particular benefit cannot be directly observed from official statistics or population surveys. Consequently, tax-benefit microsimulation models based on representative survey data are commonly used to simulate eligibility. Given the simulated eligibility, the NTR can be estimated by comparing the social benefits reported in the survey with the simulated entitlements for each household in the dataset.

Globally, NTR estimates for social benefit programmes range from 20~percent to over 80~percent, depending on programme type, administrative mechanisms, cultural attitudes, and national economic development \parencite{KoMoffitt2022}. In Europe, NTR estimates for means-tested benefits typically range between 40 and 60~percent, with some estimates nearing 80~percent \parencite{Eurofound2015}. For Germany, \textcite{Har19} and \textcite{Bru21} estimated that NTRs for social assistance for working-age persons (Unemployment Benefit II, referred to as UB~II in the following) lie in the range between 37 and 56~percent.

The non-take-up of means-tested benefits can be a relevant social, economic, and political concern, although its normative implications depend on who refrains from claiming and why. Socially, non-take-up may undermine equity by creating disparities between otherwise comparable eligible individuals who do and do not receive support. This concern is most pressing when non-take-up is concentrated among the households for whom the transfer would have the highest insurance value. By contrast, if claiming is a deliberate and informed decision, refraining from a benefit of little net value to the household need not reduce welfare and may even improve the targeting of scarce public funds towards those most in need. Policymakers may tolerate, or even deliberately maintain, a degree of administrative complexity as a screening device that deters households who value the benefit least \parencite{KoMoffitt2022}. Economically, non-take-up may reduce the effectiveness of benefit systems in combating poverty and could contribute to increased long-term expenditures on healthcare and social interventions. Politically, non-take-up may erode trust in public institutions, especially among disadvantaged groups who perceive benefit systems as either inaccessible or unfair \parencite{Eurofound2015}.

Empirical studies on the determinants of non-take-up frequently depict the decision to claim social benefits as a rational process. From this viewpoint, individuals will claim benefits only if the net utility of doing so -- factoring in both monetary and non-monetary costs -- is positive. Costs related to benefit claims can include informational and transactional expenditures (e.g., application efforts, time demands, compliance obligations), along with the social stigma associated with receiving such benefits.

Long-term factors may substantially affect the expected net utility of current benefit claims. For instance, prior interactions with the benefit system could have ambiguous effects on take-up: Although prior participation might reduce informational and transactional barriers, it can also increase perceived stigma costs due to negative past experiences. Moreover, benefit claiming may negatively correlate with a household's income potential, which is defined here as the expected average earned income. Finally, perceived negative income shocks might raise the likelihood of claiming benefits, with the impact on take-up potentially varying based on the perceived duration of the income loss.

Few empirical studies have examined the role of long-term factors, as access to longitudinal data has been limited in the past. For instance, German panel studies, such as the \textit{Socio-Economic Panel} (SOEP) and the \textit{Panel Study Labour Market and Social Security} (PASS), can track households over time. However, their usefulness is limited because low-income households often remain in the panel for only a small number of waves, which limits the observability of past benefit receipts and income histories. For these reasons, existing empirical studies on the determinants of participation in social welfare systems typically neglect the long-term factors of benefit receipt and focus only on factors that are measured simultaneously with the observed non-take-up.\footnote{For instances relating to the take-up of \emph{Unemployment Benefit II} in Germany, refer to \textcite{Bru11, Bru13, BruWie2017, Bruckmeier2018, Har19, Bru21}.} The few exceptions that do exploit income dynamics point in a consistent direction: \textcite{BlankRuggles96} find that take-up differs between the transitorily and the persistently needy, and \textcite{Paukkeri2018}, using linked Finnish register data, documents that eligible non-claimants are disproportionately households experiencing a short-term income fall who display greater income variability than claimants. Comprehensive longitudinal evidence of this kind nonetheless remains scarce, particularly for Germany.

This paper aims to investigate how long-term factors, namely ``benefit history'', ``income potential'', ``income volatility'', and ``income shocks'', relate to the receipt of \emph{Unemployment Benefit II} (UB~II) in Germany. For this purpose, survey data from the PASS (2008--2020) are linked to administrative data via the PASS-ADIAB dataset. This linkage enables the inclusion of income and benefit histories for households that have consented to the data linkage. The primary advantage of this linked dataset lies in its ability to observe long-term factors even during periods when households were not actively participating in the PASS survey. Additionally, administrative data provide corrections for inaccuracies in self-reported survey responses regarding UB~II receipt, addressing a critical source of measurement error.

The findings indicate that a longer history of receiving benefits is associated with a higher likelihood of claiming UB~II, whereas a higher income potential and a positive income shock are associated with a lower likelihood of claiming. By contrast, income volatility is not significantly associated with benefit claims. These associations are robust and economically meaningful. However, as discussed in Section~\ref{subsec:MethTakeUp}, they should be interpreted as conditional correlations that describe claiming behaviour rather than as causal effects. Overall, these findings highlight the relevance of long-term factors for understanding benefit take-up within the German basic safety net.

The structure of this paper is as follows. Section~\ref{sec:InstLit} outlines the institutional background. Section~\ref{sec:EmpApp} details the empirical methodology, including a description of the data and criteria for sample selection. Section~\ref{sec:Results} presents the estimated NTRs and estimation results from reduced-form take-up models, starting with a specification as typically found in the literature. Building on this baseline specification, long-term factors are incrementally added to assess their impact on the take-up decision. Section~\ref{sec:Conclusions} concludes with a discussion of the findings and their implications.

\section{Institutional background}
\label{sec:InstLit}

This paper examines the take-up of Germany's basic minimum income support programme, \emph{Unemployment Benefit II}\footnote{In 2023, a reform renamed UB~II to \emph{Citizen's Income} and made several adjustments, including alterations to sanction policies. Nevertheless, since the data analysed in this paper spans the years 2008 to 2020, I will continue to use the term ``UB~II'' throughout.} (UB~II). The programme assists households with at least one working-age individual who is able to work. The UB~II programme follows federal regulations and is managed by local authorities. The benefit is available when a household's net income is below a legally defined minimum income, with benefits compensating for the gap. The benefit unit is referred to as the ``community of needs'', which typically corresponds to the immediate family -- either single adults or couples, and if applicable, their dependent children under the age of 25.\footnote{Children aged 25 and above who live with their parents form a separate community of need within the household.} It consists of two components: a standard benefit amount per person and a payment covering rent and heating. The standard benefit varies depending on the number of adults in the household and is also age-dependent for the children.\footnote{As of 2025, the standard benefit for a single adult is set at 563~euros per month.} Rent and heating expenses are covered up to acceptable limits, which are determined by the average local rent level. Entitlement to UB~II is contingent upon a means test encompassing almost all types of income. The overall entitlement is diminished by family members' incomes, excluding a provision for earned income. Benefits can be higher under certain conditions, such as for single parents, pregnant women, or those with specific dietary requirements. Households surpassing a wealth threshold are ineligible, where wealth includes financial assets and property value minus mortgages. Owner-occupied homes, a car for each working-age person, and retirement provisions are usually excluded as long as they are appropriate. 

Recipients must actively seek or accept reasonable employment, although certain groups, such as those caring for children or engaged in education, are exempt from work requirements. However, unemployment itself is not a prerequisite for entitlement, and recipients can vary widely in their personal circumstances, including being low-wage earners, long-term unemployed, caregivers, or self-employed individuals. 
In 2023, about 43~percent of the employable\footnote{According to §8~(1)~Social Code Book~II, a person is deemed employable if they are not hindered due to illness or disability to work for at least three hours a day under the usual conditions of the general labour market for the foreseeable future.} beneficiaries were unemployed, while 20~percent of employable beneficiaries used the benefit to supplement insufficient earnings from employment. Others were unable to work temporarily due to reasons such as childcare responsibilities. The programme covered approximately 5.8~million individuals in 2.9~million households and paid out 42.6~billion~euros \parencite{BA2024}, establishing it as by far the largest means-tested benefit scheme in Germany.

In addition to UB~II, Germany provides two other important means-tested social benefits: \emph{Housing Benefit} (HB) and \emph{Supplementary Child Benefit} (SCB). These benefits are prioritised over UB~II with the goal of enabling households with low incomes to avoid depending on social welfare. HB and SCB are closely linked to UB~II: Once the entitlement to HB and/or SCB exceeds the entitlement to UB~II, the household becomes ineligible to receive UB~II. Consequently, HB and SCB can be received at the same time, while simultaneous receipt of HB/SCB on the one hand and UB~II on the other is prohibited.\footnote{For further details regarding how the three benefits interact, refer to \textcite{Bruckmeier2018}.}

\section{Empirical approach}
\label{sec:EmpApp}

\subsection{Modelling benefit take-up}
\label{subsec:MethTakeUp}

In recent decades, numerous empirical studies into the factors influencing means-tested benefit take-up have been carried out, see, e.g., \textcite{Blu88, BlankRuggles96, Rip01, Wil05, Whe09, Bru12, Har19, Bru21}. These studies model the take-up decision as a binary choice in which a household claims the benefit if its net utility from claiming exceeds the utility from not claiming. Following most of this literature, this study estimates take-up models in reduced form: rather than structurally estimating the household's preferences and budget constraint, the observed take-up indicator is related directly to household characteristics and the simulated benefit entitlement.\footnote{While a large number of empirical studies frame (non-)take-up of social benefits as a utility maximising choice, only a small number of those studies actually embed the take-up choice in a structural model of labour supply. For the latter, see, e.g., \textcite{Mof83, Hoy96, BluMac99, Kea98, KeaneWolpin2010, Cha13}; for a recent literature overview of structural models of benefit take-up, see \textcite{ChanMoffitt2018}. Given that the aim of this study is to explore the correlations between take-up and possible long-term factors, rather than conducting policy analyses, I have also opted for a reduced-form approach.} The net utility accounts for the fact that benefit claims are associated with monetary and non-monetary costs, which arise, for example, due to inadequate knowledge of entitlement regulations, the complexities of the claiming process, complex administrative procedures, and stigma costs, i.e., the fear of negative social perceptions concerning welfare dependency \parencite{Oor91}.

In the literature, the empirical specification for the reduced-form take-up model is usually motivated as follows \parencite[see, e.g.,][]{Blu88}. For any means-tested benefit, take-up ($P=1$) is assumed to be observed if the net level of utility from claiming the benefit exceeds the utility when not claiming the benefit; otherwise, the household will refrain from claiming the benefit ($P=0$). Formally,
\begin{equation}
P=\mathbf{I}\left( U\left( y+b\left( y,\mathbf{x}^{\ast }\right) ,\mathbf{x}\right) -C\left( \mathbf{x}\right) >U\left( y,\mathbf{x}\right) \right) , \label{eq:takeup}
\end{equation}
where $\mathbf{I}\left( \cdot \right) $ denotes the indicator function, $U\left( \cdot \right) $ represents utility, $y$ is disposable income excluding the benefit entitlement,
\begin{equation}
b\equiv b\left( y,\mathbf{x}^{\ast }\right) =\max \left( \overline{b}\left(\mathbf{x}^{\ast }\right) -y,\;0\right) ,   \label{eq:benefit}
\end{equation}
which is given by the (positive) difference of household needs $\overline{b}\left( \mathbf{x}^{\ast }\right) $ and disposable income $y$ excluding the benefit. The household needs are a function of characteristics $\mathbf{x}^{\ast }$, which are assumed to affect utility only indirectly through their impact on the benefit entitlement. \footnote{Equation~(\ref{eq:benefit}) is a stylised representation of the budget constraint: as written, the benefit closes the entire gap between needs and disposable income, which corresponds to a withdrawal rate of 100~percent. In practice, UB~II withdraws additional income at varying rates between 0~percent and 100~percent because of earned-income disregards (Section~\ref{sec:InstLit}). The entitlement used in the estimations, the relative income gap defined below, is computed by the microsimulation model under the actual statutory disregards rather than under the stylised expression in equation~(\ref{eq:benefit}).} The disutility from claiming, $C\left( \mathbf{x}\right) $, is assumed to depend on observable household characteristics $\mathbf{x}$. Besides the observed characteristics, it is probable that take-up is also influenced by unobserved characteristics. Assuming linear functions for $U\left( \cdot \right) $ and $C\left( \cdot \right) $ results in
\begin{eqnarray}
U\left( y+b\left( y,\mathbf{x}^{\ast }\right) ,\mathbf{x}\right)  &=&\alpha_{0}+\alpha _{1}\left( y+b\right) +\mathbf{\alpha }_{2}^{\prime }\mathbf{x} + \varepsilon _{1},  \nonumber \\
U\left( y,\mathbf{x}\right)  &=&\alpha _{0}+\alpha_{1}y+\mathbf{\alpha}_{2}^{\prime }\mathbf{x}+\varepsilon _{0},  \label{eq:util_costs} \\
-C\left( \mathbf{x}\right)  &=&\beta _{0}+\mathbf{\beta }_{1}^{\prime }
\mathbf{x+}\varepsilon_{C} ,  \nonumber
\end{eqnarray}
where $\varepsilon _{1}$, $\varepsilon _{0}$, and $\varepsilon_{C} $ represent the household-specific unobservables, and $\mathbf{\alpha =}\left( \alpha
_{0},\alpha _{1},\mathbf{\alpha }_{2}\right) $, $\mathbf{\beta =}\left(
\beta _{0},\mathbf{\beta }_{1}\right) $ are coefficient vectors. 

Substituting (\ref{eq:benefit}) and (\ref{eq:util_costs}) into (\ref{eq:takeup}) and assuming an i.i.d. standard normal distribution for the combined error terms\footnote{Assuming a normal distribution for the error term is a standard assumption in the literature, see, e.g., \textcite{Blu88, Duc95, Pud01}.} $\upsilon \equiv \varepsilon
_{1}-\varepsilon _{0}+\varepsilon _{C}$, i.e., $\upsilon \sim N\left(
0,1\right) $, the probability of observing take-up is given by
\begin{equation}
\Pr \left( P=1\right) =\Pr \left( \upsilon >-\left( \beta _{0}+\alpha _{1}b+\mathbf{\beta }_{1}^{\prime }\mathbf{x}\right) \right) =\Phi \left( \beta_{0}+\alpha _{1}b+\mathbf{\beta }_{1}^{\prime}\mathbf{x}\right) \label{eq:probit},
\end{equation}
where $\Phi \left( \cdot \right) $ denotes the cumulative standard normal distribution. The model is conditional on eligibility for the benefit, i.e., only households with $b>0$ enter the estimation. In contrast to $\mathbf{x}$, the true level of entitlement $b\left( y,\mathbf{x}^{\ast }\right)$ is not directly observable in the data. Therefore, $b\left( y,\mathbf{x}^{\ast }\right)$ is usually approximated using a tax-transfer microsimulation model. Read in reverse, this derivation makes explicit what a structural (and hence causal) interpretation of equation~(\ref{eq:probit}) would require: constant marginal utility of income (the linearity of $U$), an additively separable claiming cost whose unobservable is normal and homoscedastic, and, most demandingly, an entitlement $b$ that is exogenous to the unobservables collected in $\upsilon$, a condition that is unlikely to hold, since labour supply plausibly responds to the same unobserved factors that drive claiming. Accordingly, I do not read the coefficient on $b$ as the structural effect of entitlement on take-up. Like the coefficients on the long-term factors that are the focus of this study, I interpret it as a conditional association rather than as a structural or causal effect.

While this derivation of the empirical take-up model suggests the (pooled) probit estimation (\ref{eq:probit}), my preferred specification exploits the panel structure of the data and estimates a random effects (RE) probit model, which allows me to account for unobserved heterogeneity at the household level. For the RE probit, the probability of benefit take-up for household $i$ conditional on $b_{it}>0$ in period $t$ is given by
\begin{equation}   
Pr(P_{it}= 1)=Pr(\upsilon_{it}>-(\beta_0+\alpha_{1}b_{it}+\mathbf{\beta_1}'\mathbf{x}_{it}+\nu_i))=\Phi(\beta_0+\alpha_{1}b_{it}+\mathbf{\beta_1}'\mathbf{x}_{it}+\nu_i) \label{eq:REprobit},
\end{equation}
with i.i.d. $\upsilon_{it}\sim N(0,1)$, which are assumed independent of the i.i.d. random effects $\nu_i\sim N(0,\sigma_\nu^2)$. The share of the total variance contributed by the panel-level variance component is given by $\rho=\sigma_\nu^2/(\sigma_\nu^2+1)$.\footnote{Table~\ref{tab:TUests_w0_e0_excl11001} reveals that $\rho$ is large and statistically significant on the 1~percent level for all estimations, thus the pooled probit model is strongly rejected.}

While there is a consensus in the literature that the benefit entitlement $b$ is a key regressor for benefit take-up, there is much less agreement on how exactly it should be included in the estimation. In this study, I employ the approach proposed, e.g., by \textcite{Fri07}, which implements the relative income gap as a measure for $b$. The relative income gap is defined as the ratio of benefit entitlement to the household's UB~II needs. The relative income gap normalises the entitlement of a household to the interval $[0, 1]$, which allows for a more meaningful comparison of entitlements between households of different sizes.\footnote{The relative income gap enters the estimation equation (equation~(\ref{eq:full}) below) as a quadratic polynomial. This functional form is not imposed a priori: Figure~\ref{fig:income_gap_fit} first traces the relationship between the relative income gap and the take-up rate non-parametrically, by regressing the take-up indicator on a full set of income-gap category dummies. The resulting profile is approximately quadratic, which motivates the parsimonious quadratic approximation used in the take-up models.}

As a starting point for a baseline specification of a UB~II take-up model and to facilitate comparison with the existing literature, I build on the existing empirical literature for selecting the household characteristics $\mathbf{x}$, with particular reference to \textcite{Bru12}, \textcite{Har19}, and \textcite{Bru21}. I include indicators of household type because utility and claiming costs may vary depending on whether the household consists of a single person or a couple as well as the presence of children. Further socio-demographic indicators employed are age, educational level, an indicator for first- or second-generation immigrant status, and the disability status of the household head. In line with prior investigations, I also account for indicators of home ownership status, a binary indicator for Eastern Germany, and indicators regarding the size of the municipality where a household is situated. Finally, all specifications include time-fixed effects.

My baseline specification includes three additional indicators not typically found in previous studies, which proved to be highly significant in my estimated UB~II take-up models.
Firstly, as explained in the following Section~\ref{subsubsec:DataPASS}, my dataset comprises two subsamples, an administrative sample drawn from data on UB~II recipients and a general population sample. Since the non-take-up rate in the latter is expected to be substantially higher than in the former, I incorporate a dummy to indicate household membership in the general population sample.
Secondly, there exists a relatively small group of households primarily dependent on reduced earning capacity pensions (\emph{Erwerbsminderungsrenten}). Typically, these households qualify for UB~II benefits, unless the adult household members are permanently and fully incapable of working, a condition not discernible in the PASS data. Although technically eligible for UB~II benefits, such households might see themselves as pensioners, leading to exceedingly high non-take-up rates for UB~II. Therefore, I introduce an indicator for self-assessed early retirement status. 
Thirdly, a small number of households are simulated as entitled to UB~II but report having received benefits prioritised over UB~II, particularly HB and SCB. As described in Section~\ref{sec:InstLit}, concurrent receipt of HB/SCB and UB~II is legally prohibited, and benefit administering agencies actively communicate to avert simultaneous receipt. Possible causes include a misclassification of entitlement by the simulation model or household misreporting of received benefits in the survey. Unsurprisingly, these households display a very high NTR for UB~II. Consequently, I include a binary indicator for reported receipt of upstream benefits.\footnote{As a robustness check, I have excluded self-assessed early retirement households along with recipients of upstream benefits from the sample, which does not alter the results regarding the long-term determinants of take-up either qualitatively or quantitatively.}

Building on the baseline specification, I gradually incorporate the potential long-term determinants of benefit take-up, as discussed in Section~\ref{sec:Introduction}, into the take-up model. In these extended specifications, I use the lagged share of yearly days with UB~II receipts to capture ``benefit history'', the lagged equivalent real annual earned income to proxy ``income potential'', and both the first difference of annual earned income and the standard deviation of lagged real equivalent earned income to measure ``income volatility''.

Collecting these elements, the index of the most general (full) specification can be written as
\begin{align}
\Phi^{-1}\!\big(\Pr(P_{it}=1)\big) ={}& \beta_{0}+\alpha_{1}g_{it}+\alpha_{2}g_{it}^{2}+\boldsymbol{\beta}_{1}'\mathbf{x}_{it}+\sum_{k=1}^{3}\gamma_{k}\,h_{i,t-k} \nonumber\\
&+\sum_{k=1}^{3}\delta_{k}\,e_{i,t-k}+\theta\,\Delta e_{it}+\lambda\,\sigma^{e}_{it}+\psi\,\Delta e_{it}\,\sigma^{e}_{it}+\tau_{t}+\nu_{i},
\label{eq:full}
\end{align}
where $g_{it}$ is the relative income gap (entered as a quadratic), $\mathbf{x}_{it}$ the vector of baseline covariates, $h_{i,t-k}$ the lagged shares of days in UB~II receipt (benefit history), $e_{i,t-k}$ the lagged real equivalised annual earned incomes (income potential), $\Delta e_{it}$ the income shock (the year-on-year change in earned income, expressed in units of the household's current UB~II needs), $\sigma^{e}_{it}$ the earnings volatility, $\Delta e_{it}\,\sigma^{e}_{it}$ their interaction (with coefficient $\psi$), $\tau_{t}$ time fixed effects, and $\nu_{i}$ the household random effect. Earned incomes are deflated to 2020 prices and equivalised using the modified (``new'') OECD scale. The baseline specification (Section~\ref{subsubsec:ResultsTUEstsBase}) restricts $\gamma_{k}=\delta_{k}=\theta=\lambda=\psi=0$. The extended specifications M1--M3 add the benefit-history, income-potential, and income-volatility blocks cumulatively, with the income-volatility block comprising the income shock, the volatility measure, and their interaction, so that M3 contains all long-term factors jointly. Standard errors are cluster-robust at the household level throughout.

This descriptive reading deserves two further qualifications. First, the coefficients summarise how the take-up probability co-varies with household characteristics and the long-term factors, conditional on the other covariates and the household random effect. They are not causal effects, because households that differ in, say, their benefit history are also likely to differ along unobserved dimensions -- labour-market attachment, health, financial literacy, or attitudes towards welfare -- that the model cannot separate from the long-term factors themselves. Second, the random-effects estimator assumes that the household effect $\nu_{i}$ is uncorrelated with the regressors. Where this assumption is questionable, the estimates are best understood as partial correlations within the modelled error structure. I therefore phrase the findings throughout in terms of conditional associations rather than effects.

The four long-term factors capture distinct dimensions of a household's situation. Table~\ref{tab:LongTermFactors_corr} reports their pairwise correlations in the estimation sample. They are only weakly to moderately correlated, with most coefficients below $0.25$ in absolute value, so the extended specifications do not simply re-describe the same households from different angles. The strongest association is between income potential and the level-based volatility measure ($0.56$), which reflects the mechanical tendency of a level standard deviation to be larger when earnings are higher. I return to this point, and to alternative volatility measures, in Section~\ref{subsubsec:ResultsTUEstsIncomeVolatility}.

\revisedfloat{\begin{table}[!ht]
\footnotesize
\centering
\begin{threeparttable}
\caption{Correlations among the long-term factors}
\phantomsection\label{tab:LongTermFactors_corr}
\renewcommand{\arraystretch}{1.2}

\begin{tabular}{lrrrr}
\hline\hline
 & (1) & (2) & (3) & (4) \\
\hline
(1) Benefit history       &  1.000 &        &        &       \\
(2) Income potential      & -0.247 &  1.000 &        &       \\
(3) Income shock          & 0.013 & -0.140 &  1.000 &       \\
(4) Income volatility     & -0.157 & 0.562 & -0.031 & 1.000 \\
\hline
\end{tabular}

\begin{tablenotes}
\item \scriptsize \textsc{Note. ---} Pairwise Pearson correlations in the estimation sample (24,486 household-wave observations, unweighted). ``Benefit history'' is the mean lagged share of days in UB~II receipt over $t-1$ to $t-3$; ``income potential'' is the mean lagged real equivalised annual earned income over $t-1$ to $t-3$; ``income shock'' is the year-over-year change in earned income standardised by current UB~II needs; ``income volatility'' is the within-household standard deviation of real equivalised quarterly earned income over the preceding twelve quarters. Source: PASS 0620 v1, PASS-ADIAB 7520 v1, own calculations.
\end{tablenotes}

\end{threeparttable}
\end{table}
 }

\subsection{Data}
\label{subsec:Data}

\subsubsection{PASS}
\label{subsubsec:DataPASS}

The primary dataset used is the \textit{Panel Labour Market and Social Security} (PASS)\footnote{See \textcite{trappmann2010pass, Trappmann2013PASS} for a technical documentation.}, provided as a scientific use file by the Research Data Centre (FDZ) of the German Federal Employment Agency (BA) at the Institute for Employment Research (IAB). This dataset is specifically designed for the analysis of low-income households. The PASS consists of two subsamples: a sample randomly drawn from the administrative records of the Federal Employment Agency (sample \emph{Admin} in the following), capturing those who have ever received UB~II, and a general population sample (sample \emph{GenPop}) that is not conditioned on previous UB II receipt. This design ensures comprehensive coverage of the target population for analysing welfare dynamics. In order to maintain representativeness for recipients of UB~II, sample \emph{Admin} is regularly updated to incorporate new UB~II recipients.

The \emph{GenPop} sample is randomly drawn from a database of addresses of private households in Germany.\footnote{For a detailed description of the sampling design, see \textcite{GebhardtEtAl2009_PASS}.} Weights are applied to all descriptive results to correct for distortions from the sampling design, so that they are representative of the relevant population. For the behavioural take-up model, by contrast, the unweighted estimator is the standard and more efficient choice when the model is correctly specified \parencite{SolonHaiderWooldridge2015}, so all regression analyses reported in the main text are estimated unweighted.\footnote{As a robustness test, weighted estimations are reported in Appendix~\ref{sec:AppWeightedVsUnweighted}. The weighted estimations result in point estimates very similar to the unweighted estimations. Although the confidence intervals for the weighted estimates are generally larger, they do not qualitatively alter the results.}

This study specifically uses PASS version 0620 v1, encompassing the initial 14 waves from the years 2006-07 to 2020. Due to revisions in the survey tools and interview protocol post-first wave \parencite{GebhardtEtAl2009_PASS}, the analysis is restricted to waves~2 through 14, spanning the years 2008 to 2020. The PASS dataset is particularly well-suited for examining non-take-up issues as it targets potential beneficiaries residing in low-income households.\footnote{\textcite{Beste2018} find that, from wave 2 onwards, the income distribution in the PASS data resembles that of two other German household surveys, GSOEP and microcensus.} Most importantly for this study, PASS asks respondents about their current welfare benefits, in particular UB~II, and it enables a linkage of the survey data with administrative data on welfare receipt and earned incomes of the household members.

\subsubsection{PASS-ADIAB}
\label{subsubsec:DataAdiab}

The PASS dataset can be linked with the administrative records of the Federal Employment Agency on dependent employment and benefit receipt of UB~II.\footnote{Due to legal restrictions, the survey data can be linked to the administrative records only if participants have given their consent to the linkage. Therefore, interviewers request this consent during interviews. Since the consent rate is high, only 6~percent of all household-wave observations with simulated eligibility for UB~II were discarded due to either non-consent for data linkage or because they could not be identified in the administrative records.} This study employs the scientific use file PASS-ADIAB version~7520~v1, provided by the Research Data Centre (FDZ) of the IAB.\footnote{See \textcite{AntoniBethmann2019_PASS_ADIAB} for more information on PASS-ADIAB.} PASS-ADIAB encompasses detailed employment histories, detailing income, employment sectors, employers since 1975, durations of UB~II and unemployment insurance benefit receipt, registered unemployment, and participation in labour market programmes. The administrative records linked to PASS are representative for the workforce, excluding civil servants and the self-employed. Consequently, PASS-ADIAB offers valuable insights into the long-term elements that may affect benefit take-up.

The PASS survey includes questions regarding the receipt of UB~II during the month of the interview and the previous receipt history. \textcite{Bru21} illustrate that the PASS data's self-reported benefit receipt is influenced by a relatively small amount of misreporting.\footnote{In \textcite{Bru21}, correcting for benefit misreporting reduces the weighted NTR from 40~percent to 37~percent.} Within this misreporting, underreporting is more frequent than overreporting, leading to an upward bias in the estimated NTR. To address misreporting, this study corrects self-reported benefit information with the administrative PASS-ADIAB records on UB~II receipt at the time of the interview. \footnote{Both the survey-administrative linkage and the receipt correction broadly follow \textcite{Bru21}.} As these administrative records accurately reflect actual payments, this method eliminates an important source of measurement error in the analysis of benefit (non-)take-up.

A possible issue with data linkage is the chance of bias arising from selective non-consent or lack of identifiability in administrative records\footnote{The potential reasons for a linkage failure are described in detail in Appendix~B of \textcite{Bru21}.}. Regarding non-consent, the extent of underreporting benefit receipt could be underestimated if those declining to participate in data linkage are also prone to misreporting their benefit status. For instance, individuals who choose not to report their receipt of UB~II may similarly be reluctant to consent to data linkage due to worries about exposing inaccuracies in their reports. \textcite{Bru21} examine the relevance of this potential bias and find that individual characteristics have only small statistical effects on the likelihood of non-consent. With respect to the issue of non-identifiability in administrative data, the authors do not find significant correlations with household characteristics. Overall, these findings suggest that non-consent and non-identifiability likely have a minimal impact on the misreporting of benefit receipt.

\subsubsection{Sample selection}
\label{subsubsec:DataSampSel}

The estimation sample consists of household-wave observations from 2008 to 2020 that meet the following inclusion criteria: (a) the household forms a single ``community of needs'' made up of a core family unit (single adults or couples, with or without dependent children under~25); (b) the interviews required for the simulation -- in particular the partner interview, where applicable -- are available and contain consistent income information; (c) no household member has reached the statutory pension age, and neither the household head nor the partner is in education or vocational training, since these groups are generally outside the scope of UB~II; (d) the household does not belong to the refugee samples added from 2016 onwards, whose take-up is mechanically close to one and which lack a pre-sample administrative history; (e) the household is simulated to be eligible for UB~II in the respective year ($b_{it}>0$), as take-up is defined conditional on eligibility; and (f) the household head is consistent across waves and the explanatory variables and linked administrative records are available. Table~\ref{tab:Admin_Spells_Sample} quantifies how each criterion reduces the sample. The remainder of this section sets out the rationale for the individual steps.

The initial PASS sample consists of 113,841 household-wave observations for the years 2008 to 2020. However, the selection of households for the analysis is subject to several constraints. Observations were excluded from the initial sample based on i) the data requirements of the benefit simulation, ii) the focus on UB~II benefits, and iii) the specification of the take-up estimation models.

Regarding the data requirements of the benefit simulation, the absence of partner interviews accounts for the exclusion of 25,850 ($\approx$ 23~percent) of the initial household-wave observations. Other simulation-related reasons for excluding observations have relatively little impact on the sample size. First, eligibility for UB~II is not assessed at the household level but rather for the ``community of needs'', which typically includes the core family unit, namely parents and potentially their minor children, as well as children aged 18 to 24 residing in the same household. Individuals older than 24 years living in the household establish their own community of needs. Consequently, a household may comprise more than one community of needs. Second, there are rare cases in which a community of needs does not consist of a core family unit. For this analysis, I retain only those households that constitute a singular community of needs where all members are part of the core family unit, as GETTSIM v0.7.0, the tax-and-transfer microsimulation model I use for imputing entitlements (see Section~\ref{subsec:MethBenefitSimulation}), does not ensure accurate simulations for households with multiple communities of needs and non-core family communities of needs. Third, households with persons reporting inconsistent wage information are dropped, specifically where the PASS variable ``current primary status'' (\code{statakt}) indicates their employment status as ``gainful employment with an income exceeding 400 euros per month'', yet their declared income is 0 euros. Fourth, two households are dropped because of inconsistent partner information. Overall, the four additional simulation-related reasons account for the exclusion of 6,379 household-wave observations, which constitute approximately 6~percent of the initial observations.

The second class of observation exclusions arises from the focus of this study on the take-up of UB~II benefits. Given that UB~II is aimed at individuals of working age (and their dependent children), pensioners are ineligible. Thus, I drop all households with at least one member who has reached legal pension age from the sample. This results in the exclusion of 12,876 household-wave observations, accounting for over 11~percent of the initial household-wave observations. I also exclude observations where either the head of household or their partner is studying or in vocational training, as these individuals are usually excluded from UB~II benefits. Furthermore, administrative samples of refugee households (primarily Syrian and Iraqi nationals) were included in the PASS during waves 10 to 14 (from 2016 to 2020). As these households were previously recipients of benefits under the Asylum Seekers' Benefits Act (AsylbLG)\footnote{Upon arrival in Germany, asylum seekers initially obtain benefits according to the AsylbLG. Should their asylum procedure determine them to be eligible for protection, they generally receive unrestricted access to the labour market after nine months (or six months for minors residing in the household), at which point they may begin to receive UB~II benefits instead of those provided under the AsylbLG.} and only recently transitioned to UB~II after their status as asylum seekers was confirmed, the take-up rate for these households is inherently close to 100~percent. Consequently, including refugee households in the sample would potentially bias the baseline take-up estimation. Furthermore, as refugee households only recently arrived in Germany, they lack a UB~II receipt and earned income history in the PASS-ADIAB data, which would thus result in their exclusion from the take-up estimations that also control for potential long-term factors of benefit take-up. However, with only 2,086 household-wave observations, the proportion of excluded refugee households is relatively low, approximately 1.8~percent of the initial sample. Finally, the primary reason for dropping observations belonging to the second class of observation exclusions is the requirement that take-up is defined conditional on the simulated eligibility for UB~II. Consequently, 33,105 household-wave observations (29~percent of the initial sample) simulated as ineligible for UB~II are dropped at this stage. The calculation of the (non-)take-up rates is based on the remaining 28,998 household-wave observations. Table~\ref{tab:Admin_Spells_Sample} indicates that out of the 28,998 observed UB~II take-up decisions, 5,055 resulted in non-take-up according to the administrative PASS-ADIAB data, leading to an average unweighted NTR of 17.4~percent across all waves. Table~\ref{tab:NTRTable_uncorr_corr} reveals that the weighted NTR based on this administratively corrected receipt is substantially greater than the unweighted rate, reaching 32.7~percent, which is close to the estimates obtained by \textcite{Bru21}, who also use the PASS dataset.

The third category of exclusions from the data is due to the requirements of the take-up estimations. First, since the head of the household is regarded as responsible for the UB~II take-up decision, and typical take-up estimations in the literature incorporate socio-demographic characteristics of the household head as explanatory variables, only those households with a consistent head of household across all observed waves in the PASS are retained. Finally, household-wave observations with missing observations in the explanatory variables, as well as households with no available administrative data in PASS-ADIAB, must be dropped from the estimation sample.\footnote{Among the 1,779 household-wave observations lacking administrative data in PASS-ADIAB, approximately two thirds declined to consent to data linkage, while one third was excluded due to the absence of a successful match.} Of these three exclusions, the requirement of a consistent household head and the availability of the (baseline) explanatory variables are of minor importance, together removing 1,956 household-wave observations and yielding a baseline-specification sample of 27,042 household-wave observations. All further exclusions concern the linked PASS-ADIAB administrative histories from which the lagged long-term factors are constructed, as shown in the final three rows of Table~\ref{tab:Admin_Spells_Sample}. A household enters a more demanding specification only if the additional administrative history that specification requires is available for it. Specification~M1 adds the lagged UB~II receipt history and therefore retains only households linked to the administrative records, leaving 25,263 household-wave observations. Specification~M2 additionally requires the lagged real equivalent earned income, which is unavailable for a small number of further households, leaving 24,516. Specification~M3 additionally requires the income-shock and income-volatility terms, leaving 24,486 household-wave observations.

The resulting panel is unbalanced. In the full specification~M3, the 24{,}486 household-wave observations come from 8{,}753 distinct households. Table~\ref{tab:PanelBalance} shows that the number of survey waves in which a household is observed ranges from one to thirteen, with a median of two: about 42~percent of households are observed in a single wave and only a small minority in many waves, reflecting the well-known difficulty of retaining low-income households in panel surveys. The unbalancedness of the \emph{survey} panel does not, however, by itself leave the benefit and income \emph{histories} unobserved. Because these lagged long-term factors are reconstructed from the linked PASS-ADIAB records rather than from repeated survey participation (Section~\ref{subsubsec:DataAdiab}), they can be recovered even for a household that appears in only a single survey wave, and irrespective of gaps in its survey participation, provided the household is linked to the administrative records and the relevant history is actually recorded there.

\revisedfloat{\begin{table}[!ht]
\footnotesize
\centering
\begin{threeparttable}
\caption{Panel structure of the estimation sample}
\phantomsection\label{tab:PanelBalance}
\renewcommand{\arraystretch}{1.1}

\begin{tabular}{crr}
\hline\hline
Waves observed & Households & Percent \\
\hline
1 & 3,694 & 42.2 \\
2 & 1,795 & 20.5 \\
3 & 988 & 11.3 \\
4 & 647 & 7.4 \\
5 & 455 & 5.2 \\
6 & 335 & 3.8 \\
7 & 256 & 2.9 \\
8 & 196 & 2.2 \\
9 & 129 & 1.5 \\
10 & 109 & 1.2 \\
11 & 66 & 0.8 \\
12 & 51 & 0.6 \\
13 & 32 & 0.4 \\
\hline
Total & 8,753 & 100.0 \\
\hline
\end{tabular}

\begin{tablenotes}
\item \scriptsize \textsc{Note. ---} Distribution of the number of survey waves in which each household is observed in the (fully specified) estimation sample of 24,486 household-wave observations from 8,753 distinct households. The panel is unbalanced (mean 2.8 waves, median 2, maximum 13). Source: PASS 0620 v1, PASS-ADIAB 7520 v1, own calculations.
\end{tablenotes}

\end{threeparttable}
\end{table}
 }

\subsection{Benefit simulation}
\label{subsec:MethBenefitSimulation}

The simulation of income taxes, social security contributions, child benefits, and the means-tested benefits HB, SCB, and, most importantly, UB~II uses version v0.7.0 of the ``German Taxes and Transfers Simulator'' (GETTSIM), an open-source static tax-benefit microsimulation model developed by several German universities and economic research institutions.\footnote{GETTSIM models the key elements of the German tax and benefit system, such as income tax, social security contributions, child benefits, and the most important means-tested benefits UB~II, HB, and SCB. The model is implemented in Python, with the model code, the documentation, and further information on using GETTSIM available online. See \href{https://gettsim.readthedocs.io/en/v0.7.0}{https://gettsim.readthedocs.io/en/v0.7.0}.}  
Unlike most other tax-transfer simulators for Germany, GETTSIM is not tailored to a specific dataset but can be adapted for various datasets. To the best of my knowledge, this study is the first application of GETTSIM to PASS data. Users of GETTSIM must prepare their database to meet GETTSIM's input requirements. The input variables and their definitions for GETTSIM are detailed in the online documentation.\footnote{See \href{https://gettsim.readthedocs.io/en/v0.7.0/gettsim_objects/input\_variables.html}{https://gettsim.readthedocs.io/en/v0.7.0/gettsim\_objects/input\_variables.html}.} Typically, not all listed input variables in the documentation are required for a simulation; only those pertinent to the specific research question need to be provided by the user. This is feasible because GETTSIM does not automatically simulate the entire tax and transfer system but focuses solely on the policy targets defined by the user, using only the input variables needed to compute the selected targets. 

Adapting the PASS data for GETTSIM is relatively straightforward, as most of the input variables required are collected directly in the PASS questionnaire. However, assumptions are needed for some of the input variables. For example, the rent exclusive of heating costs is needed to calculate the entitlement for HB, but PASS only asks for the rent including heating costs. Here I assume a constant heating cost share of 19~percent, which is based on my evaluations of the German Socio-Economic Panel (SOEP). Another example is the wealth of the household, which is needed for the eligibility tests of UB~II, HB, and SCB. In the PASS, assets are only surveyed in the form of discrete asset classes. For each household, I assume the respective class mean value for the simulations.

I implement two adjustments to the GETTSIM code to enhance its accuracy in simulating UB~II entitlements.
Firstly, when calculating the amount of UB~II entitlement in the application process, rent reimbursement is capped at an upper limit, which depends on the local rent level and the size of the household. However, GETTSIM v0.7.0 imposes an overly strict and time-constant upper limit of 10 euros per square metre for the recognised rent. Thus, the accommodation costs included in the UB~II entitlement are typically underestimated when compared to the actual housing expenses of UB~II recipients, as reported by the Federal Labour Agency's statistics. I am therefore increasing the upper limit for the recognised rent to 15 euros per square metre, since this setting improves the fit of the simulated entitlements to the reported entitlements in the PASS.
Secondly, according to the legal regulations, individuals qualify for HB only when the household's income is adequate to meet the household's needs as specified by UB~II. Nevertheless, HB regulations provide the administering offices with a degree of discretion, allowing for HB eligibility if the considered income plus the notional HB entitlement covers at least 80~percent of the UB~II need. As this ``80~percent rule'' is not included in GETTSIM v0.7.0, I implemented it for the simulations.

\subsection{Quality of the benefit simulation}
\label{subsec:SimQuality}

Because the entire analysis rests on simulated eligibility and entitlement, it is important to establish that GETTSIM reproduces the relevant features of the data before turning to the results. Appendix~\ref{sec:AppSimQual} provides a detailed assessment. Its main points are summarised here.

First, the distribution of disposable income implied by the simulation closely tracks the distribution reported by households, both for the full sample and for households simulated as ineligible (Figure~\ref{fig:income_distribution_all} and Table~\ref{tab:SimVsRepInc}). For households simulated as eligible the simulated distribution is somewhat more compressed than the reported one, consistent with the rigidity of the statutory rules for the maximum recognised rent. For these households the median and mean of simulated and reported disposable income are nonetheless close, and the simulated and reported UB~II entitlement distributions share the same bimodal shape (Figure~\ref{fig:UBIIentitlements_kdensity_simrep}). This pattern is independently corroborated by \textcite{Bruckmeier2026}, who evaluates GETTSIM~v0.7.0 against individual administrative records of UB~II receipt (roughly 1.4~million person-month observations for 2017--2018). Because these records contain the benefit amounts actually \emph{paid out}, her assessment validates exactly the dimension that my survey data cannot: at the individual level the simulated entitlements track the recorded ones closely, with the simulated mean benefit lying within about two percent of the recorded mean. As in my data, the simulated distribution is somewhat more compressed than the recorded one, a feature she connects to the statutory treatment of the recognised rent, the same rigidity I point to for my eligible sample. Consistent with the discussion below, she also finds that the fit is tightest near the centre of the distribution and weaker for households with more complex circumstances, where small modelling and measurement errors matter most. Because her evaluation largely abstracts from error in the underlying inputs, it isolates the accuracy of the simulation model itself. In my survey-based application, measurement error in the reported inputs, such as underreporting of earned income, adds a further source of discrepancy, in line with the account of beta errors below.

Second, a natural summary measure of simulation quality is the ``type II'' or ``beta-error'' rate (BER), defined as the share of households that report receiving UB~II but are simulated to be ineligible, expressed relative to all reporting households \parencite{Bar10b,Fri07}. Averaged over 2008--2020, the weighted BER is 6.6~percent (Table~\ref{tab:NTRTable_uncorr_corr}), which is low by the standards of the literature\footnote{For example, using the same definition and also based on the German Socio-Economic Panel, \textcite{Fri07} report a beta-error rate of 12.6~percent, while \textcite{Har19} reports rates between 14 and 20~percent across waves (17.4~percent in her baseline specification).} and points to a relatively accurate simulation.\footnote{In terms of unweighted household-wave observations, 1{,}755 of the 23{,}834 household-wave observations with reported receipt are simulated ineligible, which implies an unweighted BER of 7.4~percent.}

Beta errors can arise for several reasons. A mismatch in periodicity between the simulation and the monthly benefit decision is unlikely to be the main source here: the income and wealth information entering the simulation, the self-reported benefit receipt, and the administrative receipt used to correct it all refer to the same reference point, namely the month of the interview. The more relevant reasons are measurement error and misreporting. Eligibility for UB~II rests on a detailed assessment of household income, wealth and housing costs. In particular, wealth is only observed in broad brackets rather than as an exact amount, so the statutory asset test cannot be reproduced precisely. Small errors in reported income, recognised rent or disregarded wealth can therefore push a genuinely eligible household with a small entitlement to a simulated entitlement of zero. Consistent with this, beta-error households tend to report comparatively low UB~II amounts (Figure~\ref{fig:UBIIentitlements_kdensity_simrep}) and are thus concentrated near the eligibility margin where such errors matter most. A further source is ``benefit confusion'', whereby respondents report UB~II while in fact receiving a different means-tested benefit (or none). For these reasons the BER is best read as an upper bound on the genuine simulation error, as it also absorbs administrative errors in the original eligibility assessment and false survey responses \parencite{Bru21}. Because take-up is defined conditional on simulated eligibility, beta-error households do not enter the take-up regressions. They are retained only for the assessment of simulation quality.

\section{Results}
\label{sec:Results}

\subsection{Simulated rates of non-take-up}
\label{subsec:ResultsNTR}

Table~\ref{tab:NTRTable_uncorr_corr} presents the simulated weighted non-take-up rates (NTR) for UB~II benefits across each PASS wave from 2008 through 2020, with the average NTR displayed in the last row. The results are based on the selected sample of 28,998 household-wave observations simulated as eligible for UB~II, as shown in Table~\ref{tab:Admin_Spells_Sample}. The column ``NTR self-rep.'' shows NTRs based on the self-reported benefit receipt of PASS households.
I find an average weighted self-reported NTR of 36.1~percent, which is at the lower end of the estimated NTRs for UB~II reported in the literature. However, due to the extensive sample selection, the comparability to previous findings in the literature is limited. Nevertheless, the NTRs found here are roughly in line with the result from \textcite{Bru21}, who also use PASS data. 

The column ``NTR admin.'' provides the NTR after replacing the self-reported receipt with the administrative PASS-ADIAB records, which corrects both over- and underreporting. The third column reports the difference (in percentage points) between the administrative and self-reported NTRs. Column ``NTR admin.'' shows that the administrative average NTR is 32.7~percent, 3.3~percentage points lower than the self-reported average NTR. The effect of the correction is also in line with \textcite{Bru21}.\footnote{Reporting both measures is informative in its own right: the gap between them quantifies the extent of benefit misreporting in the survey, predominantly underreporting, and is the reason the administrative correction lowers the estimated NTR.} The last two columns show the self-reported and administrative ``beta-error'' rate (BER); see Section~\ref{subsec:SimQuality} for the definition and a detailed discussion. I find an average BER of 6.6~percent (self-reported) and 6.3~percent (administrative), consistent with the relatively accurate simulation discussed in Section~\ref{subsec:SimQuality}.\footnote{
Generally, there exists an inverse relationship between the NTR and the BER: A more stringent benefits simulation results in fewer households being simulated as eligible, which, c.p., tends to decrease the NTR while increasing the BER \parencite[][]{Fri07}.}

\begin{table}[!ht]
\footnotesize
\centering
\resizebox{1.0\textwidth}{!}{\begin{threeparttable}
\caption{Weighted non-take-up and beta error rates (self-reported/administratively corrected)}
\phantomsection\label{tab:NTRTable_uncorr_corr}
\renewcommand{\arraystretch}{1.2}

\begin{tabular}{@{\hspace*{1.8em}}lrrrrrrrrrrrrrr}
\hline\hline
\cline{1-6}
\multicolumn{1}{c}{} &
  \multicolumn{1}{r}{NTR self-rep. (\%)} &
  \multicolumn{1}{r}{NTR admin. (\%)} &
  \multicolumn{1}{r}{(2)-(1) (pp)} &
  \multicolumn{1}{r}{BER self-rep. (\%)} &
  \multicolumn{1}{r}{BER admin. (\%)} \\
\cline{1-6}
\multicolumn{1}{l}{2008} &
  \multicolumn{1}{r}{34.9} &
  \multicolumn{1}{r}{29.3} &
  \multicolumn{1}{r}{-5.5} &
  \multicolumn{1}{r}{5.4} &
  \multicolumn{1}{r}{5.0} \\
\multicolumn{1}{l}{2009} &
  \multicolumn{1}{r}{32.8} &
  \multicolumn{1}{r}{28.4} &
  \multicolumn{1}{r}{-4.4} &
  \multicolumn{1}{r}{7.1} &
  \multicolumn{1}{r}{6.7} \\
\multicolumn{1}{l}{2010} &
  \multicolumn{1}{r}{32.5} &
  \multicolumn{1}{r}{27.5} &
  \multicolumn{1}{r}{-4.9} &
  \multicolumn{1}{r}{5.3} &
  \multicolumn{1}{r}{4.9} \\
\multicolumn{1}{l}{2011} &
  \multicolumn{1}{r}{33.1} &
  \multicolumn{1}{r}{28.8} &
  \multicolumn{1}{r}{-4.3} &
  \multicolumn{1}{r}{4.7} &
  \multicolumn{1}{r}{4.4} \\
\multicolumn{1}{l}{2012} &
  \multicolumn{1}{r}{37.0} &
  \multicolumn{1}{r}{34.5} &
  \multicolumn{1}{r}{-2.5} &
  \multicolumn{1}{r}{5.8} &
  \multicolumn{1}{r}{5.6} \\
\multicolumn{1}{l}{2013} &
  \multicolumn{1}{r}{36.7} &
  \multicolumn{1}{r}{33.2} &
  \multicolumn{1}{r}{-3.4} &
  \multicolumn{1}{r}{5.3} &
  \multicolumn{1}{r}{5.1} \\
\multicolumn{1}{l}{2014} &
  \multicolumn{1}{r}{35.5} &
  \multicolumn{1}{r}{33.7} &
  \multicolumn{1}{r}{-1.9} &
  \multicolumn{1}{r}{7.4} &
  \multicolumn{1}{r}{7.2} \\
\multicolumn{1}{l}{2015} &
  \multicolumn{1}{r}{35.5} &
  \multicolumn{1}{r}{34.9} &
  \multicolumn{1}{r}{-0.6} &
  \multicolumn{1}{r}{9.5} &
  \multicolumn{1}{r}{9.5} \\
\multicolumn{1}{l}{2016} &
  \multicolumn{1}{r}{36.1} &
  \multicolumn{1}{r}{33.3} &
  \multicolumn{1}{r}{-2.8} &
  \multicolumn{1}{r}{8.1} &
  \multicolumn{1}{r}{7.8} \\
\multicolumn{1}{l}{2017} &
  \multicolumn{1}{r}{42.6} &
  \multicolumn{1}{r}{37.8} &
  \multicolumn{1}{r}{-4.8} &
  \multicolumn{1}{r}{6.7} &
  \multicolumn{1}{r}{6.2} \\
\multicolumn{1}{l}{2018} &
  \multicolumn{1}{r}{40.4} &
  \multicolumn{1}{r}{38.6} &
  \multicolumn{1}{r}{-1.8} &
  \multicolumn{1}{r}{8.3} &
  \multicolumn{1}{r}{8.1} \\
\multicolumn{1}{l}{2019} &
  \multicolumn{1}{r}{39.7} &
  \multicolumn{1}{r}{38.2} &
  \multicolumn{1}{r}{-1.5} &
  \multicolumn{1}{r}{7.7} &
  \multicolumn{1}{r}{7.6} \\
\multicolumn{1}{l}{2020} &
  \multicolumn{1}{r}{41.5} &
  \multicolumn{1}{r}{39.2} &
  \multicolumn{1}{r}{-2.3} &
  \multicolumn{1}{r}{6.8} &
  \multicolumn{1}{r}{6.6} \\
\cline{1-6}
\multicolumn{1}{l}{Total} &
  \multicolumn{1}{r}{36.1} &
  \multicolumn{1}{r}{32.7} &
  \multicolumn{1}{r}{-3.3} &
  \multicolumn{1}{r}{6.6} &
  \multicolumn{1}{r}{6.3} \\
\cline{1-6}
\hline
\end{tabular}

\begin{tablenotes}
\item \scriptsize \textsc{Note. ---} NTR = non-take-up rate, self-rep. = reported receipt, admin. = reported receipt corrected with information from administrative data, BER = beta error rate, pp = percentage points. Weighted results. Source: PASS 0620 v1, PASS-ADIAB 7520 v1, GETTSIM v0.7.0, own calculations.
\end{tablenotes}

\end{threeparttable}
} \end{table}
 
\begin{figure}[p]
\caption{Simulated non-take-up of UB~II: through time and by PASS subsample}
   \label{fig:NTR_beta_combined}

   \begin{subfigure}{\textwidth}
      \centering
      \caption{Non-take-up \& ``beta error'' through time}
      \label{fig:NTR_beta}
      \includegraphics[width=1\textwidth]{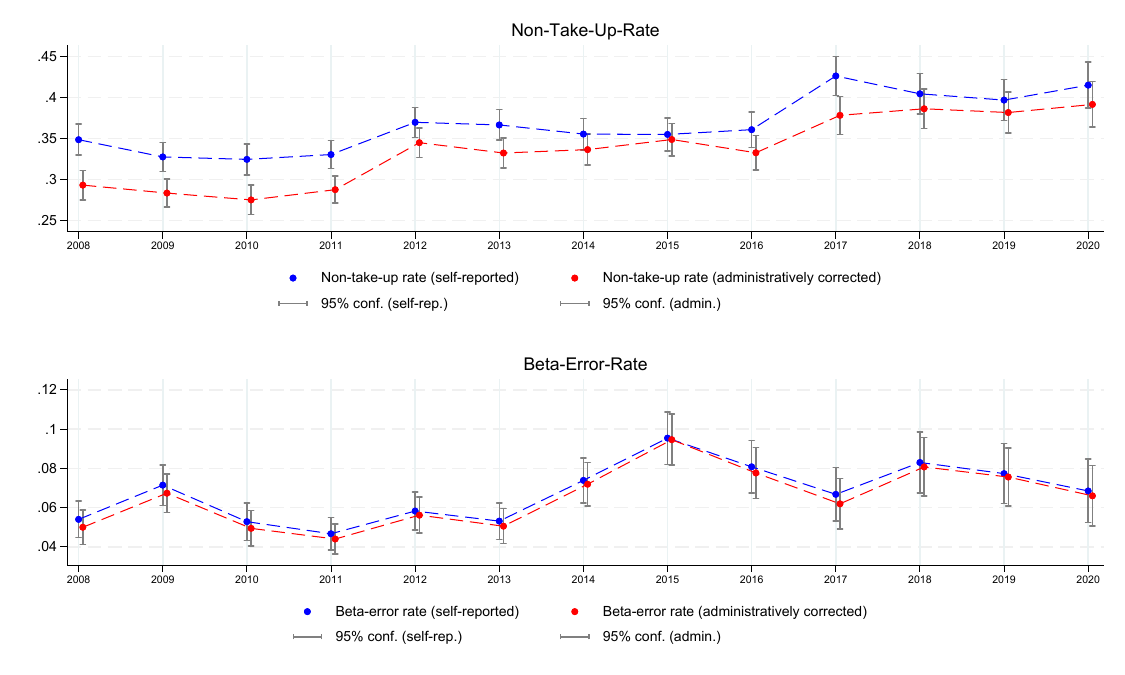}
   \end{subfigure}

   \vspace{0.5cm}

   \begin{subfigure}{\textwidth}
      \centering
      \caption{Non-take-up rate differentiated by PASS subsample}
      \label{fig:NTR_admin_microm}
      \includegraphics[width=1\textwidth]{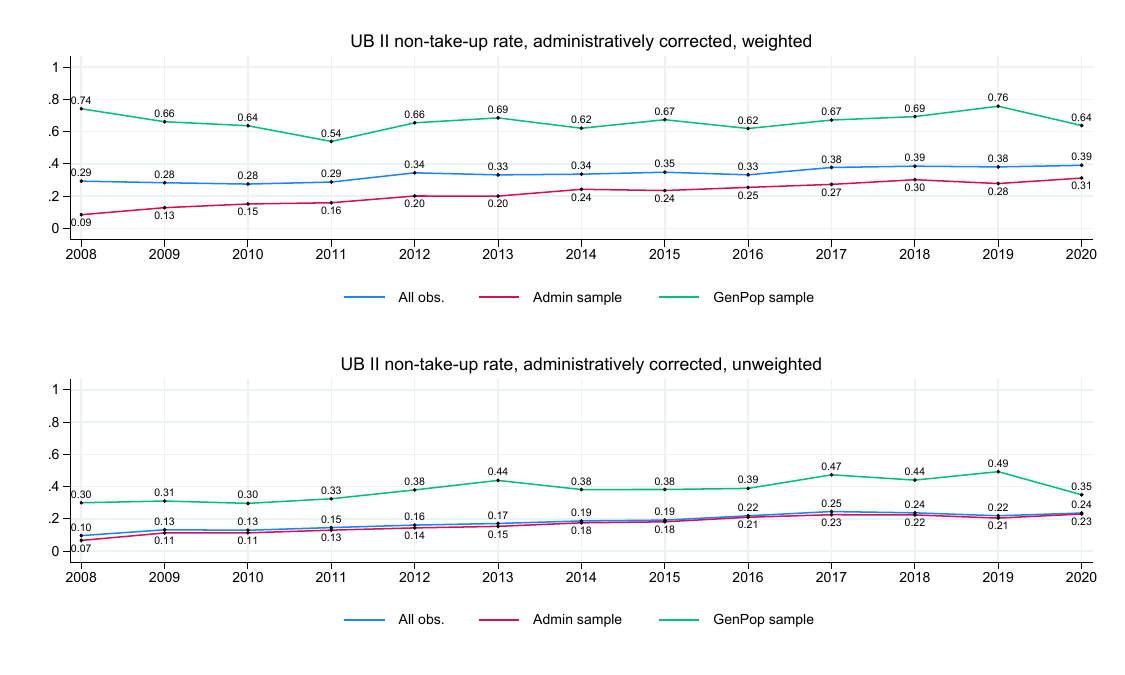}
   \end{subfigure}

   \vspace{-0.3cm}
   \floatfoot{ \scriptsize \textsc{Note. ---} \textbf{(a)} The upper panel shows the self-reported (in blue) and administratively corrected (in red) non-take-up rate over time. The lower panel shows the corresponding beta-error rates (share of households simulated ineligible to UB~II with a self-reported/administratively corrected benefit receipt). \textbf{(b)} NTRs for PASS subsamples \emph{Admin} (illustrated in red) and \emph{GenPop} (illustrated in green). The blue line indicates the mean NTR of both subsamples collectively. The upper panel displays weighted results, while the lower panel shows unweighted results. Source: Own calculations, GETTSIM v0.7.0, PASS 0620 v1, PASS-ADIAB 7520 v1.}
\end{figure}

The upper part of Figure~\ref{fig:NTR_beta} shows an upward trend in the administrative NTR over time. This trend is partly a composition effect that arises from the way the \emph{Admin} subsample is drawn. The \emph{Admin} subsample is built from repeated stock draws of current UB~II recipients from BA administrative records (each July, starting in 2006), with every subsequent wave adding only newly-entered recipient households. A household was thus a current recipient at the time of its own draw. As the panel ages, a growing fraction of the \emph{Admin} households sampled in earlier waves are former recipients who no longer claim. Among those who remain simulated as eligible, an increasing share are non-claimants, so the measured NTR drifts upward. The same sampling design accounts for the large and persistent level difference between the two subsamples in Figure~\ref{fig:NTR_admin_microm}: because the \emph{Admin} sample conditions on current receipt, its non-take-up rate lies far below that of the \emph{GenPop} sample, which is drawn from the general population and is not conditioned on past receipt. Since the \emph{GenPop} weights are typically much larger than the \emph{Admin} weights, weighting is essential for the descriptive NTRs to represent the eligible population.\footnote{For a discussion of NTRs by subgroup, see Appendix~\ref{sec:AppNTR_subgroup}.} The lower part of Figure~\ref{fig:NTR_beta} shows the development of the self-reported (blue) and administrative (red) BER over time. In contrast to the NTR, no evident trend is observable in the BER.

The administrative NTR also varies across space. Non-take-up is somewhat higher in West Germany (34~percent) than in East Germany (29~percent), and it varies non-monotonically with municipality size, from about 27~percent in mid-sized core cities to about 38~percent in peripheral districts with at least 50,000 inhabitants (see the regional breakdowns in Table~\ref{tab:NTRTable_subgroup_uncorr_corr} in Appendix~\ref{sec:AppNTR_subgroup}). A natural channel for such geographic differences is that knowledge about benefit programmes, eligibility and how to claim, is not uniform across space but diffuses through local social networks. \textcite{ChettyFriedmanSaez2013} document this dynamic for the US Earned Income Tax Credit: knowledge of the credit's incentive structure varies markedly across areas and is most strongly predicted by the local density of recipients and the availability of advice infrastructure. The same diffusion mechanism would predict lower non-take-up where claimants are more concentrated and advice is more accessible, broadly consistent with the lower NTR in larger core cities, although the pattern across municipality-size classes is non-monotonic and the channel itself cannot be tested with these data. More generally, because of the sample selection underlying the estimation sample (Section~\ref{subsubsec:DataSampSel}), the NTRs reported in this section should not be read as unbiased estimates for the general population.\footnote{Beyond the East/West distinction and the municipality-size classes, the scientific-use file of the PASS also identifies the federal state (\textit{Bundesland}) of residence. Tables~\ref{tab:NTRTable_subgroup_uncorr_corr} and~\ref{tab:covariate_means} (Appendices~\ref{sec:AppNTR_subgroup} and~\ref{sec:AppCovMeans}) therefore additionally report NTRs and covariate means by federal state. The point estimates of the NTR vary noticeably across states (roughly from 23 to 43~percent). However, a joint test of equality of the administrative NTR across federal states is not rejected ($p=0.29$), so this cross-state variation cannot be distinguished from sampling noise. Consistent with this, the estimation-sample shares of take-up and non-take-up households differ only marginally within most federal states, and replacing the East/West indicator with a full set of federal-state dummies in the take-up regressions leaves the results essentially unchanged while yielding individually insignificant state coefficients. I therefore retain the parsimonious East/West indicator in the take-up estimations. Finer geographic units (like districts) are not available in the data, so a district-fixed-effects analysis is not feasible with these data. The East/West indicator and the municipality-size dummies are included as controls in all take-up regressions.}

\subsection{Take-up estimations}
\label{subsec:ResultsTUEsts}

The next section discusses the estimation outcomes for a baseline specification of UB~II take-up, as conventionally found in the literature. Building on this baseline specification, the potential long-term determinants of take-up outlined in Section~\ref{subsec:MethTakeUp} are incrementally added to the model in the following sections. Covariate means are shown in Table~\ref{tab:covariate_means} in Appendix~\ref{sec:AppCovMeans}.

\subsubsection{Baseline specification}
\label{subsubsec:ResultsTUEstsBase}

\begin{figure}[!htb]
\caption{Take-up regression: baseline specification\label{fig:est_base}}
\includegraphics[width=1\textwidth]{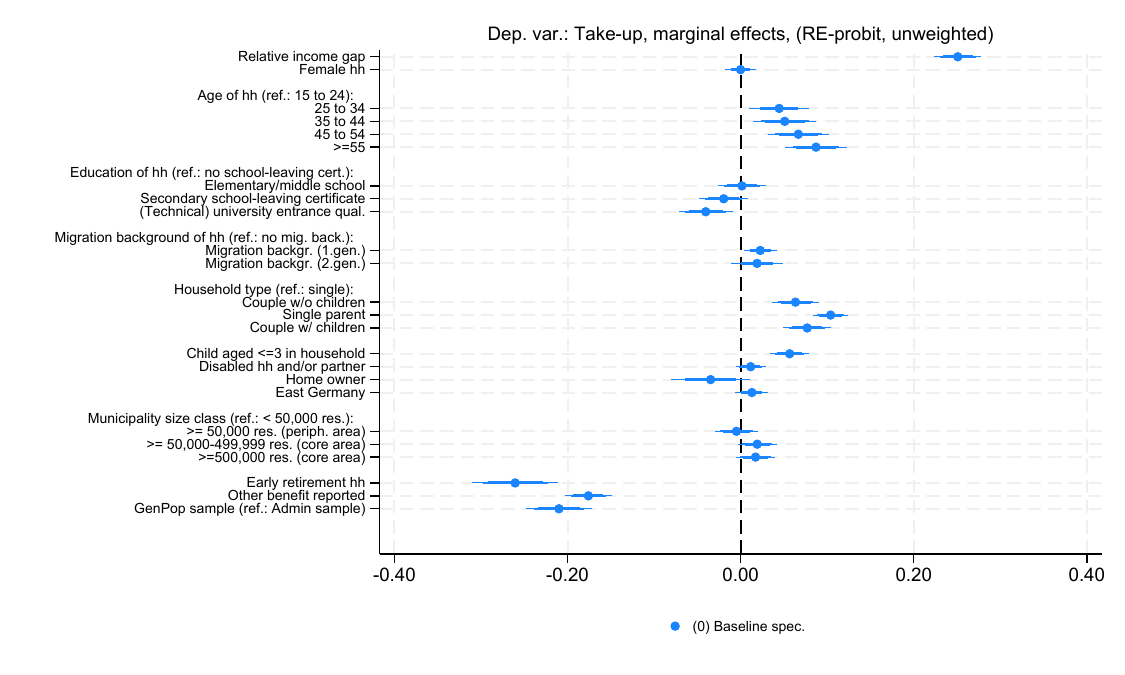}
\vspace{-1cm}\floatfoot{\scriptsize \textsc{Note. ---} Random effects probit model with binary indicator for UB~II take-up (take-up = 1) as dependent variable. Dots mark the point estimate of the respective marginal effect, while the corresponding horizontal lines represent the confidence intervals at the 99/95/90~percent level (thin/medium/thick line). The model contains time fixed effects. Unweighted results. hh=head of household. Source: Own calculations, GETTSIM v0.7.0, PASS 0620 v1, PASS-ADIAB 7520 v1.}
\end{figure}

Figure~\ref{fig:est_base} illustrates the marginal effects for the baseline specification of the UB~II take-up model.\footnote{Table~\ref{tab:covariate_means} compares the means of the covariates in the baseline specification for take-up and non-take-up households.} In both this and subsequent models, the dependent variable is a binary indicator denoting the administrative receipt of UB~II benefits. All estimations are conditioned on households eligible for UB~II benefits as per the GETTSIM simulation in the corresponding PASS wave. My preferred estimator takes advantage of the panel structure of the PASS to estimate a probit model with random effects (RE probit), enabling control for unobserved heterogeneity at the household level. Additionally, this and all subsequent take-up models include time fixed effects.\footnote{This and subsequent estimations are unweighted. As a robustness check, a comparison with weighted estimations is shown in Figure \ref{fig:est_base_weighted_vs_unweighted}. The comparison demonstrates that while the point estimates are generally similar, confidence intervals are substantially larger for the weighted estimations.} An estimation table of the baseline and the extended models is presented in Table~\ref{tab:TUests_w0_e0_excl11001} in Appendix~\ref{sec:AppEstResults}.

\begin{figure}[!htb]
\caption{Relative income gap vs. take-up rate\label{fig:income_gap_fit}}
\includegraphics[width=1\textwidth]{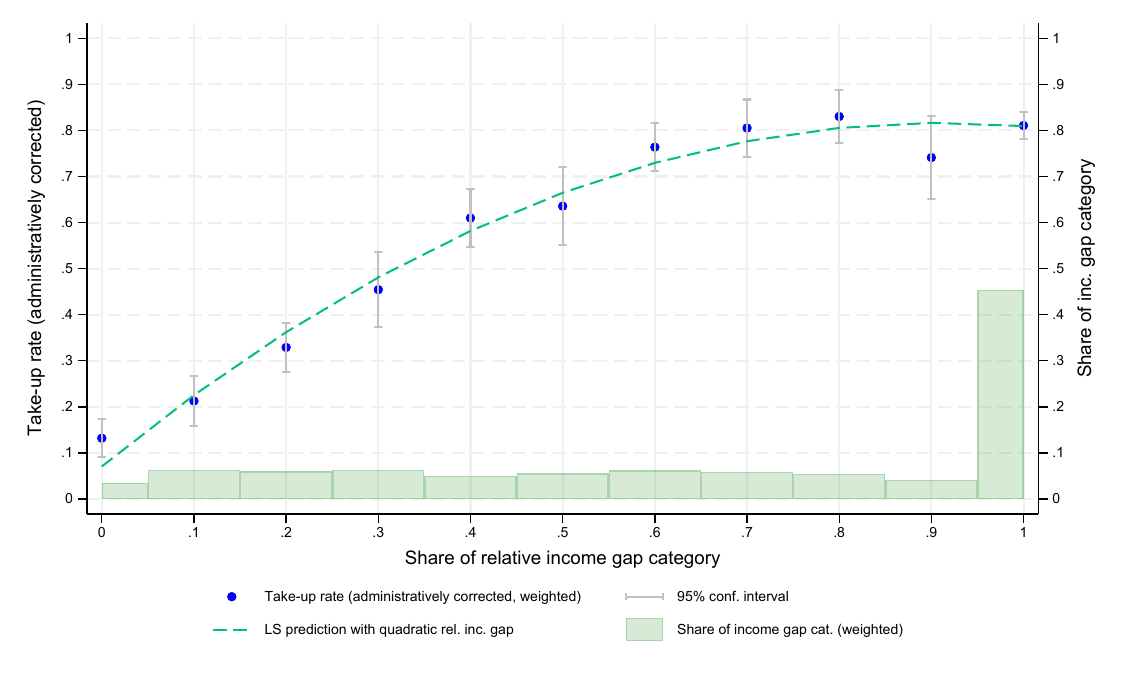}
\vspace{-1cm}\floatfoot{\scriptsize \textsc{Note. ---} The simulated income gap is discretised in eleven categories. The blue points show the results of a weighted linear regression of the administrative take-up indicator on a constant for each category. The dashed green line shows the fit of a quadratic polynomial in the relative income categories. The green vertical bars show the weighted share of household-wave observations in each category. Source: Own calculations, GETTSIM v0.7.0, PASS 0620 v1, PASS-ADIAB 7520 v1.}   
\end{figure}

The central indicator for measuring the utility from receiving benefits is the simulated level of entitlement to UB~II benefits, which enters the estimates as the relative income gap defined in Section~\ref{subsec:MethTakeUp}. Normalising the entitlement to the $[0, 1]$ interval also simplifies the comparison of its marginal effects with those of the categorical variables in the estimation model. The income gap enters as a quadratic polynomial, motivated by the non-linear relationship between the probability of claiming and the income gap shown in Figure~\ref{fig:income_gap_fit}\footnote{The figure also shows the distribution of the relative income gap. Approximately 45~percent of the households observed have an income gap of (almost) 100~percent, which means that these households have no other income that is deducted from their UB~II entitlement.}. For the basic specification without long-term or dynamic factors, the average marginal effect of the relative income gap is 0.25, i.e., a one-percentage-point increase in the income gap is associated with a 0.25-percentage-point higher probability of claiming, on average. The estimate is of a similar order of magnitude to comparable estimates in the literature.

Regarding the categorical covariates in the baseline specification, Figure~\ref{fig:est_base} shows that the probability of claiming benefits increases with increasing age and with a decreasing highest level of education. Household type plays a substantial role, with couples, both with and without children, as well as single parents being more likely to claim benefits than single households. The presence of young children (aged 3 or younger) within the household is positively associated with take-up. Other factors such as sex, migration background, disability status, home ownership, and regional characteristics are insignificant or borderline insignificant, with small marginal effects. The three categorical covariates displayed last in Figure~\ref{fig:est_base} show notably strong negative associations with take-up, with marginal effects of around $-20$~percentage points, which was expected given the extremely high administrative NTRs between 66 and 88~percent for the corresponding subgroups (``GenPop samples'', ``early retirement hh'', and ``other benefit reported''), as shown in Table~\ref{tab:NTRTable_subgroup_uncorr_corr} in Appendix~\ref{sec:AppNTR_subgroup}.

\subsubsection{Extended specification M1: Adding UB~II benefit history}
\label{subsubsec:ResultsTUEstsUBIIexp}

The PASS-ADIAB records the UB~II receipts as spell data with daily accuracy. I aggregate this data at the level of days of receipt per calendar quarter and measure time relative to the calendar quarter of the respective household interview in the PASS data. As a measure of a household's experience with UB~II, I use the shares of days per year on which UB~II benefits were received in each of the three years (twelve quarters) prior to the quarter of the respective PASS interview.\footnote{In the subsequent model extensions, the same timing convention is used to construct measures of income potential, income shocks and income volatility from the PASS-ADIAB data.}

Table~\ref{tab:covariate_means} in Appendix~\ref{sec:AppCovMeans} shows highly significant differences in the share of days with UB~II receipt between take-up and non-take-up households, with an average share over all three years of 62~percent for take-up and only 14~percent for non-take-up households. The difference in shares is largest for the year (four quarters) immediately preceding the current quarter of the interview. This indicates that households presently receiving the benefit are likely those who have claimed it in the past year(s) too, whereas those not currently claiming were either not eligible previously or chose not to claim the benefit even though they were entitled.

Table~\ref{tab:TakeUpTypes} also demonstrates a relatively stable longitudinal take-up behaviour of households. The table presents all distinct households within the sample that were eligible at least once during the period from 2008 to 2020. I categorise these households into those that never claim UB~II (``never takers''), those that claim the benefit intermittently but not consistently (``sometimes takers''), and those that consistently claim the benefit (``always takers''). Examining the final two columns with weighted results indicates that over 87~percent consistently either always claim or never claim the benefit, while less than 13~percent alter their take-up behaviour at least once throughout my observation period. The large difference between the unweighted and weighted shares for the ``never takers'' and ``always takers'' reflects that ``never takers'' are strongly overrepresented in the \emph{GenPop} subsample, where average sample weights are also notably higher than in the \emph{Admin} subsample of the PASS data.

\begin{table}[!ht]
\footnotesize
\centering
\resizebox{0.8\textwidth}{!}{\begin{threeparttable}
\caption{Households by intertemporal take-up types}
\phantomsection\label{tab:TakeUpTypes}
\renewcommand{\arraystretch}{1.2}

\begin{tabular}{@{\hspace*{1.8em}}lrrrrrrrrrrrrrr}
\hline\hline
\cline{1-5}
\multicolumn{1}{r}{} &
  \multicolumn{2}{c}{Housholds} &
  \multicolumn{2}{c}{Households (weighted)} \\
\cline{1-5}
\multicolumn{1}{l}{Take-up type} &
  \multicolumn{1}{r}{} &
  \multicolumn{1}{r}{} &
  \multicolumn{1}{r}{} &
  \multicolumn{1}{r}{} \\
\multicolumn{1}{l}{\hspace{1em}``Never taker''} &
  \multicolumn{1}{r}{1,482} &
  \multicolumn{1}{r}{(14.2\%)} &
  \multicolumn{1}{r}{5,560,874} &
  \multicolumn{1}{r}{(36.7\%)} \\
\multicolumn{1}{l}{\hspace{1em}``Sometimes taker''} &
  \multicolumn{1}{r}{1,274} &
  \multicolumn{1}{r}{(12.2\%)} &
  \multicolumn{1}{r}{1,917,016} &
  \multicolumn{1}{r}{(12.7\%)} \\
\multicolumn{1}{l}{\hspace{1em}``Always taker''} &
  \multicolumn{1}{r}{7,700} &
  \multicolumn{1}{r}{(73.6\%)} &
  \multicolumn{1}{r}{7,673,938} &
  \multicolumn{1}{r}{(50.6\%)} \\
\cline{1-5}
\multicolumn{1}{l}{Total} &
  \multicolumn{1}{r}{10,456} &
  \multicolumn{1}{r}{(100.0\%)} &
  \multicolumn{1}{r}{15,151,828} &
  \multicolumn{1}{r}{(100.0\%)} \\
\cline{1-5}
\hline
\end{tabular}

\begin{tablenotes}
\item \scriptsize \textsc{Note. ---} The sample contains households that are simulated as entitled at least once between 2008 and 2020. Take-up types are defined by either never claiming the benefit (\textquotedblleft never taker\textquotedblright), claiming the benefit at least once but not always (\textquotedblleft sometimes taker\textquotedblright), and always claiming the benefit (\textquotedblleft always taker\textquotedblright). Source: PASS 0620 v1, PASS-ADIAB 7520 v1, GETTSIM v0.7.0, own calculations.
\end{tablenotes}

\end{threeparttable}
} \end{table}
 
\begin{figure}[!htb]
\caption{Take-up regression: adding UB~II experience (M1) \label{fig:est_M1}}
\includegraphics[width=1\textwidth]{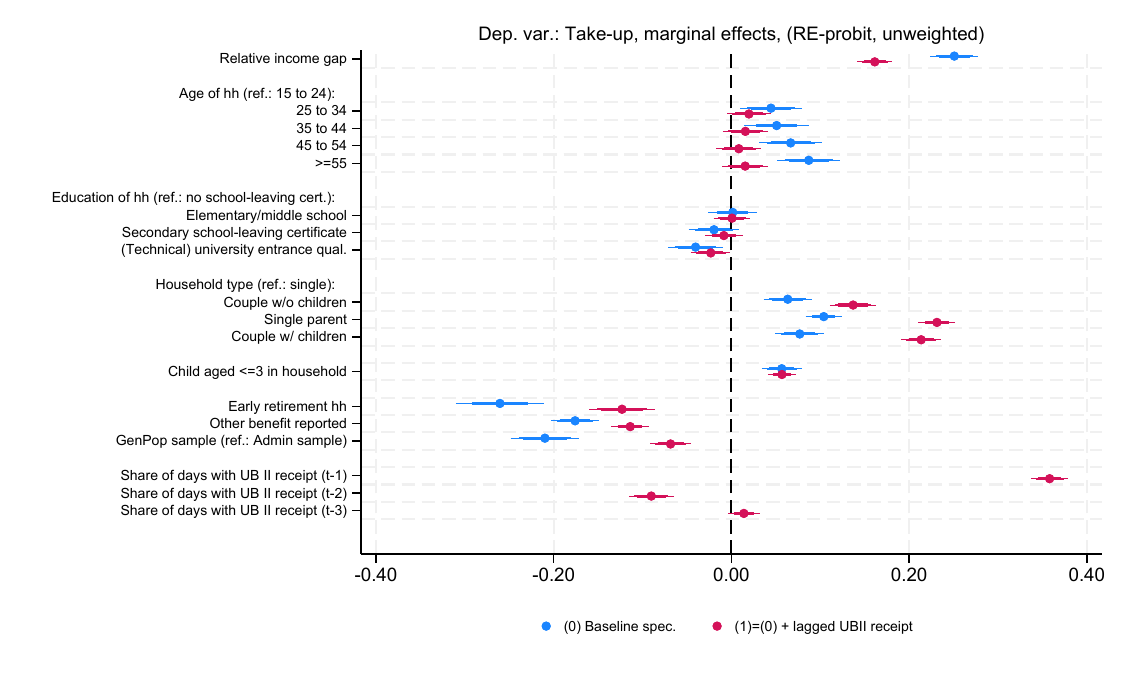}
\vspace{-1cm}\floatfoot{\scriptsize \textsc{Note. ---} Random effects probit model with binary indicator for UB~II take-up (take-up = 1) as dependent variable. Dots mark the point estimate of the respective marginal effect, while the corresponding horizontal lines represent the confidence intervals at the 99/95/90~percent level (thin/medium/thick line). The model contains time fixed effects. Unweighted results. hh=head of household. The baseline specification (extended specification) is shown in blue (red). Insignificant effects are not included in the figure to improve readability. See Table~\ref{tab:TUests_w0_e0_excl11001} in Appendix~\ref{sec:AppEstResults} for the full specification. Source: Own calculations, GETTSIM v0.7.0, PASS 0620 v1, PASS-ADIAB 7520 v1.}   
\end{figure}

Figure \ref{fig:est_M1} shows that benefit receipt in the previous year, in particular, has a large marginal effect on the current probability of take-up: the average marginal effect of the lagged share of days in receipt in $t-1$ is 0.36, so a one-percentage-point increase in that share is associated with a 0.36-percentage-point higher probability of current take-up. The share in $t-2$ enters with a smaller, negative marginal effect ($-0.09$), while the $t-3$ effect, although statistically significant, is negligible in size ($0.01$).\footnote{The negative $t-2$ coefficient is the partial association holding the other lags fixed: among households with the same receipt share last year, a higher share two years earlier indicates a declining benefit trajectory, and such households are somewhat less likely to be claiming now. The three lagged shares are in any case strongly collinear (their pairwise correlations range from about $0.64$ to $0.84$ in the estimation sample), so the split across individual years is not the object of interest; the interpretable quantity is their sum.} The long-run association of a permanent one-percentage-point increase in the share of days in receipt is therefore about $0.28$ percentage points.\footnote{The long-run association corresponds to the sum of the marginal effects of the three lagged shares.} Overall, the results are consistent with the hypothesis that previous experience with UB~II is associated with lower current take-up costs and hence a higher propensity to claim the benefit.

Adding the variables related to UB~II experience into the model substantially changes the marginal effects of certain regressors present in the baseline specification (see also Table~\ref{tab:TUests_w0_e0_excl11001}). For example, the marginal effect of the relative income gap falls from 0.25 to 0.16, with most age group and education indicators losing statistical significance. The marginal effects of the group indicators ``GenPop samples'', ``early retirement hh'', and ``other benefit reported'' are also considerably smaller than their values in the baseline specification. Finally, while the absolute values of most marginal effects in the baseline specification decrease upon including the UB~II experience indicators, the marginal effects of household type indicators increase to two or three times their baseline specification values.

\subsubsection{Extended specification M2: Adding earned income history}
\label{subsubsec:ResultsTUEstsEarnedIncomeHist}

The income potential of households is measured using the real equivalised earned income of the three years (12 quarters) prior to the calendar quarter of the current PASS household interview.\footnote{Income is measured in 2020 prices and equivalised using the new OECD scale.} Table~\ref{tab:covariate_means} indicates that the mean earned income of non-take-up households is four to five times higher than the earned income of take-up households in the three years prior to the current entitlement.

\begin{figure}[!htb]
\caption{Take-up regression: adding earned income history (M2)\label{fig:est_M2}}
\includegraphics[width=1\textwidth]{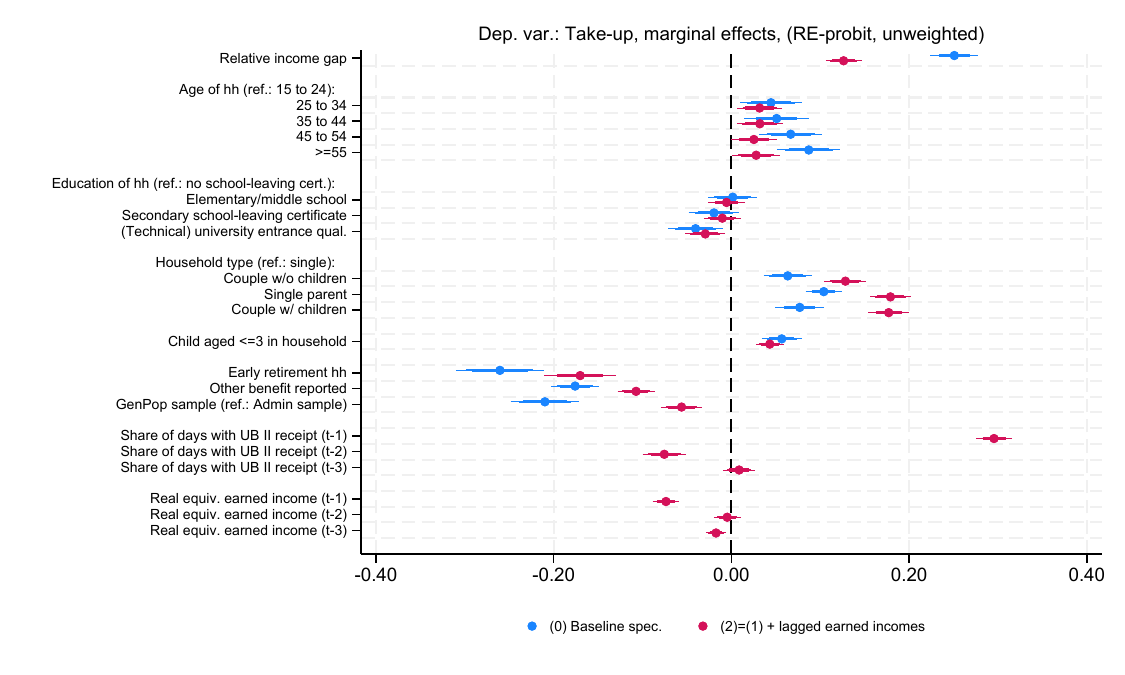}
\vspace{-1cm}\floatfoot{\scriptsize \textsc{Note. ---} Random effects probit model with binary indicator for UB~II take-up (take-up = 1) as dependent variable. Dots mark the point estimate of the respective marginal effect, while the corresponding horizontal lines represent the confidence intervals at the 99/95/90~percent level (thin/medium/thick line). Real equivalised earned income is measured in 1,000 euros/month in 2020 prices using the new OECD scale. The model contains time fixed effects. Unweighted results. hh=head of household. The baseline specification (extended specification) is shown in blue (red). Insignificant effects are not included in the figure to improve readability. See Table~\ref{tab:TUests_w0_e0_excl11001} in Appendix~\ref{sec:AppEstResults} for the full specification. Source: Own calculations, GETTSIM v0.7.0, PASS 0620 v1, PASS-ADIAB 7520 v1.}   
\end{figure}

Figure \ref{fig:est_M2} shows that the probability of claiming UB~II at the time of the current interview is significantly lower for households with higher previously earned income, although the magnitude of the association is small: a permanent increase in equivalised earned income by 1,000 euros per month is associated with a roughly 10-percentage-point lower take-up probability in the long run.\footnote{The sum of the three lagged income marginal effects is $-0.096$; see Table~\ref{tab:TUests_w0_e0_excl11001}.} This long-run association is accounted for primarily by income earned in the year immediately preceding the interview ($t-1$). The marginal effects of income two and three years earlier are small. Two readings are consistent with this pattern. On the one hand, the results provide little evidence that households weigh their longer-run earning potential when deciding whether to claim, and instead suggest that recent income is more closely associated with current claiming, consistent with a role for financial reserves that allow households to postpone an application. On the other hand, income in $t-1$ may simply be a good proxy for a household's income potential, with incomes in $t-2$ and $t-3$ adding little explanatory power once $t-1$ is accounted for. The data cannot sharply distinguish these interpretations. Both point to recent earned income, rather than the longer income history, as the relevant margin.

\subsubsection{Extended specification M3: Adding income volatility}
\label{subsubsec:ResultsTUEstsIncomeVolatility}

The final model extension incorporates income shocks and income volatility. Income shocks are measured as the year-over-year change in earned income over the four quarters preceding the current calendar quarter of the PASS interview, standardised by the current household UB~II needs (comprising regular benefits and housing expenses), so that one unit corresponds to a change in earned income equal to those needs.\footnote{For example, a value of $0.5$ denotes an income gain worth half of the household's UB~II needs.} Income volatility is measured by the within-household standard deviation of quarterly earned income across the 12 quarters leading up to the current interview. Table~\ref{tab:covariate_means} shows a statistically highly significant difference in the mean income shocks between households that take up benefits and those that do not; the shock is positive for the former and negative for the latter. Furthermore, the standard deviation of within-household quarterly earned income is nearly double for non-take-up households in comparison with take-up households.

\begin{figure}[!htb]
\caption{Take-up regression: adding income volatility (M3)\label{fig:est_M3}}
\includegraphics[width=1\textwidth]{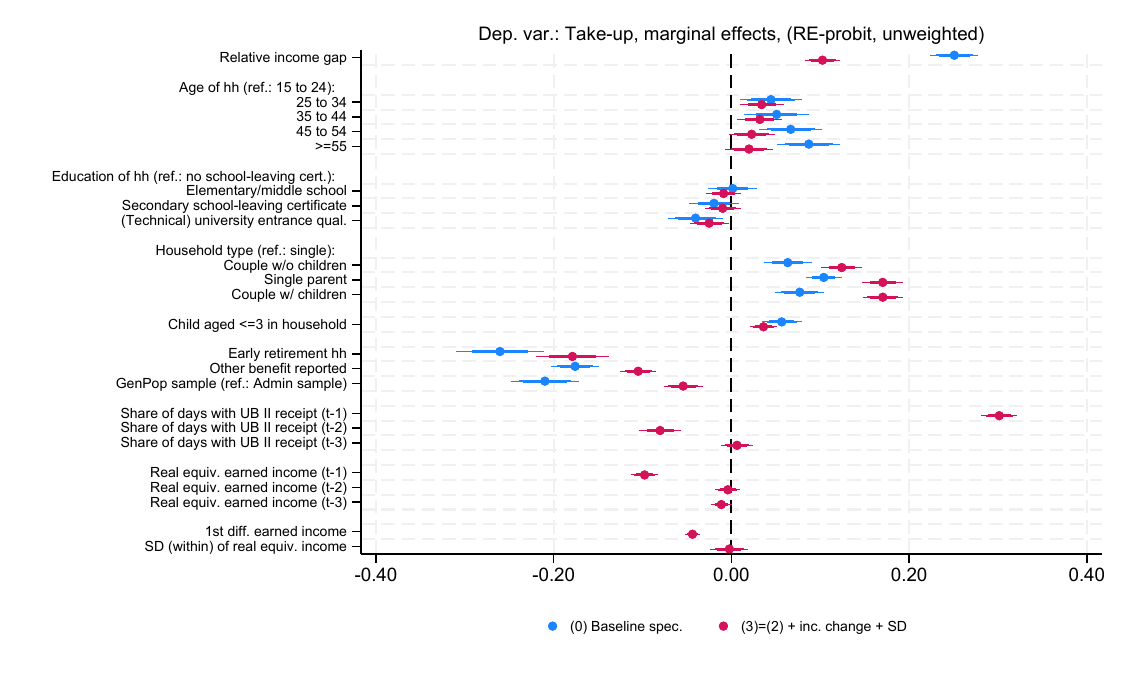}
\vspace{-1cm}\floatfoot{\scriptsize \textsc{Note. ---} Random effects probit model with binary indicator for UB~II take-up (take-up = 1) as dependent variable. Dots mark the point estimate of the respective marginal effect, while the corresponding horizontal lines represent the confidence intervals at the 99/95/90~percent level (thin/medium/thick line). Real equivalised earned income is measured in 1,000 euros/month in 2020 prices using the new OECD scale. 1st diff. earned income=year-over-year change in income preceding the quarter of the PASS interview. SD (within) of real equiv. income=within-household standard deviation of real equivalised earned income in the twelve quarters preceding the current interview. The model contains time fixed effects. Unweighted results. hh=head of household. The baseline specification (extended specification) is shown in blue (red). Insignificant effects are not included in the figure to improve readability. See Table~\ref{tab:TUests_w0_e0_excl11001} in Appendix~\ref{sec:AppEstResults} for the full specification. Source: Own calculations, GETTSIM v0.7.0, PASS 0620 v1, PASS-ADIAB 7520 v1.}   
\end{figure}

Figure \ref{fig:est_M3} shows that the income shock has a statistically significant marginal effect on take-up: the probability of claiming is lower when income has recently risen. The association is small in magnitude: a positive shock of one unit is associated with a $4.4$-percentage-point lower probability of claiming. Finally, I do not find a statistically significant association between the variation in income and the probability of claiming the benefit (Table~\ref{tab:VolatilityRobustness}). This holds whether volatility is measured in levels or via the inverse-hyperbolic-sine transform.\footnote{A level-based standard deviation tends to be mechanically larger for higher-income households, which is why it is the long-term factor most strongly correlated with income potential (Table~\ref{tab:LongTermFactors_corr}). Therefore, I re-estimated M3 using the standard deviation of the inverse-hyperbolic-sine (asinh) transform of quarterly earnings, which behaves like a logarithm but is defined at zero earnings. With the asinh-transform, the volatility term remains statistically insignificant, while the income shock retains its sign, magnitude, and significance (Table~\ref{tab:VolatilityRobustness}).}

These findings suggest that take-up shows no systematic association with long-run earned income variations. Instead, taken together with the results for the \emph{level} of previous earned income from model~M2, they indicate that current take-up is associated mainly with short-term (negative) income shocks. This may reflect the reality that households entitled to UB~II typically possess limited savings, which limits their ability to ``smooth over'' an adverse income shock.\footnote{As a robustness check, Figure \ref{fig:est_M3_weighted_vs_unweighted} shows the marginal effects for the weighted extended estimation model M3 compared to the unweighted estimations. Although the point estimators of the marginal effects in the weighted estimation do not differ significantly from the unweighted results, the former exhibit markedly wider confidence intervals. Nevertheless, the statistical significance of the additional variables in models M1 to M3 remains largely unchanged when compared to the unweighted estimations.} This pattern echoes \textcite{Paukkeri2018}, who finds for Finnish social assistance and housing benefits that eligible non-claimants are disproportionately households undergoing a transient income drop rather than households with persistently variable incomes. The panel structure used here separates the recent income shock, the level of past earned income, and long-run volatility, and it is the first two, rather than volatility, that co-vary with take-up.

\section{Conclusions}
\label{sec:Conclusions}

This paper has examined how long-term factors are associated with the take-up of Unemployment Benefit~II (UB~II) in Germany, using longitudinal data that integrate survey responses with administrative records. The analysis provides evidence on how prior benefit receipt, income potential, income volatility, and income shocks relate to the likelihood of claiming UB~II, dimensions largely absent from previous empirical models.

The PASS-ADIAB administrative data for the period 2008 to 2020 enables tracking household histories beyond survey participation and correcting for misreported benefit receipt. To the best of my knowledge, this is also the first application of the open-source tax-and-transfer simulation model GETTSIM to the PASS dataset.

The results indicate that a longer history of prior benefit receipt is associated with a higher probability of current take-up, consistent with reduced informational and transactional barriers for households familiar with the system. Higher household income potential, measured as average past earned income, is associated with a lower likelihood of claiming UB~II, consistent with households with stronger labour market prospects perceiving lower net utility from participation. Recent positive income shocks are associated with lower take-up, whereas long-term income volatility has no significant association, suggesting that take-up is associated mainly with immediate financial changes rather than longer-run fluctuations. Incorporating these long-term and dynamic factors changes the marginal effects of conventional explanatory variables, in some cases substantially, indicating that standard empirical models that omit such factors may be misspecified and may give a misleading account of the correlates of household claiming behaviour. These findings extend the literature by demonstrating the relevance of long-term factors for take-up within the German basic safety net.

Several caveats apply. As set out in Section~\ref{subsec:MethTakeUp}, the estimates describe robust conditional associations rather than causal effects: households that differ in their benefit history or income trajectory are also likely to differ along unobserved dimensions that the model cannot separate from the long-term factors, and the random-effects estimator assumes that the household-specific effect is uncorrelated with the regressors. Establishing causal effects would require exogenous variation in benefit history or income dynamics, which the present data do not provide. The contribution here is to document patterns that existing, purely contemporaneous models miss.

With this caveat in mind, the findings suggest several tentative implications for policy. First, the strong positive association between past and current take-up points to substantial persistence in claiming, which suggests that lowering the fixed informational and transactional costs of a first or renewed claim -- for example through simplified applications, automatic continuation of awards, or targeted outreach to newly eligible households -- could have effects that compound over time. Second, the sensitivity of take-up to recent income changes, rather than to longer-run income levels or volatility, suggests that providing timely information at moments of income loss (job separation, reductions in working hours, or the exhaustion of unemployment-insurance entitlement) may reach households precisely when eligibility arises. Third, the pronounced non-take-up among households that report receiving upstream benefits, housing benefit and the supplementary child benefit, points to ``benefit confusion'' at the boundary between programmes, where clearer cross-programme communication could reduce inadvertent non-take-up. Because the underlying estimates are correlational, these implications are best read as hypotheses for targeted evaluation rather than as established policy effects.

\textbf\printbibliography \clearpage

\begin{appendix}

\begin{landscape}
\section{Sample Selection}\label{sec:AppSampSel}

\setcounter{table}{0}\setcounter{figure}{0}
\renewcommand\thetable{\thesection\arabic{table}} 
\renewcommand\thefigure{\thesection\arabic{figure}} 

\updatedfloat{\begin{table}[!ht]
\footnotesize
\centering
\resizebox{1.0\textwidth}{!}{\begin{threeparttable}
\caption{Sample selection: household-wave observations}
\phantomsection\label{tab:Admin_Spells_Sample}
\renewcommand{\arraystretch}{1.0}

\begin{tabular}{@{\hspace*{1.8em}}lrrrrrrrrrrrrrr}
\hline\hline
\cline{1-15}
\multicolumn{1}{c}{} &
  \multicolumn{14}{c}{Year} \\
\multicolumn{1}{c}{} &
  \multicolumn{1}{r}{2008} &
  \multicolumn{1}{r}{2009} &
  \multicolumn{1}{r}{2010} &
  \multicolumn{1}{r}{2011} &
  \multicolumn{1}{r}{2012} &
  \multicolumn{1}{r}{2013} &
  \multicolumn{1}{r}{2014} &
  \multicolumn{1}{r}{2015} &
  \multicolumn{1}{r}{2016} &
  \multicolumn{1}{r}{2017} &
  \multicolumn{1}{r}{2018} &
  \multicolumn{1}{r}{2019} &
  \multicolumn{1}{r}{2020} &
  \multicolumn{1}{r}{Total} \\
\cline{1-15}
\multicolumn{1}{l}{Initial sample} &
  \multicolumn{1}{r}{8,183} &
  \multicolumn{1}{r}{9,249} &
  \multicolumn{1}{r}{7,662} &
  \multicolumn{1}{r}{10,092} &
  \multicolumn{1}{r}{9,373} &
  \multicolumn{1}{r}{9,363} &
  \multicolumn{1}{r}{8,845} &
  \multicolumn{1}{r}{8,758} &
  \multicolumn{1}{r}{8,364} &
  \multicolumn{1}{r}{9,155} &
  \multicolumn{1}{r}{8,943} &
  \multicolumn{1}{r}{8,347} &
  \multicolumn{1}{r}{7,507} &
  \multicolumn{1}{r}{113,841} \\
\cline{1-15}
\multicolumn{1}{l}{i) Exclusions due to data requirements of the benefit simulation} &
  \multicolumn{1}{r}{} &
  \multicolumn{1}{r}{} &
  \multicolumn{1}{r}{} &
  \multicolumn{1}{r}{} &
  \multicolumn{1}{r}{} &
  \multicolumn{1}{r}{} &
  \multicolumn{1}{r}{} &
  \multicolumn{1}{r}{} &
  \multicolumn{1}{r}{} &
  \multicolumn{1}{r}{} &
  \multicolumn{1}{r}{} &
  \multicolumn{1}{r}{} &
  \multicolumn{1}{r}{} &
  \multicolumn{1}{r}{} \\
\multicolumn{1}{l}{\hspace{1em}Excl. hh with missing interviews} &
  \multicolumn{1}{r}{6,132} &
  \multicolumn{1}{r}{6,620} &
  \multicolumn{1}{r}{6,126} &
  \multicolumn{1}{r}{7,898} &
  \multicolumn{1}{r}{7,553} &
  \multicolumn{1}{r}{7,502} &
  \multicolumn{1}{r}{7,098} &
  \multicolumn{1}{r}{7,030} &
  \multicolumn{1}{r}{6,586} &
  \multicolumn{1}{r}{6,904} &
  \multicolumn{1}{r}{6,841} &
  \multicolumn{1}{r}{6,279} &
  \multicolumn{1}{r}{5,422} &
  \multicolumn{1}{r}{87,991} \\
\multicolumn{1}{l}{\hspace{1em}Excl. hh with $>$1 community of needs} &
  \multicolumn{1}{r}{5,932} &
  \multicolumn{1}{r}{6,428} &
  \multicolumn{1}{r}{5,909} &
  \multicolumn{1}{r}{7,583} &
  \multicolumn{1}{r}{7,211} &
  \multicolumn{1}{r}{7,189} &
  \multicolumn{1}{r}{6,805} &
  \multicolumn{1}{r}{6,748} &
  \multicolumn{1}{r}{6,297} &
  \multicolumn{1}{r}{6,612} &
  \multicolumn{1}{r}{6,564} &
  \multicolumn{1}{r}{6,060} &
  \multicolumn{1}{r}{5,272} &
  \multicolumn{1}{r}{84,610} \\
\multicolumn{1}{l}{\hspace{1em}Excl. non-core family hh} &
  \multicolumn{1}{r}{5,836} &
  \multicolumn{1}{r}{6,343} &
  \multicolumn{1}{r}{5,823} &
  \multicolumn{1}{r}{7,473} &
  \multicolumn{1}{r}{7,122} &
  \multicolumn{1}{r}{7,095} &
  \multicolumn{1}{r}{6,731} &
  \multicolumn{1}{r}{6,674} &
  \multicolumn{1}{r}{6,232} &
  \multicolumn{1}{r}{6,541} &
  \multicolumn{1}{r}{6,497} &
  \multicolumn{1}{r}{6,016} &
  \multicolumn{1}{r}{5,243} &
  \multicolumn{1}{r}{83,626} \\
\multicolumn{1}{l}{\hspace{1em}Excl. hh with inconsistent wage information} &
  \multicolumn{1}{r}{5,552} &
  \multicolumn{1}{r}{6,074} &
  \multicolumn{1}{r}{5,690} &
  \multicolumn{1}{r}{7,323} &
  \multicolumn{1}{r}{6,955} &
  \multicolumn{1}{r}{6,941} &
  \multicolumn{1}{r}{6,591} &
  \multicolumn{1}{r}{6,525} &
  \multicolumn{1}{r}{6,094} &
  \multicolumn{1}{r}{6,430} &
  \multicolumn{1}{r}{6,396} &
  \multicolumn{1}{r}{5,902} &
  \multicolumn{1}{r}{5,141} &
  \multicolumn{1}{r}{81,614} \\
\multicolumn{1}{l}{\hspace{1em}Excl. hh with inconsistent partner information} &
  \multicolumn{1}{r}{5,552} &
  \multicolumn{1}{r}{6,074} &
  \multicolumn{1}{r}{5,690} &
  \multicolumn{1}{r}{7,322} &
  \multicolumn{1}{r}{6,955} &
  \multicolumn{1}{r}{6,941} &
  \multicolumn{1}{r}{6,591} &
  \multicolumn{1}{r}{6,525} &
  \multicolumn{1}{r}{6,094} &
  \multicolumn{1}{r}{6,429} &
  \multicolumn{1}{r}{6,396} &
  \multicolumn{1}{r}{5,902} &
  \multicolumn{1}{r}{5,141} &
  \multicolumn{1}{r}{81,612} \\
\cline{1-15}
\multicolumn{1}{l}{ii) Exclusions due to the focus on UB II takeup} &
  \multicolumn{1}{r}{} &
  \multicolumn{1}{r}{} &
  \multicolumn{1}{r}{} &
  \multicolumn{1}{r}{} &
  \multicolumn{1}{r}{} &
  \multicolumn{1}{r}{} &
  \multicolumn{1}{r}{} &
  \multicolumn{1}{r}{} &
  \multicolumn{1}{r}{} &
  \multicolumn{1}{r}{} &
  \multicolumn{1}{r}{} &
  \multicolumn{1}{r}{} &
  \multicolumn{1}{r}{} &
  \multicolumn{1}{r}{} \\
\multicolumn{1}{l}{\hspace{1em}Excl. hh with pensioners} &
  \multicolumn{1}{r}{4,828} &
  \multicolumn{1}{r}{5,313} &
  \multicolumn{1}{r}{4,992} &
  \multicolumn{1}{r}{6,244} &
  \multicolumn{1}{r}{5,891} &
  \multicolumn{1}{r}{5,851} &
  \multicolumn{1}{r}{5,481} &
  \multicolumn{1}{r}{5,380} &
  \multicolumn{1}{r}{5,295} &
  \multicolumn{1}{r}{5,348} &
  \multicolumn{1}{r}{5,302} &
  \multicolumn{1}{r}{4,777} &
  \multicolumn{1}{r}{4,084} &
  \multicolumn{1}{r}{68,786} \\
\multicolumn{1}{l}{\hspace{1em}Excl. hh with head/partner ineligible to UB II (student, etc.)} &
  \multicolumn{1}{r}{4,566} &
  \multicolumn{1}{r}{5,030} &
  \multicolumn{1}{r}{4,713} &
  \multicolumn{1}{r}{5,859} &
  \multicolumn{1}{r}{5,530} &
  \multicolumn{1}{r}{5,489} &
  \multicolumn{1}{r}{5,109} &
  \multicolumn{1}{r}{4,994} &
  \multicolumn{1}{r}{4,895} &
  \multicolumn{1}{r}{4,993} &
  \multicolumn{1}{r}{4,902} &
  \multicolumn{1}{r}{4,389} &
  \multicolumn{1}{r}{3,720} &
  \multicolumn{1}{r}{64,189} \\
\multicolumn{1}{l}{\hspace{1em}Excl. hh from refugee samples} &
  \multicolumn{1}{r}{4,566} &
  \multicolumn{1}{r}{5,030} &
  \multicolumn{1}{r}{4,713} &
  \multicolumn{1}{r}{5,859} &
  \multicolumn{1}{r}{5,530} &
  \multicolumn{1}{r}{5,489} &
  \multicolumn{1}{r}{5,109} &
  \multicolumn{1}{r}{4,994} &
  \multicolumn{1}{r}{4,628} &
  \multicolumn{1}{r}{4,600} &
  \multicolumn{1}{r}{4,289} &
  \multicolumn{1}{r}{3,950} &
  \multicolumn{1}{r}{3,346} &
  \multicolumn{1}{r}{62,103} \\
\multicolumn{1}{l}{\hspace{1em}Excl. hh simulated ineligible to UB II} &
  \multicolumn{1}{r}{2,509} &
  \multicolumn{1}{r}{2,753} &
  \multicolumn{1}{r}{2,440} &
  \multicolumn{1}{r}{3,033} &
  \multicolumn{1}{r}{2,769} &
  \multicolumn{1}{r}{2,687} &
  \multicolumn{1}{r}{2,470} &
  \multicolumn{1}{r}{2,306} &
  \multicolumn{1}{r}{1,988} &
  \multicolumn{1}{r}{1,733} &
  \multicolumn{1}{r}{1,594} &
  \multicolumn{1}{r}{1,495} &
  \multicolumn{1}{r}{1,221} &
  \multicolumn{1}{r}{28,998} \\
\cline{1-15}
\multicolumn{1}{l}{Thereof: take-up of UB II} &
  \multicolumn{1}{r}{} &
  \multicolumn{1}{r}{} &
  \multicolumn{1}{r}{} &
  \multicolumn{1}{r}{} &
  \multicolumn{1}{r}{} &
  \multicolumn{1}{r}{} &
  \multicolumn{1}{r}{} &
  \multicolumn{1}{r}{} &
  \multicolumn{1}{r}{} &
  \multicolumn{1}{r}{} &
  \multicolumn{1}{r}{} &
  \multicolumn{1}{r}{} &
  \multicolumn{1}{r}{} &
  \multicolumn{1}{r}{} \\
\multicolumn{1}{l}{\hspace{1em}No} &
  \multicolumn{1}{r}{246} &
  \multicolumn{1}{r}{365} &
  \multicolumn{1}{r}{317} &
  \multicolumn{1}{r}{447} &
  \multicolumn{1}{r}{449} &
  \multicolumn{1}{r}{461} &
  \multicolumn{1}{r}{465} &
  \multicolumn{1}{r}{443} &
  \multicolumn{1}{r}{438} &
  \multicolumn{1}{r}{426} &
  \multicolumn{1}{r}{379} &
  \multicolumn{1}{r}{328} &
  \multicolumn{1}{r}{291} &
  \multicolumn{1}{r}{5,055} \\
\multicolumn{1}{l}{\hspace{1em}Yes} &
  \multicolumn{1}{r}{2,263} &
  \multicolumn{1}{r}{2,388} &
  \multicolumn{1}{r}{2,123} &
  \multicolumn{1}{r}{2,586} &
  \multicolumn{1}{r}{2,320} &
  \multicolumn{1}{r}{2,226} &
  \multicolumn{1}{r}{2,005} &
  \multicolumn{1}{r}{1,863} &
  \multicolumn{1}{r}{1,550} &
  \multicolumn{1}{r}{1,307} &
  \multicolumn{1}{r}{1,215} &
  \multicolumn{1}{r}{1,167} &
  \multicolumn{1}{r}{930} &
  \multicolumn{1}{r}{23,943} \\
\cline{1-15}
\multicolumn{1}{l}{iii) Exclusions due to the specification of the take-up models} &
  \multicolumn{1}{r}{} &
  \multicolumn{1}{r}{} &
  \multicolumn{1}{r}{} &
  \multicolumn{1}{r}{} &
  \multicolumn{1}{r}{} &
  \multicolumn{1}{r}{} &
  \multicolumn{1}{r}{} &
  \multicolumn{1}{r}{} &
  \multicolumn{1}{r}{} &
  \multicolumn{1}{r}{} &
  \multicolumn{1}{r}{} &
  \multicolumn{1}{r}{} &
  \multicolumn{1}{r}{} &
  \multicolumn{1}{r}{} \\
\multicolumn{1}{l}{\hspace{1em}Excl. hh with changing head of household} &
  \multicolumn{1}{r}{2,454} &
  \multicolumn{1}{r}{2,693} &
  \multicolumn{1}{r}{2,370} &
  \multicolumn{1}{r}{2,947} &
  \multicolumn{1}{r}{2,678} &
  \multicolumn{1}{r}{2,609} &
  \multicolumn{1}{r}{2,391} &
  \multicolumn{1}{r}{2,232} &
  \multicolumn{1}{r}{1,927} &
  \multicolumn{1}{r}{1,692} &
  \multicolumn{1}{r}{1,556} &
  \multicolumn{1}{r}{1,470} &
  \multicolumn{1}{r}{1,204} &
  \multicolumn{1}{r}{28,223} \\
\multicolumn{1}{l}{\hspace{1em}Excl. hh with missings in regressors (Baseline)} &
  \multicolumn{1}{r}{2,365} &
  \multicolumn{1}{r}{2,589} &
  \multicolumn{1}{r}{2,277} &
  \multicolumn{1}{r}{2,838} &
  \multicolumn{1}{r}{2,579} &
  \multicolumn{1}{r}{2,506} &
  \multicolumn{1}{r}{2,284} &
  \multicolumn{1}{r}{2,116} &
  \multicolumn{1}{r}{1,833} &
  \multicolumn{1}{r}{1,608} &
  \multicolumn{1}{r}{1,485} &
  \multicolumn{1}{r}{1,402} &
  \multicolumn{1}{r}{1,160} &
  \multicolumn{1}{r}{27,042} \\
\multicolumn{1}{l}{\hspace{1em}Excl. hh with missing receipt history (M1)} &
  \multicolumn{1}{r}{2,190} &
  \multicolumn{1}{r}{2,410} &
  \multicolumn{1}{r}{2,158} &
  \multicolumn{1}{r}{2,637} &
  \multicolumn{1}{r}{2,430} &
  \multicolumn{1}{r}{2,359} &
  \multicolumn{1}{r}{2,148} &
  \multicolumn{1}{r}{2,011} &
  \multicolumn{1}{r}{1,740} &
  \multicolumn{1}{r}{1,502} &
  \multicolumn{1}{r}{1,364} &
  \multicolumn{1}{r}{1,270} &
  \multicolumn{1}{r}{1,044} &
  \multicolumn{1}{r}{25,263} \\
\multicolumn{1}{l}{\hspace{1em}Excl. hh with missing lagged earned income (M2)} &
  \multicolumn{1}{r}{2,136} &
  \multicolumn{1}{r}{2,340} &
  \multicolumn{1}{r}{2,102} &
  \multicolumn{1}{r}{2,578} &
  \multicolumn{1}{r}{2,376} &
  \multicolumn{1}{r}{2,302} &
  \multicolumn{1}{r}{2,096} &
  \multicolumn{1}{r}{1,958} &
  \multicolumn{1}{r}{1,688} &
  \multicolumn{1}{r}{1,454} &
  \multicolumn{1}{r}{1,307} &
  \multicolumn{1}{r}{1,194} &
  \multicolumn{1}{r}{985} &
  \multicolumn{1}{r}{24,516} \\
\multicolumn{1}{l}{\hspace{1em}Excl. hh with missing income shock or volatility (M3)} &
  \multicolumn{1}{r}{2,131} &
  \multicolumn{1}{r}{2,334} &
  \multicolumn{1}{r}{2,097} &
  \multicolumn{1}{r}{2,572} &
  \multicolumn{1}{r}{2,372} &
  \multicolumn{1}{r}{2,299} &
  \multicolumn{1}{r}{2,095} &
  \multicolumn{1}{r}{1,958} &
  \multicolumn{1}{r}{1,688} &
  \multicolumn{1}{r}{1,454} &
  \multicolumn{1}{r}{1,307} &
  \multicolumn{1}{r}{1,194} &
  \multicolumn{1}{r}{985} &
  \multicolumn{1}{r}{24,486} \\
\cline{1-15}
\hline
\end{tabular}

\begin{tablenotes}
\item \scriptsize \textsc{Note. ---} UB II=unemployment benefit II, hh=household. Source: PASS 0620 v1, PASS-ADIAB 7520 v1, own calculations.
\end{tablenotes}

\end{threeparttable}
} \end{table}
 }
\end{landscape}

\clearpage
\section{Non-take-up rate by subgroup}\label{sec:AppNTR_subgroup}

\setcounter{table}{0}\setcounter{figure}{0}
\renewcommand\thetable{\thesection\arabic{table}} 
\renewcommand\thefigure{\thesection\arabic{figure}} 

Table~\ref{tab:NTRTable_subgroup_uncorr_corr} presents the simulated weighted group-specific UB~II NTRs averaged over the period 2008 to 2020. Column 1 details the self-reported NTRs, while column 2 displays the administrative NTRs, corrected for both underreporting and overreporting. The table highlights significant variability in NTRs across subgroups. Initially examining the self-reported rates, the highest non-take-up rate is observed in households with self-assessed early retirement, at 88.8~percent, whereas households with children aged up to three years exhibit the lowest benefit non-take-up rates at 17.5~percent. Although the NTRs by subgroup are fairly similar to those identified in other studies \parencite[see, e.g.,][]{Bru12}, comparability is limited due to the selection procedures outlined in Section~\ref{subsubsec:DataSampSel}, which render the estimation sample unrepresentative of the entire population.

The impact of correcting misreported UB~II receipts on the NTRs exhibits substantial variability across subgroups, as illustrated in Column 3 of Table~\ref{tab:NTRTable_subgroup_uncorr_corr}. After correction, the NTRs decrease across all subgroups, with reductions ranging from $-0.4$ to $-7.8$~percentage points. The largest decrease in the NTR occurs for households reporting receipt of upstream benefits (HB and/or SCB). This suggests that a significant part of the NTR for this group can be attributed to ``benefit confusion'' (see Section~\ref{subsec:SimQuality}). Additionally, the analysis reveals that the magnitude of correction effects diminishes in absolute terms as the age of the head of household increases and their education level decreases.

\begin{ThreePartTable}
\footnotesize
\centering
\renewcommand{\arraystretch}{1.0}

\begin{TableNotes}
\item \scriptsize \textsc{Note. ---} NTR = non-take-up rate, self-rep. = reported receipt, admin. = reported receipt corrected with information from administrative data. hh = head of household. Weighted results. Source: PASS 0620 v1, PASS-ADIAB 7520 v1, GETTSIM v0.7.0, own calculations.
\end{TableNotes}

\begin{longtable}{@{\hspace*{1.8em}}lrrr}
\caption{Non-take-up rates (self-reported/administratively corrected) by subgroups (weighted results)}\phantomsection\label{tab:NTRTable_subgroup_uncorr_corr}\\
\hline\hline
\multicolumn{1}{c}{} &
  \multicolumn{1}{r}{NTR self-rep. (\%)} &
  \multicolumn{1}{r}{NTR admin. (\%)} &
  \multicolumn{1}{r}{(2)-(1) (pp)} \\
\hline
\endfirsthead

\caption[]{(continued from previous page)}\\
\hline
\multicolumn{1}{c}{} &
  \multicolumn{1}{r}{NTR self-rep. (\%)} &
  \multicolumn{1}{r}{NTR admin. (\%)} &
  \multicolumn{1}{r}{(2)-(1) (pp)} \\
\hline
\endhead
\cmidrule{4-4}
\multicolumn{4}{r}{\textit{continued}}
\endfoot

\hline
\insertTableNotes\\
\endlastfoot

\multicolumn{1}{l}{Sex of hh (ref: male hh)} &
  \multicolumn{1}{r}{} &
  \multicolumn{1}{r}{} &
  \multicolumn{1}{r}{} \\
\multicolumn{1}{l}{\hspace{1em}Male hh} &
  \multicolumn{1}{r}{32.4} &
  \multicolumn{1}{r}{30.1} &
  \multicolumn{1}{r}{-2.3} \\
\multicolumn{1}{l}{\hspace{1em}Female hh} &
  \multicolumn{1}{r}{39.8} &
  \multicolumn{1}{r}{35.4} &
  \multicolumn{1}{r}{-4.4} \\
\multicolumn{1}{l}{Region of residence (ref.: West Germany)} &
  \multicolumn{1}{r}{} &
  \multicolumn{1}{r}{} &
  \multicolumn{1}{r}{} \\
\multicolumn{1}{l}{\hspace{1em}West Germany} &
  \multicolumn{1}{r}{37.8} &
  \multicolumn{1}{r}{34.3} &
  \multicolumn{1}{r}{-3.5} \\
\multicolumn{1}{l}{\hspace{1em}East Germany} &
  \multicolumn{1}{r}{32.3} &
  \multicolumn{1}{r}{29.3} &
  \multicolumn{1}{r}{-3.0} \\
\multicolumn{1}{l}{Home ownership} &
  \multicolumn{1}{r}{} &
  \multicolumn{1}{r}{} &
  \multicolumn{1}{r}{} \\
\multicolumn{1}{l}{\hspace{1em}No home owner} &
  \multicolumn{1}{r}{35.2} &
  \multicolumn{1}{r}{31.8} &
  \multicolumn{1}{r}{-3.4} \\
\multicolumn{1}{l}{\hspace{1em}Home owner} &
  \multicolumn{1}{r}{53.8} &
  \multicolumn{1}{r}{51.6} &
  \multicolumn{1}{r}{-2.2} \\
\multicolumn{1}{l}{Presence of child aged $<$=3 in household} &
  \multicolumn{1}{r}{} &
  \multicolumn{1}{r}{} &
  \multicolumn{1}{r}{} \\
\multicolumn{1}{l}{\hspace{1em}No child aged $<$=3 in household} &
  \multicolumn{1}{r}{37.5} &
  \multicolumn{1}{r}{34.3} &
  \multicolumn{1}{r}{-3.2} \\
\multicolumn{1}{l}{\hspace{1em}Child aged $<$=3 in household} &
  \multicolumn{1}{r}{17.5} &
  \multicolumn{1}{r}{12.8} &
  \multicolumn{1}{r}{-4.7} \\
\multicolumn{1}{l}{Disabled hh and/or partner} &
  \multicolumn{1}{r}{} &
  \multicolumn{1}{r}{} &
  \multicolumn{1}{r}{} \\
\multicolumn{1}{l}{\hspace{1em}Non-disabled hh and/or partner} &
  \multicolumn{1}{r}{33.0} &
  \multicolumn{1}{r}{29.1} &
  \multicolumn{1}{r}{-3.9} \\
\multicolumn{1}{l}{\hspace{1em}Disabled hh and/or partner} &
  \multicolumn{1}{r}{46.5} &
  \multicolumn{1}{r}{45.1} &
  \multicolumn{1}{r}{-1.3} \\
\multicolumn{1}{l}{Early retirement status} &
  \multicolumn{1}{r}{} &
  \multicolumn{1}{r}{} &
  \multicolumn{1}{r}{} \\
\multicolumn{1}{l}{\hspace{1em}No early retirement hh} &
  \multicolumn{1}{r}{30.2} &
  \multicolumn{1}{r}{26.6} &
  \multicolumn{1}{r}{-3.6} \\
\multicolumn{1}{l}{\hspace{1em}Early retirement hh} &
  \multicolumn{1}{r}{88.8} &
  \multicolumn{1}{r}{87.8} &
  \multicolumn{1}{r}{-1.0} \\
\multicolumn{1}{l}{Reported receipt of other benefit than UB II} &
  \multicolumn{1}{r}{} &
  \multicolumn{1}{r}{} &
  \multicolumn{1}{r}{} \\
\multicolumn{1}{l}{\hspace{1em}No other benefit reported} &
  \multicolumn{1}{r}{28.5} &
  \multicolumn{1}{r}{25.9} &
  \multicolumn{1}{r}{-2.6} \\
\multicolumn{1}{l}{\hspace{1em}Other benefit reported} &
  \multicolumn{1}{r}{80.8} &
  \multicolumn{1}{r}{73.0} &
  \multicolumn{1}{r}{-7.8} \\
\multicolumn{1}{l}{GenPop subsample (ref.: Admin sample)} &
  \multicolumn{1}{r}{} &
  \multicolumn{1}{r}{} &
  \multicolumn{1}{r}{} \\
\multicolumn{1}{l}{\hspace{1em}Admin sample} &
  \multicolumn{1}{r}{23.7} &
  \multicolumn{1}{r}{20.4} &
  \multicolumn{1}{r}{-3.3} \\
\multicolumn{1}{l}{\hspace{1em}GenPop sample (ref.: Admin sample)} &
  \multicolumn{1}{r}{68.8} &
  \multicolumn{1}{r}{65.5} &
  \multicolumn{1}{r}{-3.3} \\
\multicolumn{1}{l}{Migr. background of hh (ref.: no migr. backgr.)} &
  \multicolumn{1}{r}{} &
  \multicolumn{1}{r}{} &
  \multicolumn{1}{r}{} \\
\multicolumn{1}{l}{\hspace{1em}No migration backgr.} &
  \multicolumn{1}{r}{37.4} &
  \multicolumn{1}{r}{34.6} &
  \multicolumn{1}{r}{-2.9} \\
\multicolumn{1}{l}{\hspace{1em}Migration backgr. (1.gen.)} &
  \multicolumn{1}{r}{30.9} &
  \multicolumn{1}{r}{26.3} &
  \multicolumn{1}{r}{-4.6} \\
\multicolumn{1}{l}{\hspace{1em}Migration backgr. (2.gen.)} &
  \multicolumn{1}{r}{35.4} &
  \multicolumn{1}{r}{31.2} &
  \multicolumn{1}{r}{-4.1} \\
\multicolumn{1}{l}{Age of hh (ref.: 15-24)} &
  \multicolumn{1}{r}{} &
  \multicolumn{1}{r}{} &
  \multicolumn{1}{r}{} \\
\multicolumn{1}{l}{\hspace{1em}15 to 24} &
  \multicolumn{1}{r}{43.6} &
  \multicolumn{1}{r}{37.6} &
  \multicolumn{1}{r}{-6.0} \\
\multicolumn{1}{l}{\hspace{1em}25 to 34} &
  \multicolumn{1}{r}{29.4} &
  \multicolumn{1}{r}{25.1} &
  \multicolumn{1}{r}{-4.3} \\
\multicolumn{1}{l}{\hspace{1em}35 to 44} &
  \multicolumn{1}{r}{34.0} &
  \multicolumn{1}{r}{30.1} &
  \multicolumn{1}{r}{-3.9} \\
\multicolumn{1}{l}{\hspace{1em}45 to 54} &
  \multicolumn{1}{r}{39.4} &
  \multicolumn{1}{r}{35.8} &
  \multicolumn{1}{r}{-3.6} \\
\multicolumn{1}{l}{\hspace{1em}$>$=55} &
  \multicolumn{1}{r}{38.1} &
  \multicolumn{1}{r}{36.8} &
  \multicolumn{1}{r}{-1.3} \\
\multicolumn{1}{l}{Education level of hh (ref.: no school leaving cert.)} &
  \multicolumn{1}{r}{} &
  \multicolumn{1}{r}{} &
  \multicolumn{1}{r}{} \\
\multicolumn{1}{l}{\hspace{1em}W/o school-leaving certificate} &
  \multicolumn{1}{r}{36.6} &
  \multicolumn{1}{r}{36.2} &
  \multicolumn{1}{r}{-0.4} \\
\multicolumn{1}{l}{\hspace{1em}Elementary/middle school} &
  \multicolumn{1}{r}{32.4} &
  \multicolumn{1}{r}{29.7} &
  \multicolumn{1}{r}{-2.7} \\
\multicolumn{1}{l}{\hspace{1em}Secondary school-leaving certificate} &
  \multicolumn{1}{r}{35.7} &
  \multicolumn{1}{r}{31.6} &
  \multicolumn{1}{r}{-4.2} \\
\multicolumn{1}{l}{\hspace{1em}(Technical) university entrance qual.} &
  \multicolumn{1}{r}{43.5} &
  \multicolumn{1}{r}{38.4} &
  \multicolumn{1}{r}{-5.1} \\
\multicolumn{1}{l}{Household type (ref.: single)} &
  \multicolumn{1}{r}{} &
  \multicolumn{1}{r}{} &
  \multicolumn{1}{r}{} \\
\multicolumn{1}{l}{\hspace{1em}1-Person household} &
  \multicolumn{1}{r}{38.0} &
  \multicolumn{1}{r}{35.2} &
  \multicolumn{1}{r}{-2.8} \\
\multicolumn{1}{l}{\hspace{1em}Couple w/o children} &
  \multicolumn{1}{r}{60.4} &
  \multicolumn{1}{r}{55.6} &
  \multicolumn{1}{r}{-4.8} \\
\multicolumn{1}{l}{\hspace{1em}Single parent household} &
  \multicolumn{1}{r}{20.0} &
  \multicolumn{1}{r}{15.4} &
  \multicolumn{1}{r}{-4.6} \\
\multicolumn{1}{l}{\hspace{1em}Couple w/ children} &
  \multicolumn{1}{r}{33.5} &
  \multicolumn{1}{r}{28.3} &
  \multicolumn{1}{r}{-5.2} \\
\multicolumn{1}{l}{Federal state of residence (ref.: Nordrhein-Westfalen)} &
  \multicolumn{1}{r}{} &
  \multicolumn{1}{r}{} &
  \multicolumn{1}{r}{} \\
\multicolumn{1}{l}{\hspace{1em}1. Schleswig-Holstein} &
  \multicolumn{1}{r}{47.3} &
  \multicolumn{1}{r}{42.6} &
  \multicolumn{1}{r}{-4.6} \\
\multicolumn{1}{l}{\hspace{1em}2. Hamburg} &
  \multicolumn{1}{r}{35.4} &
  \multicolumn{1}{r}{30.5} &
  \multicolumn{1}{r}{-4.9} \\
\multicolumn{1}{l}{\hspace{1em}3. Niedersachsen} &
  \multicolumn{1}{r}{35.4} &
  \multicolumn{1}{r}{34.7} &
  \multicolumn{1}{r}{-0.7} \\
\multicolumn{1}{l}{\hspace{1em}4. Bremen} &
  \multicolumn{1}{r}{25.1} &
  \multicolumn{1}{r}{21.7} &
  \multicolumn{1}{r}{-3.4} \\
\multicolumn{1}{l}{\hspace{1em}5. Nordrhein-Westfalen} &
  \multicolumn{1}{r}{35.1} &
  \multicolumn{1}{r}{31.6} &
  \multicolumn{1}{r}{-3.5} \\
\multicolumn{1}{l}{\hspace{1em}6. Hessen} &
  \multicolumn{1}{r}{36.4} &
  \multicolumn{1}{r}{32.0} &
  \multicolumn{1}{r}{-4.5} \\
\multicolumn{1}{l}{\hspace{1em}7. Rheinland-Pfalz} &
  \multicolumn{1}{r}{38.8} &
  \multicolumn{1}{r}{35.1} &
  \multicolumn{1}{r}{-3.7} \\
\multicolumn{1}{l}{\hspace{1em}8. Baden-Wuerttemberg} &
  \multicolumn{1}{r}{41.1} &
  \multicolumn{1}{r}{35.1} &
  \multicolumn{1}{r}{-6.0} \\
\multicolumn{1}{l}{\hspace{1em}9. Bayern} &
  \multicolumn{1}{r}{44.9} &
  \multicolumn{1}{r}{42.7} &
  \multicolumn{1}{r}{-2.3} \\
\multicolumn{1}{l}{\hspace{1em}10. Saarland} &
  \multicolumn{1}{r}{31.9} &
  \multicolumn{1}{r}{27.1} &
  \multicolumn{1}{r}{-4.7} \\
\multicolumn{1}{l}{\hspace{1em}11. Berlin} &
  \multicolumn{1}{r}{32.2} &
  \multicolumn{1}{r}{30.2} &
  \multicolumn{1}{r}{-2.0} \\
\multicolumn{1}{l}{\hspace{1em}12. Brandenburg} &
  \multicolumn{1}{r}{35.3} &
  \multicolumn{1}{r}{32.8} &
  \multicolumn{1}{r}{-2.5} \\
\multicolumn{1}{l}{\hspace{1em}13. Mecklenburg-Vorpommern} &
  \multicolumn{1}{r}{30.1} &
  \multicolumn{1}{r}{28.3} &
  \multicolumn{1}{r}{-1.7} \\
\multicolumn{1}{l}{\hspace{1em}14. Sachsen} &
  \multicolumn{1}{r}{34.9} &
  \multicolumn{1}{r}{31.1} &
  \multicolumn{1}{r}{-3.7} \\
\multicolumn{1}{l}{\hspace{1em}15. Sachsen-Anhalt} &
  \multicolumn{1}{r}{31.0} &
  \multicolumn{1}{r}{26.3} &
  \multicolumn{1}{r}{-4.8} \\
\multicolumn{1}{l}{\hspace{1em}16. Thueringen} &
  \multicolumn{1}{r}{26.6} &
  \multicolumn{1}{r}{22.5} &
  \multicolumn{1}{r}{-4.2} \\
\multicolumn{1}{l}{Population size category (ref.: $<$ 50,000 residents)} &
  \multicolumn{1}{r}{} &
  \multicolumn{1}{r}{} &
  \multicolumn{1}{r}{} \\
\multicolumn{1}{l}{\hspace{1em}$<$  50,000 residents} &
  \multicolumn{1}{r}{37.9} &
  \multicolumn{1}{r}{35.3} &
  \multicolumn{1}{r}{-2.6} \\
\multicolumn{1}{l}{\hspace{1em}$>$= 50,000 res. (periph. area)} &
  \multicolumn{1}{r}{41.0} &
  \multicolumn{1}{r}{37.6} &
  \multicolumn{1}{r}{-3.4} \\
\multicolumn{1}{l}{\hspace{1em}$>$= 50,000-499,999 res. (core area)} &
  \multicolumn{1}{r}{30.4} &
  \multicolumn{1}{r}{26.6} &
  \multicolumn{1}{r}{-3.8} \\
\multicolumn{1}{l}{\hspace{1em}$>$=500,000 res. (core area)} &
  \multicolumn{1}{r}{36.1} &
  \multicolumn{1}{r}{32.7} &
  \multicolumn{1}{r}{-3.4} \\
\cline{1-4}
\end{longtable}

\end{ThreePartTable}

\section{Descriptive results for evaluating the simulation quality\label{sec:AppSimQual}}

\setcounter{table}{0}\setcounter{figure}{0}
\renewcommand\thetable{\thesection\arabic{table}} 
\renewcommand\thefigure{\thesection\arabic{figure}} 

\begin{figure}[tbp]
\caption{Distributions of simulated and reported disposable income (kernel densities)}
   \label{fig:income_distribution_all}
    
    \includegraphics[width=.9\textwidth]{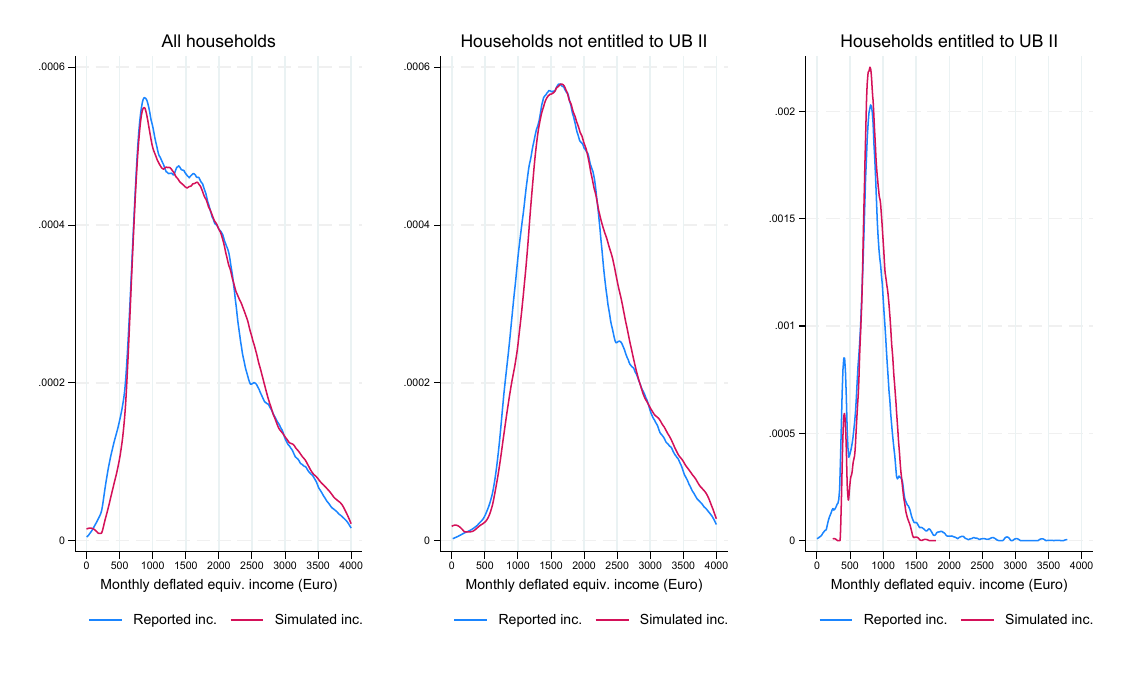}

   \floatfoot{ \scriptsize \textsc{Note. ---} The red (blue) lines show the distributions of the simulated (reported) disposable income separately for all households, UB~II eligible households, and UB~II ineligible households over the period 2008 - 2020. Incomes are deflated (in 2020 prices) and equivalised using the new OECD scale. Weighted results. Source: GETTSIM v0.7.0, PASS 0620 v1, PASS-ADIAB 7520 v1, own calculations.}   
\end{figure}

Figure~\ref{fig:income_distribution_all} shows a comparison between the distributions of disposable income as simulated using GETTSIM (in red) and the disposable income as reported by the PASS households (in blue). Distributions are separately shown for all households (left), those not simulated as eligible for UB~II (middle), and those simulated as eligible (right). The plots encompass all household-wave observations within the estimation sample from 2008 to 2020. To ensure comparability of disposable incomes across periods and households, incomes are adjusted for inflation (to 2020 prices) and normalised using the new OECD scale. For all households, as well as for those not eligible for UB~II, the distributions align rather closely. However, the simulated distribution of disposable incomes for eligible households exhibits less variance compared to the reported incomes of this group. This outcome likely arises from the rigidity in modelling the rules for maximum recognised rent according to UB~II within GETTSIM.

Table~\ref{tab:SimVsRepInc} confirms the visual observation: While the standard deviation of simulated income exceeds the standard deviation of reported income for households not entitled to UB~II (2,484 euros vs. 1,837 euros), the opposite is true for entitled households (209 euros vs. 370 euros). For eligible households, the median and mean of the simulated disposable income are only slightly greater than those of the reported income. In contrast, for households simulated as ineligible for UB~II, the simulated mean income is noticeably higher (by 10~percent) compared to the mean reported disposable income.

\begin{figure}[tbp]
\caption{Distribution of simulated UB~II entitlements by take-up (kernel densities)}
   \label{fig:UBIIentitlements_kdensity_simrep}
    
    \includegraphics[width=0.9\textwidth]{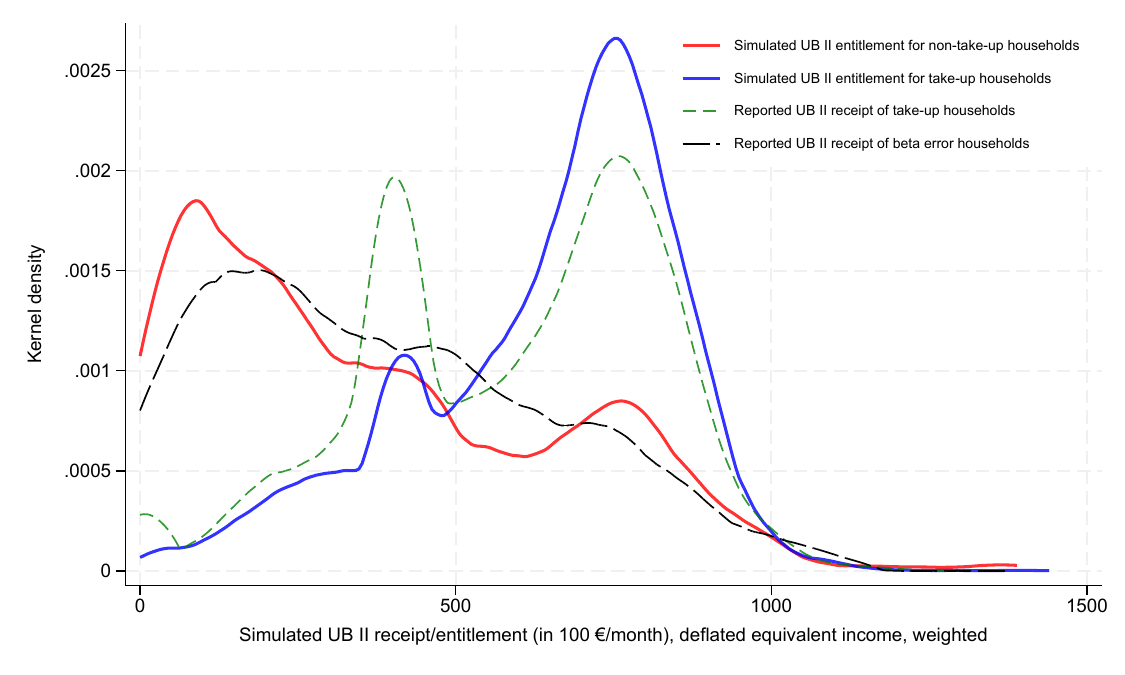}

   \floatfoot{ \scriptsize \textsc{Note. ---} The red (blue) line shows the distribution of the simulated UB~II entitlements for non-take-up (take-up) households over the period 2008 - 2020. Incomes are deflated (in 2020 prices) and equivalised using the new OECD scale. The green (blacked) dashed line shows the distribution of the reported UB~II receipts for take-up (beta error) households. Weighted results. Source: GETTSIM v0.7.0, PASS 0620 v1, PASS-ADIAB 7520 v1, own calculations.}   
\end{figure}

As discussed in Section~\ref{subsec:MethTakeUp}, the level of entitlement is expected to be a key regressor in an empirical take-up estimation. An unbiased estimation of the impact of the level of entitlements and other regressors on the take-up decision thus relies on an accurate simulation of the benefit entitlements. Figure~\ref{fig:UBIIentitlements_kdensity_simrep} shows the distribution of simulated benefit entitlements, distinguishing between take-up and non-take-up households. Consistent with expectations, households that claimed their benefits exhibit a distribution of entitlements skewed to the right compared to non-take-up households, with a median of 644 euros for the period 2008 to 2020 (equivalised deflated income). The distribution for non-take-up households shows a large proportion of households with low entitlements, with a median value of 372 euros.

Although it is not possible to verify the simulated UB~II entitlements for non-take-up households, the simulated entitlements for take-up households can, in theory, be compared to actual entitlements. Ideally, simulated entitlements should be compared to actual payments as per administrative data. However, while the administrative PASS-ADIAB data precisely records the periods during which households have received UB~II, it does not provide information on the payment amounts. Nevertheless, simulated entitlements can be compared to the UB~II receipts reported in the PASS, depicted as a dashed green line in Figure~\ref{fig:UBIIentitlements_kdensity_simrep}. Although both the distributions of simulated and reported entitlements for households that claim benefits are bimodal, and the positions of the two modes within these distributions correspond quite well, the reported benefits show greater (lesser) probability mass at the first (second) peak compared to the simulated benefits. Finally, the black dashed line illustrates the distribution of reported UB~II receipts for beta error households. The distribution indicates that a significant majority of beta error households report relatively low UB~II receipts, increasing the likelihood that the simulation classifies these households as ineligible.

\begin{table}[!ht]
\footnotesize
\centering
\resizebox{1.0\textwidth}{!}{\begin{threeparttable}
\caption{Simulated vs. reported disposable income}
\phantomsection\label{tab:SimVsRepInc}
\renewcommand{\arraystretch}{1.2}

\begin{tabular}{@{\hspace*{1.8em}}lrrrrrrrrr}
\hline\hline
\cline{1-10}
\multicolumn{1}{c}{} &
  \multicolumn{9}{c}{Reported UB II receipt (administratively corrected)} \\
\multicolumn{1}{c}{} &
  \multicolumn{3}{c}{No} &
  \multicolumn{3}{c}{Yes} &
  \multicolumn{3}{c}{Total} \\
\multicolumn{1}{c}{} &
  \multicolumn{1}{r}{Median} &
  \multicolumn{1}{r}{Mean} &
  \multicolumn{1}{r}{(sd)} &
  \multicolumn{1}{r}{Median} &
  \multicolumn{1}{r}{Mean} &
  \multicolumn{1}{r}{(sd)} &
  \multicolumn{1}{|r}{Median} &
  \multicolumn{1}{r}{Mean} &
  \multicolumn{1}{r}{(sd)} \\
\cline{1-10}
\multicolumn{1}{l}{Simulated UB II entitlement} &
  \multicolumn{1}{r}{} &
  \multicolumn{1}{r}{} &
  \multicolumn{1}{r}{} &
  \multicolumn{1}{r}{} &
  \multicolumn{1}{r}{} &
  \multicolumn{1}{r}{} &
  \multicolumn{1}{|r}{} &
  \multicolumn{1}{r}{} &
  \multicolumn{1}{r}{} \\
\multicolumn{1}{l}{\hspace{1em}No} &
  \multicolumn{1}{r}{} &
  \multicolumn{1}{r}{} &
  \multicolumn{1}{r}{} &
  \multicolumn{1}{r}{} &
  \multicolumn{1}{r}{} &
  \multicolumn{1}{r}{} &
  \multicolumn{1}{|r}{} &
  \multicolumn{1}{r}{} &
  \multicolumn{1}{r}{} \\
\multicolumn{1}{l}{\hspace{2em}Reported household income (in Euro)} &
  \multicolumn{1}{r}{1,862} &
  \multicolumn{1}{r}{2,116} &
  \multicolumn{1}{r}{(1,843)} &
  \multicolumn{1}{r}{956} &
  \multicolumn{1}{r}{980} &
  \multicolumn{1}{r}{(441)} &
  \multicolumn{1}{|r}{1,852} &
  \multicolumn{1}{r}{2,103} &
  \multicolumn{1}{r}{(1,837)} \\
\multicolumn{1}{l}{\hspace{2em}Simulated household income (in Euro)} &
  \multicolumn{1}{r}{1,959} &
  \multicolumn{1}{r}{2,336} &
  \multicolumn{1}{r}{(2,491)} &
  \multicolumn{1}{r}{1,133} &
  \multicolumn{1}{r}{1,096} &
  \multicolumn{1}{r}{(1,296)} &
  \multicolumn{1}{|r}{1,948} &
  \multicolumn{1}{r}{2,321} &
  \multicolumn{1}{r}{(2,484)} \\
\multicolumn{1}{l}{\hspace{1em}Yes} &
  \multicolumn{1}{r}{} &
  \multicolumn{1}{r}{} &
  \multicolumn{1}{r}{} &
  \multicolumn{1}{r}{} &
  \multicolumn{1}{r}{} &
  \multicolumn{1}{r}{} &
  \multicolumn{1}{|r}{} &
  \multicolumn{1}{r}{} &
  \multicolumn{1}{r}{} \\
\multicolumn{1}{l}{\hspace{2em}Reported household income (in Euro)} &
  \multicolumn{1}{r}{918} &
  \multicolumn{1}{r}{965} &
  \multicolumn{1}{r}{(496)} &
  \multicolumn{1}{r}{794} &
  \multicolumn{1}{r}{780} &
  \multicolumn{1}{r}{(269)} &
  \multicolumn{1}{|r}{820} &
  \multicolumn{1}{r}{841} &
  \multicolumn{1}{r}{(370)} \\
\multicolumn{1}{l}{\hspace{2em}Simulated household income (in Euro)} &
  \multicolumn{1}{r}{901} &
  \multicolumn{1}{r}{906} &
  \multicolumn{1}{r}{(247)} &
  \multicolumn{1}{r}{842} &
  \multicolumn{1}{r}{847} &
  \multicolumn{1}{r}{(184)} &
  \multicolumn{1}{|r}{855} &
  \multicolumn{1}{r}{866} &
  \multicolumn{1}{r}{(209)} \\
\cline{1-10}
\multicolumn{1}{l}{\hspace{1em}Total} &
  \multicolumn{1}{r}{} &
  \multicolumn{1}{r}{} &
  \multicolumn{1}{r}{} &
  \multicolumn{1}{r}{} &
  \multicolumn{1}{r}{} &
  \multicolumn{1}{r}{} &
  \multicolumn{1}{|r}{} &
  \multicolumn{1}{r}{} &
  \multicolumn{1}{r}{} \\
\multicolumn{1}{l}{\hspace{2em}Reported household income (in Euro)} &
  \multicolumn{1}{r}{1,790} &
  \multicolumn{1}{r}{2,026} &
  \multicolumn{1}{r}{(1,801)} &
  \multicolumn{1}{r}{800} &
  \multicolumn{1}{r}{792} &
  \multicolumn{1}{r}{(287)} &
  \multicolumn{1}{|r}{1,618} &
  \multicolumn{1}{r}{1,845} &
  \multicolumn{1}{r}{(1,724)} \\
\multicolumn{1}{l}{\hspace{2em}Simulated household income (in Euro)} &
  \multicolumn{1}{r}{1,868} &
  \multicolumn{1}{r}{2,223} &
  \multicolumn{1}{r}{(2,423)} &
  \multicolumn{1}{r}{851} &
  \multicolumn{1}{r}{863} &
  \multicolumn{1}{r}{(375)} &
  \multicolumn{1}{|r}{1,691} &
  \multicolumn{1}{r}{2,023} &
  \multicolumn{1}{r}{(2,293)} \\
\cline{1-10}
\hline
\end{tabular}

\begin{tablenotes}
\item \footnotesize \textsc{Note. ---} PASS waves 2008 - 2020, overall disposable incomes by entitlement and receipt, median and mean. sd = standard deviation. Income deflated with consumer price index (base: 2020) and equivalised using the new OECD scale. Weighted results. Source: PASS 0620 v1, PASS-ADIAB 7520 v1, GETTSIM v0.7.0, own calculations.
\end{tablenotes}

\end{threeparttable}
} \end{table}

\section{Covariate means\label{sec:AppCovMeans}}

\setcounter{table}{0}\setcounter{figure}{0}
\renewcommand\thetable{\thesection\arabic{table}} 
\renewcommand\thefigure{\thesection\arabic{figure}} 

Table~\ref{tab:covariate_means} reports the means of the covariates separately for take-up and non-take-up households (administrative receipt). The units of the continuous long-term factors are as follows. The lagged shares of days in UB~II receipt lie between 0 and 1 (a value of, say, $0.62$ means that benefits were received on $62$~percent of the days of the respective year, on average). The lagged earned incomes are real equivalised monthly amounts in units of 1,000~euros (2020 prices, modified OECD scale). The income shock (``difference of earned income'') is expressed in units of the household's current UB~II needs, so that the value of $0.04$ for all households corresponds to an average year-on-year income gain worth about $4$~percent of current UB~II needs. The within-household standard deviation of earned income is likewise measured in units of 1,000~euros of real equivalised monthly income.

\begin{ThreePartTable}
\footnotesize
\centering
\renewcommand{\arraystretch}{0.9}

\begin{TableNotes}
\item \scriptsize \textsc{Note. ---} *** \(p<.01\), ** \(p<.05\), * \(p<.1\). hh=head of household. All=all household year observations simulated as entitled to UB~II, TU=take-up households, NTU=non-take-up households, TU-NTU=difference between columns TU and NTU. Asterisks denote significance of the difference. Take-up indicator based on administrative (PASS-ADIAB) receipt. Weighted results. Source: PASS 0620 v1, PASS-ADIAB 7520 v1, own calculations.
\end{TableNotes}

\begin{longtable}{@{\hspace*{1.8em}}lrrrrrrrrrrrrrr}
\caption{Covariate means, PASS waves 2008 - 2020}\phantomsection\label{tab:covariate_means}\\
\hline\hline
\multicolumn{1}{c}{} &
  \multicolumn{1}{r}{All} &
  \multicolumn{1}{r}{TU} &
  \multicolumn{1}{r}{NTU} &
  \multicolumn{2}{r}{TU-NTU} \\
  
\hline
\endfirsthead

\caption[]{(continued from previous page)}\\
\hline
\multicolumn{1}{c}{} &
  \multicolumn{1}{r}{All} &
  \multicolumn{1}{r}{TU} &
  \multicolumn{1}{r}{NTU} &
  \multicolumn{2}{r}{TU-NTU} \\
  
\hline
\endhead
\cmidrule{6-6}
\multicolumn{6}{r}{\textit{continued}}
\endfoot

\hline
\insertTableNotes\\
\endlastfoot

\multicolumn{1}{l}{Relative income gap} &
  \multicolumn{1}{r}{0.70} &
  \multicolumn{1}{r}{0.80} &
  \multicolumn{1}{r}{0.50} &
  \multicolumn{1}{r@{}}{0.31} &
  \multicolumn{1}{@{}l}{***} \\
\multicolumn{1}{l}{Sex of hh (ref: male hh)} &
  \multicolumn{1}{r}{0.49} &
  \multicolumn{1}{r}{0.47} &
  \multicolumn{1}{r}{0.54} &
  \multicolumn{1}{r@{}}{-0.06} &
  \multicolumn{1}{@{}l}{*} \\
\multicolumn{1}{l}{Region of residence (ref.: West Germany)} &
  \multicolumn{1}{r}{0.31} &
  \multicolumn{1}{r}{0.33} &
  \multicolumn{1}{r}{0.28} &
  \multicolumn{1}{r@{}}{0.05} &
  \multicolumn{1}{@{}l}{*} \\
\multicolumn{1}{l}{Home ownership} &
  \multicolumn{1}{r}{0.05} &
  \multicolumn{1}{r}{0.03} &
  \multicolumn{1}{r}{0.07} &
  \multicolumn{1}{r@{}}{-0.04} &
  \multicolumn{1}{@{}l}{**} \\
\multicolumn{1}{l}{Presence of child aged $<$=3 in household} &
  \multicolumn{1}{r}{0.07} &
  \multicolumn{1}{r}{0.09} &
  \multicolumn{1}{r}{0.03} &
  \multicolumn{1}{r@{}}{0.07} &
  \multicolumn{1}{@{}l}{***} \\
\multicolumn{1}{l}{Disabled hh and/or partner} &
  \multicolumn{1}{r}{0.23} &
  \multicolumn{1}{r}{0.18} &
  \multicolumn{1}{r}{0.31} &
  \multicolumn{1}{r@{}}{-0.13} &
  \multicolumn{1}{@{}l}{***} \\
\multicolumn{1}{l}{Early retirement status} &
  \multicolumn{1}{r}{0.10} &
  \multicolumn{1}{r}{0.02} &
  \multicolumn{1}{r}{0.27} &
  \multicolumn{1}{r@{}}{-0.25} &
  \multicolumn{1}{@{}l}{***} \\
\multicolumn{1}{l}{Reported receipt of other benefit than UB II} &
  \multicolumn{1}{r}{0.14} &
  \multicolumn{1}{r}{0.06} &
  \multicolumn{1}{r}{0.32} &
  \multicolumn{1}{r@{}}{-0.26} &
  \multicolumn{1}{@{}l}{***} \\
\multicolumn{1}{l}{GenPop subsample (ref.: Admin sample)} &
  \multicolumn{1}{r}{0.27} &
  \multicolumn{1}{r}{0.14} &
  \multicolumn{1}{r}{0.55} &
  \multicolumn{1}{r@{}}{-0.41} &
  \multicolumn{1}{@{}l}{***} \\
\multicolumn{1}{l}{} &
  \multicolumn{1}{r}{} &
  \multicolumn{1}{r}{} &
  \multicolumn{1}{r}{} &
  \multicolumn{1}{r@{}}{} &
  \multicolumn{1}{@{}l}{} \\
\multicolumn{1}{l}{Migr. background of hh (ref.: no migr. backgr.)} &
  \multicolumn{1}{r}{} &
  \multicolumn{1}{r}{} &
  \multicolumn{1}{r}{} &
  \multicolumn{1}{r@{}}{} &
  \multicolumn{1}{@{}l}{} \\
\multicolumn{1}{l}{\hspace{1em}No migration backgr.} &
  \multicolumn{1}{r}{0.77} &
  \multicolumn{1}{r}{0.75} &
  \multicolumn{1}{r}{0.80} &
  \multicolumn{1}{r@{}}{-0.06} &
  \multicolumn{1}{@{}l}{***} \\
\multicolumn{1}{l}{\hspace{1em}Migration backgr. (1.gen.)} &
  \multicolumn{1}{r}{0.17} &
  \multicolumn{1}{r}{0.19} &
  \multicolumn{1}{r}{0.14} &
  \multicolumn{1}{r@{}}{0.05} &
  \multicolumn{1}{@{}l}{***} \\
\multicolumn{1}{l}{\hspace{1em}Migration backgr. (2.gen.)} &
  \multicolumn{1}{r}{0.06} &
  \multicolumn{1}{r}{0.06} &
  \multicolumn{1}{r}{0.06} &
  \multicolumn{1}{r@{}}{0.00} &
  \multicolumn{1}{@{}l}{} \\
\multicolumn{1}{l}{} &
  \multicolumn{1}{r}{} &
  \multicolumn{1}{r}{} &
  \multicolumn{1}{r}{} &
  \multicolumn{1}{r@{}}{} &
  \multicolumn{1}{@{}l}{} \\
\multicolumn{1}{l}{Age of hh (ref.: 15-24)} &
  \multicolumn{1}{r}{} &
  \multicolumn{1}{r}{} &
  \multicolumn{1}{r}{} &
  \multicolumn{1}{r@{}}{} &
  \multicolumn{1}{@{}l}{} \\
\multicolumn{1}{l}{\hspace{1em}15 to 24} &
  \multicolumn{1}{r}{0.06} &
  \multicolumn{1}{r}{0.05} &
  \multicolumn{1}{r}{0.07} &
  \multicolumn{1}{r@{}}{-0.01} &
  \multicolumn{1}{@{}l}{} \\
\multicolumn{1}{l}{\hspace{1em}25 to 34} &
  \multicolumn{1}{r}{0.21} &
  \multicolumn{1}{r}{0.23} &
  \multicolumn{1}{r}{0.16} &
  \multicolumn{1}{r@{}}{0.07} &
  \multicolumn{1}{@{}l}{***} \\
\multicolumn{1}{l}{\hspace{1em}35 to 44} &
  \multicolumn{1}{r}{0.21} &
  \multicolumn{1}{r}{0.21} &
  \multicolumn{1}{r}{0.19} &
  \multicolumn{1}{r@{}}{0.02} &
  \multicolumn{1}{@{}l}{} \\
\multicolumn{1}{l}{\hspace{1em}45 to 54} &
  \multicolumn{1}{r}{0.27} &
  \multicolumn{1}{r}{0.26} &
  \multicolumn{1}{r}{0.29} &
  \multicolumn{1}{r@{}}{-0.04} &
  \multicolumn{1}{@{}l}{*} \\
\multicolumn{1}{l}{\hspace{1em}$>$=55} &
  \multicolumn{1}{r}{0.26} &
  \multicolumn{1}{r}{0.24} &
  \multicolumn{1}{r}{0.29} &
  \multicolumn{1}{r@{}}{-0.05} &
  \multicolumn{1}{@{}l}{*} \\
\multicolumn{1}{l}{} &
  \multicolumn{1}{r}{} &
  \multicolumn{1}{r}{} &
  \multicolumn{1}{r}{} &
  \multicolumn{1}{r@{}}{} &
  \multicolumn{1}{@{}l}{} \\
\multicolumn{1}{l}{Education level of hh (ref.: no school leaving cert.)} &
  \multicolumn{1}{r}{} &
  \multicolumn{1}{r}{} &
  \multicolumn{1}{r}{} &
  \multicolumn{1}{r@{}}{} &
  \multicolumn{1}{@{}l}{} \\
\multicolumn{1}{l}{\hspace{1em}W/o school-leaving certificate} &
  \multicolumn{1}{r}{0.12} &
  \multicolumn{1}{r}{0.11} &
  \multicolumn{1}{r}{0.13} &
  \multicolumn{1}{r@{}}{-0.02} &
  \multicolumn{1}{@{}l}{} \\
\multicolumn{1}{l}{\hspace{1em}Elementary/middle school} &
  \multicolumn{1}{r}{0.37} &
  \multicolumn{1}{r}{0.39} &
  \multicolumn{1}{r}{0.34} &
  \multicolumn{1}{r@{}}{0.05} &
  \multicolumn{1}{@{}l}{*} \\
\multicolumn{1}{l}{\hspace{1em}Secondary school-leaving certificate} &
  \multicolumn{1}{r}{0.32} &
  \multicolumn{1}{r}{0.33} &
  \multicolumn{1}{r}{0.31} &
  \multicolumn{1}{r@{}}{0.02} &
  \multicolumn{1}{@{}l}{} \\
\multicolumn{1}{l}{\hspace{1em}(Technical) university entrance qual.} &
  \multicolumn{1}{r}{0.19} &
  \multicolumn{1}{r}{0.17} &
  \multicolumn{1}{r}{0.22} &
  \multicolumn{1}{r@{}}{-0.05} &
  \multicolumn{1}{@{}l}{**} \\
\multicolumn{1}{l}{} &
  \multicolumn{1}{r}{} &
  \multicolumn{1}{r}{} &
  \multicolumn{1}{r}{} &
  \multicolumn{1}{r@{}}{} &
  \multicolumn{1}{@{}l}{} \\
\multicolumn{1}{l}{Household type (ref.: single)} &
  \multicolumn{1}{r}{} &
  \multicolumn{1}{r}{} &
  \multicolumn{1}{r}{} &
  \multicolumn{1}{r@{}}{} &
  \multicolumn{1}{@{}l}{} \\
\multicolumn{1}{l}{\hspace{1em}1-Person household} &
  \multicolumn{1}{r}{0.71} &
  \multicolumn{1}{r}{0.68} &
  \multicolumn{1}{r}{0.76} &
  \multicolumn{1}{r@{}}{-0.08} &
  \multicolumn{1}{@{}l}{***} \\
\multicolumn{1}{l}{\hspace{1em}Couple w/o children} &
  \multicolumn{1}{r}{0.06} &
  \multicolumn{1}{r}{0.04} &
  \multicolumn{1}{r}{0.10} &
  \multicolumn{1}{r@{}}{-0.06} &
  \multicolumn{1}{@{}l}{***} \\
\multicolumn{1}{l}{\hspace{1em}Single parent household} &
  \multicolumn{1}{r}{0.16} &
  \multicolumn{1}{r}{0.20} &
  \multicolumn{1}{r}{0.07} &
  \multicolumn{1}{r@{}}{0.13} &
  \multicolumn{1}{@{}l}{***} \\
\multicolumn{1}{l}{\hspace{1em}Couple w/ children} &
  \multicolumn{1}{r}{0.07} &
  \multicolumn{1}{r}{0.08} &
  \multicolumn{1}{r}{0.06} &
  \multicolumn{1}{r@{}}{0.01} &
  \multicolumn{1}{@{}l}{} \\
\multicolumn{1}{l}{} &
  \multicolumn{1}{r}{} &
  \multicolumn{1}{r}{} &
  \multicolumn{1}{r}{} &
  \multicolumn{1}{r@{}}{} &
  \multicolumn{1}{@{}l}{} \\
\multicolumn{1}{l}{Federal state of residence (ref.: Nordrhein-Westfalen)} &
  \multicolumn{1}{r}{} &
  \multicolumn{1}{r}{} &
  \multicolumn{1}{r}{} &
  \multicolumn{1}{r@{}}{} &
  \multicolumn{1}{@{}l}{} \\
\multicolumn{1}{l}{\hspace{1em}1. Schleswig-Holstein} &
  \multicolumn{1}{r}{0.04} &
  \multicolumn{1}{r}{0.04} &
  \multicolumn{1}{r}{0.06} &
  \multicolumn{1}{r@{}}{-0.02} &
  \multicolumn{1}{@{}l}{} \\
\multicolumn{1}{l}{\hspace{1em}2. Hamburg} &
  \multicolumn{1}{r}{0.03} &
  \multicolumn{1}{r}{0.03} &
  \multicolumn{1}{r}{0.03} &
  \multicolumn{1}{r@{}}{0.00} &
  \multicolumn{1}{@{}l}{} \\
\multicolumn{1}{l}{\hspace{1em}3. Niedersachsen} &
  \multicolumn{1}{r}{0.10} &
  \multicolumn{1}{r}{0.10} &
  \multicolumn{1}{r}{0.11} &
  \multicolumn{1}{r@{}}{-0.01} &
  \multicolumn{1}{@{}l}{} \\
\multicolumn{1}{l}{\hspace{1em}4. Bremen} &
  \multicolumn{1}{r}{0.01} &
  \multicolumn{1}{r}{0.01} &
  \multicolumn{1}{r}{0.01} &
  \multicolumn{1}{r@{}}{0.01} &
  \multicolumn{1}{@{}l}{***} \\
\multicolumn{1}{l}{\hspace{1em}5. Nordrhein-Westfalen} &
  \multicolumn{1}{r}{0.24} &
  \multicolumn{1}{r}{0.25} &
  \multicolumn{1}{r}{0.24} &
  \multicolumn{1}{r@{}}{0.01} &
  \multicolumn{1}{@{}l}{} \\
\multicolumn{1}{l}{\hspace{1em}6. Hessen} &
  \multicolumn{1}{r}{0.06} &
  \multicolumn{1}{r}{0.06} &
  \multicolumn{1}{r}{0.06} &
  \multicolumn{1}{r@{}}{0.00} &
  \multicolumn{1}{@{}l}{} \\
\multicolumn{1}{l}{\hspace{1em}7. Rheinland-Pfalz} &
  \multicolumn{1}{r}{0.03} &
  \multicolumn{1}{r}{0.03} &
  \multicolumn{1}{r}{0.03} &
  \multicolumn{1}{r@{}}{-0.00} &
  \multicolumn{1}{@{}l}{} \\
\multicolumn{1}{l}{\hspace{1em}8. Baden-Wuerttemberg} &
  \multicolumn{1}{r}{0.08} &
  \multicolumn{1}{r}{0.08} &
  \multicolumn{1}{r}{0.08} &
  \multicolumn{1}{r@{}}{-0.01} &
  \multicolumn{1}{@{}l}{} \\
\multicolumn{1}{l}{\hspace{1em}9. Bayern} &
  \multicolumn{1}{r}{0.08} &
  \multicolumn{1}{r}{0.07} &
  \multicolumn{1}{r}{0.10} &
  \multicolumn{1}{r@{}}{-0.04} &
  \multicolumn{1}{@{}l}{**} \\
\multicolumn{1}{l}{\hspace{1em}10. Saarland} &
  \multicolumn{1}{r}{0.01} &
  \multicolumn{1}{r}{0.01} &
  \multicolumn{1}{r}{0.01} &
  \multicolumn{1}{r@{}}{0.00} &
  \multicolumn{1}{@{}l}{} \\
\multicolumn{1}{l}{\hspace{1em}11. Berlin} &
  \multicolumn{1}{r}{0.09} &
  \multicolumn{1}{r}{0.09} &
  \multicolumn{1}{r}{0.08} &
  \multicolumn{1}{r@{}}{0.01} &
  \multicolumn{1}{@{}l}{} \\
\multicolumn{1}{l}{\hspace{1em}12. Brandenburg} &
  \multicolumn{1}{r}{0.04} &
  \multicolumn{1}{r}{0.04} &
  \multicolumn{1}{r}{0.04} &
  \multicolumn{1}{r@{}}{-0.00} &
  \multicolumn{1}{@{}l}{} \\
\multicolumn{1}{l}{\hspace{1em}13. Mecklenburg-Vorpommern} &
  \multicolumn{1}{r}{0.03} &
  \multicolumn{1}{r}{0.03} &
  \multicolumn{1}{r}{0.03} &
  \multicolumn{1}{r@{}}{0.01} &
  \multicolumn{1}{@{}l}{} \\
\multicolumn{1}{l}{\hspace{1em}14. Sachsen} &
  \multicolumn{1}{r}{0.07} &
  \multicolumn{1}{r}{0.07} &
  \multicolumn{1}{r}{0.06} &
  \multicolumn{1}{r@{}}{0.00} &
  \multicolumn{1}{@{}l}{} \\
\multicolumn{1}{l}{\hspace{1em}15. Sachsen-Anhalt} &
  \multicolumn{1}{r}{0.05} &
  \multicolumn{1}{r}{0.05} &
  \multicolumn{1}{r}{0.04} &
  \multicolumn{1}{r@{}}{0.01} &
  \multicolumn{1}{@{}l}{} \\
\multicolumn{1}{l}{\hspace{1em}16. Thueringen} &
  \multicolumn{1}{r}{0.03} &
  \multicolumn{1}{r}{0.03} &
  \multicolumn{1}{r}{0.02} &
  \multicolumn{1}{r@{}}{0.01} &
  \multicolumn{1}{@{}l}{*} \\
\multicolumn{1}{l}{} &
  \multicolumn{1}{r}{} &
  \multicolumn{1}{r}{} &
  \multicolumn{1}{r}{} &
  \multicolumn{1}{r@{}}{} &
  \multicolumn{1}{@{}l}{} \\
\multicolumn{1}{l}{Population size category (ref.: $<$ 50,000 residents)} &
  \multicolumn{1}{r}{} &
  \multicolumn{1}{r}{} &
  \multicolumn{1}{r}{} &
  \multicolumn{1}{r@{}}{} &
  \multicolumn{1}{@{}l}{} \\
\multicolumn{1}{l}{\hspace{1em}$<$  50,000 residents} &
  \multicolumn{1}{r}{0.19} &
  \multicolumn{1}{r}{0.18} &
  \multicolumn{1}{r}{0.20} &
  \multicolumn{1}{r@{}}{-0.02} &
  \multicolumn{1}{@{}l}{} \\
\multicolumn{1}{l}{\hspace{1em}$>$= 50,000 res. (periph. area)} &
  \multicolumn{1}{r}{0.23} &
  \multicolumn{1}{r}{0.21} &
  \multicolumn{1}{r}{0.27} &
  \multicolumn{1}{r@{}}{-0.05} &
  \multicolumn{1}{@{}l}{*} \\
\multicolumn{1}{l}{\hspace{1em}$>$= 50,000-499,999 res. (core area)} &
  \multicolumn{1}{r}{0.26} &
  \multicolumn{1}{r}{0.28} &
  \multicolumn{1}{r}{0.21} &
  \multicolumn{1}{r@{}}{0.07} &
  \multicolumn{1}{@{}l}{***} \\
\multicolumn{1}{l}{\hspace{1em}$>$=500,000 res. (core area)} &
  \multicolumn{1}{r}{0.32} &
  \multicolumn{1}{r}{0.32} &
  \multicolumn{1}{r}{0.32} &
  \multicolumn{1}{r@{}}{0.00} &
  \multicolumn{1}{@{}l}{} \\
\multicolumn{1}{l}{} &
  \multicolumn{1}{r}{} &
  \multicolumn{1}{r}{} &
  \multicolumn{1}{r}{} &
  \multicolumn{1}{r@{}}{} &
  \multicolumn{1}{@{}l}{} \\
\multicolumn{1}{l}{Lagged share of days per year with UB II receipt (t-1)} &
  \multicolumn{1}{r}{0.48} &
  \multicolumn{1}{r}{0.67} &
  \multicolumn{1}{r}{0.07} &
  \multicolumn{1}{r@{}}{0.61} &
  \multicolumn{1}{@{}l}{***} \\
\multicolumn{1}{l}{Lagged share of days per year with UB II receipt (t-2)} &
  \multicolumn{1}{r}{0.47} &
  \multicolumn{1}{r}{0.62} &
  \multicolumn{1}{r}{0.15} &
  \multicolumn{1}{r@{}}{0.47} &
  \multicolumn{1}{@{}l}{***} \\
\multicolumn{1}{l}{Lagged share of days per year with UB II receipt (t-3)} &
  \multicolumn{1}{r}{0.44} &
  \multicolumn{1}{r}{0.56} &
  \multicolumn{1}{r}{0.19} &
  \multicolumn{1}{r@{}}{0.38} &
  \multicolumn{1}{@{}l}{***} \\
\multicolumn{1}{l}{Lagged equivalent real annual income from work (t-1, in 1,000 Euros/month)} &
  \multicolumn{1}{r}{0.36} &
  \multicolumn{1}{r}{0.15} &
  \multicolumn{1}{r}{0.81} &
  \multicolumn{1}{r@{}}{-0.66} &
  \multicolumn{1}{@{}l}{***} \\
\multicolumn{1}{l}{Lagged equivalent real annual income from work (t-2, in 1,000 Euros/month)} &
  \multicolumn{1}{r}{0.37} &
  \multicolumn{1}{r}{0.17} &
  \multicolumn{1}{r}{0.80} &
  \multicolumn{1}{r@{}}{-0.63} &
  \multicolumn{1}{@{}l}{***} \\
\multicolumn{1}{l}{Lagged equivalent real annual income from work (t-3, in 1,000 Euros/month)} &
  \multicolumn{1}{r}{0.38} &
  \multicolumn{1}{r}{0.20} &
  \multicolumn{1}{r}{0.77} &
  \multicolumn{1}{r@{}}{-0.57} &
  \multicolumn{1}{@{}l}{***} \\
\multicolumn{1}{l}{Difference of earned income (normalised by UB II needs)} &
  \multicolumn{1}{r}{0.04} &
  \multicolumn{1}{r}{0.06} &
  \multicolumn{1}{r}{-0.02} &
  \multicolumn{1}{r@{}}{0.08} &
  \multicolumn{1}{@{}l}{***} \\
\multicolumn{1}{l}{Standard deviation (within) of lagged real equiv. income} &
  \multicolumn{1}{r}{0.19} &
  \multicolumn{1}{r}{0.15} &
  \multicolumn{1}{r}{0.26} &
  \multicolumn{1}{r@{}}{-0.11} &
  \multicolumn{1}{@{}l}{***} \\
\cline{1-6}
\end{longtable}

\end{ThreePartTable}
 
\begin{landscape}
\section{Estimation results for take-up models\label{sec:AppEstResults}}

\setcounter{table}{0}\setcounter{figure}{0}
\renewcommand\thetable{\thesection\arabic{table}} 
\renewcommand\thefigure{\thesection\arabic{figure}} 

\begin{ThreePartTable}
\footnotesize
\centering
\renewcommand{\arraystretch}{0.9}

\begin{TableNotes}
\item \scriptsize \textsc{Note. ---} *** \(p<.01\), ** \(p<.05\), * \(p<.1\). Baseline=baseline model, M1-M3=extended specifications. Standard errors, cluster-robust at the household level, in parentheses. Unweighted estimation. hh=head of household. Real equivalent earned incomes in 1,000 Euros/month, deflated using 2020 prices and equivalised using the new OECD scale. Panel-\(\rho\) denotes the share of the total variance contributed by the panel level variance component. Source: PASS 0620 v1, PASS-ADIAB 7520 v1, own calculations.
\end{TableNotes}

\begin{longtable}{@{\hspace*{1.8em}}lrrrrrrrr}
\caption{Take-up estimates (average marginal effects, unweighted estimation)}\phantomsection\label{tab:TUests_w0_e0_excl11001}\\
\hline\hline
\multicolumn{1}{r}{} &
  \multicolumn{2}{c}{Baseline} &
  \multicolumn{2}{c}{M1} &
  \multicolumn{2}{c}{M2} &
  \multicolumn{2}{c}{M3} \\
\hline
\endfirsthead

\caption[]{(continued from previous page)}\\
\hline
\multicolumn{1}{r}{} &
  \multicolumn{2}{c}{Baseline} &
  \multicolumn{2}{c}{M1} &
  \multicolumn{2}{c}{M2} &
  \multicolumn{2}{c}{M3} \\
\hline
\endhead
\cmidrule{9-9}
\multicolumn{9}{r}{\textit{continued}}
\endfoot

\hline
\insertTableNotes\\
\endlastfoot

\multicolumn{1}{l}{Relative income gap} &
  \multicolumn{1}{r@{}}{0.251} &
  \multicolumn{1}{@{}l}{***} &
  \multicolumn{1}{r@{}}{0.161} &
  \multicolumn{1}{@{}l}{***} &
  \multicolumn{1}{r@{}}{0.126} &
  \multicolumn{1}{@{}l}{***} &
  \multicolumn{1}{r@{}}{0.103} &
  \multicolumn{1}{@{}l}{***} \\
\multicolumn{1}{l}{} &
  \multicolumn{1}{r@{}}{(0.010)} &
  \multicolumn{1}{@{}l}{} &
  \multicolumn{1}{r@{}}{(0.008)} &
  \multicolumn{1}{@{}l}{} &
  \multicolumn{1}{r@{}}{(0.008)} &
  \multicolumn{1}{@{}l}{} &
  \multicolumn{1}{r@{}}{(0.008)} &
  \multicolumn{1}{@{}l}{} \\
\multicolumn{1}{l}{Sex of hh (ref: male hh)} &
  \multicolumn{1}{r@{}}{-0.000} &
  \multicolumn{1}{@{}l}{} &
  \multicolumn{1}{r@{}}{0.003} &
  \multicolumn{1}{@{}l}{} &
  \multicolumn{1}{r@{}}{0.007} &
  \multicolumn{1}{@{}l}{} &
  \multicolumn{1}{r@{}}{0.007} &
  \multicolumn{1}{@{}l}{} \\
\multicolumn{1}{l}{} &
  \multicolumn{1}{r@{}}{(0.007)} &
  \multicolumn{1}{@{}l}{} &
  \multicolumn{1}{r@{}}{(0.005)} &
  \multicolumn{1}{@{}l}{} &
  \multicolumn{1}{r@{}}{(0.005)} &
  \multicolumn{1}{@{}l}{} &
  \multicolumn{1}{r@{}}{(0.005)} &
  \multicolumn{1}{@{}l}{} \\
\multicolumn{1}{l}{Age of hh (ref.: 15-24)} &
  \multicolumn{1}{r@{}}{} &
  \multicolumn{1}{@{}l}{} &
  \multicolumn{1}{r@{}}{} &
  \multicolumn{1}{@{}l}{} &
  \multicolumn{1}{r@{}}{} &
  \multicolumn{1}{@{}l}{} &
  \multicolumn{1}{r@{}}{} &
  \multicolumn{1}{@{}l}{} \\
\multicolumn{1}{l}{\hspace{1em}25 to 34} &
  \multicolumn{1}{r@{}}{0.045} &
  \multicolumn{1}{@{}l}{***} &
  \multicolumn{1}{r@{}}{0.020} &
  \multicolumn{1}{@{}l}{**} &
  \multicolumn{1}{r@{}}{0.032} &
  \multicolumn{1}{@{}l}{***} &
  \multicolumn{1}{r@{}}{0.034} &
  \multicolumn{1}{@{}l}{***} \\
\multicolumn{1}{l}{} &
  \multicolumn{1}{r@{}}{(0.014)} &
  \multicolumn{1}{@{}l}{} &
  \multicolumn{1}{r@{}}{(0.010)} &
  \multicolumn{1}{@{}l}{} &
  \multicolumn{1}{r@{}}{(0.010)} &
  \multicolumn{1}{@{}l}{} &
  \multicolumn{1}{r@{}}{(0.010)} &
  \multicolumn{1}{@{}l}{} \\
\multicolumn{1}{l}{\hspace{1em}35 to 44} &
  \multicolumn{1}{r@{}}{0.051} &
  \multicolumn{1}{@{}l}{***} &
  \multicolumn{1}{r@{}}{0.016} &
  \multicolumn{1}{@{}l}{} &
  \multicolumn{1}{r@{}}{0.032} &
  \multicolumn{1}{@{}l}{***} &
  \multicolumn{1}{r@{}}{0.032} &
  \multicolumn{1}{@{}l}{***} \\
\multicolumn{1}{l}{} &
  \multicolumn{1}{r@{}}{(0.014)} &
  \multicolumn{1}{@{}l}{} &
  \multicolumn{1}{r@{}}{(0.010)} &
  \multicolumn{1}{@{}l}{} &
  \multicolumn{1}{r@{}}{(0.010)} &
  \multicolumn{1}{@{}l}{} &
  \multicolumn{1}{r@{}}{(0.010)} &
  \multicolumn{1}{@{}l}{} \\
\multicolumn{1}{l}{\hspace{1em}45 to 54} &
  \multicolumn{1}{r@{}}{0.067} &
  \multicolumn{1}{@{}l}{***} &
  \multicolumn{1}{r@{}}{0.008} &
  \multicolumn{1}{@{}l}{} &
  \multicolumn{1}{r@{}}{0.025} &
  \multicolumn{1}{@{}l}{**} &
  \multicolumn{1}{r@{}}{0.023} &
  \multicolumn{1}{@{}l}{**} \\
\multicolumn{1}{l}{} &
  \multicolumn{1}{r@{}}{(0.014)} &
  \multicolumn{1}{@{}l}{} &
  \multicolumn{1}{r@{}}{(0.010)} &
  \multicolumn{1}{@{}l}{} &
  \multicolumn{1}{r@{}}{(0.010)} &
  \multicolumn{1}{@{}l}{} &
  \multicolumn{1}{r@{}}{(0.010)} &
  \multicolumn{1}{@{}l}{} \\
\multicolumn{1}{l}{\hspace{1em}$>$=55} &
  \multicolumn{1}{r@{}}{0.087} &
  \multicolumn{1}{@{}l}{***} &
  \multicolumn{1}{r@{}}{0.016} &
  \multicolumn{1}{@{}l}{} &
  \multicolumn{1}{r@{}}{0.028} &
  \multicolumn{1}{@{}l}{***} &
  \multicolumn{1}{r@{}}{0.020} &
  \multicolumn{1}{@{}l}{*} \\
\multicolumn{1}{l}{} &
  \multicolumn{1}{r@{}}{(0.014)} &
  \multicolumn{1}{@{}l}{} &
  \multicolumn{1}{r@{}}{(0.010)} &
  \multicolumn{1}{@{}l}{} &
  \multicolumn{1}{r@{}}{(0.010)} &
  \multicolumn{1}{@{}l}{} &
  \multicolumn{1}{r@{}}{(0.010)} &
  \multicolumn{1}{@{}l}{} \\
\multicolumn{1}{l}{Education level of hh (ref.: no school leaving cert.)} &
  \multicolumn{1}{r@{}}{} &
  \multicolumn{1}{@{}l}{} &
  \multicolumn{1}{r@{}}{} &
  \multicolumn{1}{@{}l}{} &
  \multicolumn{1}{r@{}}{} &
  \multicolumn{1}{@{}l}{} &
  \multicolumn{1}{r@{}}{} &
  \multicolumn{1}{@{}l}{} \\
\multicolumn{1}{l}{\hspace{1em}Elementary/middle school} &
  \multicolumn{1}{r@{}}{0.002} &
  \multicolumn{1}{@{}l}{} &
  \multicolumn{1}{r@{}}{0.001} &
  \multicolumn{1}{@{}l}{} &
  \multicolumn{1}{r@{}}{-0.005} &
  \multicolumn{1}{@{}l}{} &
  \multicolumn{1}{r@{}}{-0.009} &
  \multicolumn{1}{@{}l}{} \\
\multicolumn{1}{l}{} &
  \multicolumn{1}{r@{}}{(0.011)} &
  \multicolumn{1}{@{}l}{} &
  \multicolumn{1}{r@{}}{(0.008)} &
  \multicolumn{1}{@{}l}{} &
  \multicolumn{1}{r@{}}{(0.008)} &
  \multicolumn{1}{@{}l}{} &
  \multicolumn{1}{r@{}}{(0.008)} &
  \multicolumn{1}{@{}l}{} \\
\multicolumn{1}{l}{\hspace{1em}Secondary school-leaving certificate} &
  \multicolumn{1}{r@{}}{-0.020} &
  \multicolumn{1}{@{}l}{*} &
  \multicolumn{1}{r@{}}{-0.008} &
  \multicolumn{1}{@{}l}{} &
  \multicolumn{1}{r@{}}{-0.010} &
  \multicolumn{1}{@{}l}{} &
  \multicolumn{1}{r@{}}{-0.010} &
  \multicolumn{1}{@{}l}{} \\
\multicolumn{1}{l}{} &
  \multicolumn{1}{r@{}}{(0.011)} &
  \multicolumn{1}{@{}l}{} &
  \multicolumn{1}{r@{}}{(0.008)} &
  \multicolumn{1}{@{}l}{} &
  \multicolumn{1}{r@{}}{(0.008)} &
  \multicolumn{1}{@{}l}{} &
  \multicolumn{1}{r@{}}{(0.008)} &
  \multicolumn{1}{@{}l}{} \\
\multicolumn{1}{l}{\hspace{1em}(Technical) university entrance qual.} &
  \multicolumn{1}{r@{}}{-0.040} &
  \multicolumn{1}{@{}l}{***} &
  \multicolumn{1}{r@{}}{-0.023} &
  \multicolumn{1}{@{}l}{***} &
  \multicolumn{1}{r@{}}{-0.029} &
  \multicolumn{1}{@{}l}{***} &
  \multicolumn{1}{r@{}}{-0.025} &
  \multicolumn{1}{@{}l}{***} \\
\multicolumn{1}{l}{} &
  \multicolumn{1}{r@{}}{(0.012)} &
  \multicolumn{1}{@{}l}{} &
  \multicolumn{1}{r@{}}{(0.009)} &
  \multicolumn{1}{@{}l}{} &
  \multicolumn{1}{r@{}}{(0.009)} &
  \multicolumn{1}{@{}l}{} &
  \multicolumn{1}{r@{}}{(0.008)} &
  \multicolumn{1}{@{}l}{} \\
\multicolumn{1}{l}{Migr. background of hh (ref.: no migr. backgr.)} &
  \multicolumn{1}{r@{}}{} &
  \multicolumn{1}{@{}l}{} &
  \multicolumn{1}{r@{}}{} &
  \multicolumn{1}{@{}l}{} &
  \multicolumn{1}{r@{}}{} &
  \multicolumn{1}{@{}l}{} &
  \multicolumn{1}{r@{}}{} &
  \multicolumn{1}{@{}l}{} \\
\multicolumn{1}{l}{\hspace{1em}Migration backgr. (1.gen.)} &
  \multicolumn{1}{r@{}}{0.023} &
  \multicolumn{1}{@{}l}{***} &
  \multicolumn{1}{r@{}}{0.018} &
  \multicolumn{1}{@{}l}{***} &
  \multicolumn{1}{r@{}}{0.013} &
  \multicolumn{1}{@{}l}{**} &
  \multicolumn{1}{r@{}}{0.015} &
  \multicolumn{1}{@{}l}{***} \\
\multicolumn{1}{l}{} &
  \multicolumn{1}{r@{}}{(0.007)} &
  \multicolumn{1}{@{}l}{} &
  \multicolumn{1}{r@{}}{(0.006)} &
  \multicolumn{1}{@{}l}{} &
  \multicolumn{1}{r@{}}{(0.005)} &
  \multicolumn{1}{@{}l}{} &
  \multicolumn{1}{r@{}}{(0.005)} &
  \multicolumn{1}{@{}l}{} \\
\multicolumn{1}{l}{\hspace{1em}Migration backgr. (2.gen.)} &
  \multicolumn{1}{r@{}}{0.019} &
  \multicolumn{1}{@{}l}{*} &
  \multicolumn{1}{r@{}}{0.003} &
  \multicolumn{1}{@{}l}{} &
  \multicolumn{1}{r@{}}{0.004} &
  \multicolumn{1}{@{}l}{} &
  \multicolumn{1}{r@{}}{0.004} &
  \multicolumn{1}{@{}l}{} \\
\multicolumn{1}{l}{} &
  \multicolumn{1}{r@{}}{(0.011)} &
  \multicolumn{1}{@{}l}{} &
  \multicolumn{1}{r@{}}{(0.008)} &
  \multicolumn{1}{@{}l}{} &
  \multicolumn{1}{r@{}}{(0.008)} &
  \multicolumn{1}{@{}l}{} &
  \multicolumn{1}{r@{}}{(0.007)} &
  \multicolumn{1}{@{}l}{} \\
\multicolumn{1}{l}{Household type (ref.: single)} &
  \multicolumn{1}{r@{}}{} &
  \multicolumn{1}{@{}l}{} &
  \multicolumn{1}{r@{}}{} &
  \multicolumn{1}{@{}l}{} &
  \multicolumn{1}{r@{}}{} &
  \multicolumn{1}{@{}l}{} &
  \multicolumn{1}{r@{}}{} &
  \multicolumn{1}{@{}l}{} \\
\multicolumn{1}{l}{\hspace{1em}Couple w/o children} &
  \multicolumn{1}{r@{}}{0.063} &
  \multicolumn{1}{@{}l}{***} &
  \multicolumn{1}{r@{}}{0.137} &
  \multicolumn{1}{@{}l}{***} &
  \multicolumn{1}{r@{}}{0.128} &
  \multicolumn{1}{@{}l}{***} &
  \multicolumn{1}{r@{}}{0.124} &
  \multicolumn{1}{@{}l}{***} \\
\multicolumn{1}{l}{} &
  \multicolumn{1}{r@{}}{(0.011)} &
  \multicolumn{1}{@{}l}{} &
  \multicolumn{1}{r@{}}{(0.010)} &
  \multicolumn{1}{@{}l}{} &
  \multicolumn{1}{r@{}}{(0.009)} &
  \multicolumn{1}{@{}l}{} &
  \multicolumn{1}{r@{}}{(0.009)} &
  \multicolumn{1}{@{}l}{} \\
\multicolumn{1}{l}{\hspace{1em}Single parent household} &
  \multicolumn{1}{r@{}}{0.104} &
  \multicolumn{1}{@{}l}{***} &
  \multicolumn{1}{r@{}}{0.231} &
  \multicolumn{1}{@{}l}{***} &
  \multicolumn{1}{r@{}}{0.179} &
  \multicolumn{1}{@{}l}{***} &
  \multicolumn{1}{r@{}}{0.170} &
  \multicolumn{1}{@{}l}{***} \\
\multicolumn{1}{l}{} &
  \multicolumn{1}{r@{}}{(0.008)} &
  \multicolumn{1}{@{}l}{} &
  \multicolumn{1}{r@{}}{(0.008)} &
  \multicolumn{1}{@{}l}{} &
  \multicolumn{1}{r@{}}{(0.009)} &
  \multicolumn{1}{@{}l}{} &
  \multicolumn{1}{r@{}}{(0.009)} &
  \multicolumn{1}{@{}l}{} \\
\multicolumn{1}{l}{\hspace{1em}Couple w/ children} &
  \multicolumn{1}{r@{}}{0.077} &
  \multicolumn{1}{@{}l}{***} &
  \multicolumn{1}{r@{}}{0.214} &
  \multicolumn{1}{@{}l}{***} &
  \multicolumn{1}{r@{}}{0.177} &
  \multicolumn{1}{@{}l}{***} &
  \multicolumn{1}{r@{}}{0.170} &
  \multicolumn{1}{@{}l}{***} \\
\multicolumn{1}{l}{} &
  \multicolumn{1}{r@{}}{(0.011)} &
  \multicolumn{1}{@{}l}{} &
  \multicolumn{1}{r@{}}{(0.009)} &
  \multicolumn{1}{@{}l}{} &
  \multicolumn{1}{r@{}}{(0.009)} &
  \multicolumn{1}{@{}l}{} &
  \multicolumn{1}{r@{}}{(0.009)} &
  \multicolumn{1}{@{}l}{} \\
\multicolumn{1}{l}{Presence of child aged $<$=3 in household} &
  \multicolumn{1}{r@{}}{0.057} &
  \multicolumn{1}{@{}l}{***} &
  \multicolumn{1}{r@{}}{0.057} &
  \multicolumn{1}{@{}l}{***} &
  \multicolumn{1}{r@{}}{0.043} &
  \multicolumn{1}{@{}l}{***} &
  \multicolumn{1}{r@{}}{0.036} &
  \multicolumn{1}{@{}l}{***} \\
\multicolumn{1}{l}{} &
  \multicolumn{1}{r@{}}{(0.009)} &
  \multicolumn{1}{@{}l}{} &
  \multicolumn{1}{r@{}}{(0.006)} &
  \multicolumn{1}{@{}l}{} &
  \multicolumn{1}{r@{}}{(0.006)} &
  \multicolumn{1}{@{}l}{} &
  \multicolumn{1}{r@{}}{(0.006)} &
  \multicolumn{1}{@{}l}{} \\
\multicolumn{1}{l}{Disabled hh and/or partner} &
  \multicolumn{1}{r@{}}{0.012} &
  \multicolumn{1}{@{}l}{*} &
  \multicolumn{1}{r@{}}{-0.002} &
  \multicolumn{1}{@{}l}{} &
  \multicolumn{1}{r@{}}{-0.011} &
  \multicolumn{1}{@{}l}{**} &
  \multicolumn{1}{r@{}}{-0.013} &
  \multicolumn{1}{@{}l}{**} \\
\multicolumn{1}{l}{} &
  \multicolumn{1}{r@{}}{(0.007)} &
  \multicolumn{1}{@{}l}{} &
  \multicolumn{1}{r@{}}{(0.005)} &
  \multicolumn{1}{@{}l}{} &
  \multicolumn{1}{r@{}}{(0.006)} &
  \multicolumn{1}{@{}l}{} &
  \multicolumn{1}{r@{}}{(0.005)} &
  \multicolumn{1}{@{}l}{} \\
\multicolumn{1}{l}{Home ownership} &
  \multicolumn{1}{r@{}}{-0.035} &
  \multicolumn{1}{@{}l}{*} &
  \multicolumn{1}{r@{}}{-0.014} &
  \multicolumn{1}{@{}l}{} &
  \multicolumn{1}{r@{}}{-0.018} &
  \multicolumn{1}{@{}l}{*} &
  \multicolumn{1}{r@{}}{-0.022} &
  \multicolumn{1}{@{}l}{**} \\
\multicolumn{1}{l}{} &
  \multicolumn{1}{r@{}}{(0.018)} &
  \multicolumn{1}{@{}l}{} &
  \multicolumn{1}{r@{}}{(0.010)} &
  \multicolumn{1}{@{}l}{} &
  \multicolumn{1}{r@{}}{(0.010)} &
  \multicolumn{1}{@{}l}{} &
  \multicolumn{1}{r@{}}{(0.010)} &
  \multicolumn{1}{@{}l}{} \\
\multicolumn{1}{l}{Region of residence (ref.: West Germany)} &
  \multicolumn{1}{r@{}}{0.013} &
  \multicolumn{1}{@{}l}{*} &
  \multicolumn{1}{r@{}}{0.006} &
  \multicolumn{1}{@{}l}{} &
  \multicolumn{1}{r@{}}{0.004} &
  \multicolumn{1}{@{}l}{} &
  \multicolumn{1}{r@{}}{0.003} &
  \multicolumn{1}{@{}l}{} \\
\multicolumn{1}{l}{} &
  \multicolumn{1}{r@{}}{(0.007)} &
  \multicolumn{1}{@{}l}{} &
  \multicolumn{1}{r@{}}{(0.005)} &
  \multicolumn{1}{@{}l}{} &
  \multicolumn{1}{r@{}}{(0.005)} &
  \multicolumn{1}{@{}l}{} &
  \multicolumn{1}{r@{}}{(0.005)} &
  \multicolumn{1}{@{}l}{} \\
\multicolumn{1}{l}{Population size category (ref.: $<$ 50,000 residents)} &
  \multicolumn{1}{r@{}}{} &
  \multicolumn{1}{@{}l}{} &
  \multicolumn{1}{r@{}}{} &
  \multicolumn{1}{@{}l}{} &
  \multicolumn{1}{r@{}}{} &
  \multicolumn{1}{@{}l}{} &
  \multicolumn{1}{r@{}}{} &
  \multicolumn{1}{@{}l}{} \\
\multicolumn{1}{l}{\hspace{1em}$>$= 50,000 res. (periph. area)} &
  \multicolumn{1}{r@{}}{-0.005} &
  \multicolumn{1}{@{}l}{} &
  \multicolumn{1}{r@{}}{-0.000} &
  \multicolumn{1}{@{}l}{} &
  \multicolumn{1}{r@{}}{0.002} &
  \multicolumn{1}{@{}l}{} &
  \multicolumn{1}{r@{}}{0.001} &
  \multicolumn{1}{@{}l}{} \\
\multicolumn{1}{l}{} &
  \multicolumn{1}{r@{}}{(0.010)} &
  \multicolumn{1}{@{}l}{} &
  \multicolumn{1}{r@{}}{(0.006)} &
  \multicolumn{1}{@{}l}{} &
  \multicolumn{1}{r@{}}{(0.006)} &
  \multicolumn{1}{@{}l}{} &
  \multicolumn{1}{r@{}}{(0.006)} &
  \multicolumn{1}{@{}l}{} \\
\multicolumn{1}{l}{\hspace{1em}$>$= 50,000-499,999 res. (core area)} &
  \multicolumn{1}{r@{}}{0.019} &
  \multicolumn{1}{@{}l}{**} &
  \multicolumn{1}{r@{}}{0.019} &
  \multicolumn{1}{@{}l}{***} &
  \multicolumn{1}{r@{}}{0.020} &
  \multicolumn{1}{@{}l}{***} &
  \multicolumn{1}{r@{}}{0.017} &
  \multicolumn{1}{@{}l}{***} \\
\multicolumn{1}{l}{} &
  \multicolumn{1}{r@{}}{(0.009)} &
  \multicolumn{1}{@{}l}{} &
  \multicolumn{1}{r@{}}{(0.006)} &
  \multicolumn{1}{@{}l}{} &
  \multicolumn{1}{r@{}}{(0.006)} &
  \multicolumn{1}{@{}l}{} &
  \multicolumn{1}{r@{}}{(0.006)} &
  \multicolumn{1}{@{}l}{} \\
\multicolumn{1}{l}{\hspace{1em}$>$=500,000 res. (core area)} &
  \multicolumn{1}{r@{}}{0.017} &
  \multicolumn{1}{@{}l}{*} &
  \multicolumn{1}{r@{}}{0.010} &
  \multicolumn{1}{@{}l}{*} &
  \multicolumn{1}{r@{}}{0.013} &
  \multicolumn{1}{@{}l}{**} &
  \multicolumn{1}{r@{}}{0.012} &
  \multicolumn{1}{@{}l}{**} \\
\multicolumn{1}{l}{} &
  \multicolumn{1}{r@{}}{(0.009)} &
  \multicolumn{1}{@{}l}{} &
  \multicolumn{1}{r@{}}{(0.006)} &
  \multicolumn{1}{@{}l}{} &
  \multicolumn{1}{r@{}}{(0.006)} &
  \multicolumn{1}{@{}l}{} &
  \multicolumn{1}{r@{}}{(0.006)} &
  \multicolumn{1}{@{}l}{} \\
\multicolumn{1}{l}{Early retirement status} &
  \multicolumn{1}{r@{}}{-0.260} &
  \multicolumn{1}{@{}l}{***} &
  \multicolumn{1}{r@{}}{-0.123} &
  \multicolumn{1}{@{}l}{***} &
  \multicolumn{1}{r@{}}{-0.170} &
  \multicolumn{1}{@{}l}{***} &
  \multicolumn{1}{r@{}}{-0.179} &
  \multicolumn{1}{@{}l}{***} \\
\multicolumn{1}{l}{} &
  \multicolumn{1}{r@{}}{(0.019)} &
  \multicolumn{1}{@{}l}{} &
  \multicolumn{1}{r@{}}{(0.014)} &
  \multicolumn{1}{@{}l}{} &
  \multicolumn{1}{r@{}}{(0.016)} &
  \multicolumn{1}{@{}l}{} &
  \multicolumn{1}{r@{}}{(0.016)} &
  \multicolumn{1}{@{}l}{} \\
\multicolumn{1}{l}{Reported receipt of other benefit than UB II} &
  \multicolumn{1}{r@{}}{-0.176} &
  \multicolumn{1}{@{}l}{***} &
  \multicolumn{1}{r@{}}{-0.114} &
  \multicolumn{1}{@{}l}{***} &
  \multicolumn{1}{r@{}}{-0.107} &
  \multicolumn{1}{@{}l}{***} &
  \multicolumn{1}{r@{}}{-0.105} &
  \multicolumn{1}{@{}l}{***} \\
\multicolumn{1}{l}{} &
  \multicolumn{1}{r@{}}{(0.011)} &
  \multicolumn{1}{@{}l}{} &
  \multicolumn{1}{r@{}}{(0.008)} &
  \multicolumn{1}{@{}l}{} &
  \multicolumn{1}{r@{}}{(0.008)} &
  \multicolumn{1}{@{}l}{} &
  \multicolumn{1}{r@{}}{(0.008)} &
  \multicolumn{1}{@{}l}{} \\
\multicolumn{1}{l}{GenPop subsample (ref.: Admin sample)} &
  \multicolumn{1}{r@{}}{-0.210} &
  \multicolumn{1}{@{}l}{***} &
  \multicolumn{1}{r@{}}{-0.068} &
  \multicolumn{1}{@{}l}{***} &
  \multicolumn{1}{r@{}}{-0.056} &
  \multicolumn{1}{@{}l}{***} &
  \multicolumn{1}{r@{}}{-0.054} &
  \multicolumn{1}{@{}l}{***} \\
\multicolumn{1}{l}{} &
  \multicolumn{1}{r@{}}{(0.015)} &
  \multicolumn{1}{@{}l}{} &
  \multicolumn{1}{r@{}}{(0.009)} &
  \multicolumn{1}{@{}l}{} &
  \multicolumn{1}{r@{}}{(0.009)} &
  \multicolumn{1}{@{}l}{} &
  \multicolumn{1}{r@{}}{(0.009)} &
  \multicolumn{1}{@{}l}{} \\
\multicolumn{1}{l}{Lagged share of days per year with UB II receipt (t-1)} &
  \multicolumn{1}{r@{}}{} &
  \multicolumn{1}{@{}l}{} &
  \multicolumn{1}{r@{}}{0.358} &
  \multicolumn{1}{@{}l}{***} &
  \multicolumn{1}{r@{}}{0.296} &
  \multicolumn{1}{@{}l}{***} &
  \multicolumn{1}{r@{}}{0.301} &
  \multicolumn{1}{@{}l}{***} \\
\multicolumn{1}{l}{} &
  \multicolumn{1}{r@{}}{} &
  \multicolumn{1}{@{}l}{} &
  \multicolumn{1}{r@{}}{(0.008)} &
  \multicolumn{1}{@{}l}{} &
  \multicolumn{1}{r@{}}{(0.008)} &
  \multicolumn{1}{@{}l}{} &
  \multicolumn{1}{r@{}}{(0.008)} &
  \multicolumn{1}{@{}l}{} \\
\multicolumn{1}{l}{Lagged share of days per year with UB II receipt (t-2)} &
  \multicolumn{1}{r@{}}{} &
  \multicolumn{1}{@{}l}{} &
  \multicolumn{1}{r@{}}{-0.090} &
  \multicolumn{1}{@{}l}{***} &
  \multicolumn{1}{r@{}}{-0.075} &
  \multicolumn{1}{@{}l}{***} &
  \multicolumn{1}{r@{}}{-0.080} &
  \multicolumn{1}{@{}l}{***} \\
\multicolumn{1}{l}{} &
  \multicolumn{1}{r@{}}{} &
  \multicolumn{1}{@{}l}{} &
  \multicolumn{1}{r@{}}{(0.010)} &
  \multicolumn{1}{@{}l}{} &
  \multicolumn{1}{r@{}}{(0.009)} &
  \multicolumn{1}{@{}l}{} &
  \multicolumn{1}{r@{}}{(0.009)} &
  \multicolumn{1}{@{}l}{} \\
\multicolumn{1}{l}{Lagged share of days per year with UB II receipt (t-3)} &
  \multicolumn{1}{r@{}}{} &
  \multicolumn{1}{@{}l}{} &
  \multicolumn{1}{r@{}}{0.014} &
  \multicolumn{1}{@{}l}{**} &
  \multicolumn{1}{r@{}}{0.009} &
  \multicolumn{1}{@{}l}{} &
  \multicolumn{1}{r@{}}{0.006} &
  \multicolumn{1}{@{}l}{} \\
\multicolumn{1}{l}{} &
  \multicolumn{1}{r@{}}{} &
  \multicolumn{1}{@{}l}{} &
  \multicolumn{1}{r@{}}{(0.007)} &
  \multicolumn{1}{@{}l}{} &
  \multicolumn{1}{r@{}}{(0.007)} &
  \multicolumn{1}{@{}l}{} &
  \multicolumn{1}{r@{}}{(0.007)} &
  \multicolumn{1}{@{}l}{} \\
\multicolumn{1}{l}{Real equivalent annual earned income (t-1)} &
  \multicolumn{1}{r@{}}{} &
  \multicolumn{1}{@{}l}{} &
  \multicolumn{1}{r@{}}{} &
  \multicolumn{1}{@{}l}{} &
  \multicolumn{1}{r@{}}{-0.074} &
  \multicolumn{1}{@{}l}{***} &
  \multicolumn{1}{r@{}}{-0.098} &
  \multicolumn{1}{@{}l}{***} \\
\multicolumn{1}{l}{} &
  \multicolumn{1}{r@{}}{} &
  \multicolumn{1}{@{}l}{} &
  \multicolumn{1}{r@{}}{} &
  \multicolumn{1}{@{}l}{} &
  \multicolumn{1}{r@{}}{(0.006)} &
  \multicolumn{1}{@{}l}{} &
  \multicolumn{1}{r@{}}{(0.006)} &
  \multicolumn{1}{@{}l}{} \\
\multicolumn{1}{l}{Real equivalent annual earned income (t-2)} &
  \multicolumn{1}{r@{}}{} &
  \multicolumn{1}{@{}l}{} &
  \multicolumn{1}{r@{}}{} &
  \multicolumn{1}{@{}l}{} &
  \multicolumn{1}{r@{}}{-0.005} &
  \multicolumn{1}{@{}l}{} &
  \multicolumn{1}{r@{}}{-0.004} &
  \multicolumn{1}{@{}l}{} \\
\multicolumn{1}{l}{} &
  \multicolumn{1}{r@{}}{} &
  \multicolumn{1}{@{}l}{} &
  \multicolumn{1}{r@{}}{} &
  \multicolumn{1}{@{}l}{} &
  \multicolumn{1}{r@{}}{(0.006)} &
  \multicolumn{1}{@{}l}{} &
  \multicolumn{1}{r@{}}{(0.005)} &
  \multicolumn{1}{@{}l}{} \\
\multicolumn{1}{l}{Real equivalent annual earned income (t-3)} &
  \multicolumn{1}{r@{}}{} &
  \multicolumn{1}{@{}l}{} &
  \multicolumn{1}{r@{}}{} &
  \multicolumn{1}{@{}l}{} &
  \multicolumn{1}{r@{}}{-0.017} &
  \multicolumn{1}{@{}l}{***} &
  \multicolumn{1}{r@{}}{-0.011} &
  \multicolumn{1}{@{}l}{**} \\
\multicolumn{1}{l}{} &
  \multicolumn{1}{r@{}}{} &
  \multicolumn{1}{@{}l}{} &
  \multicolumn{1}{r@{}}{} &
  \multicolumn{1}{@{}l}{} &
  \multicolumn{1}{r@{}}{(0.004)} &
  \multicolumn{1}{@{}l}{} &
  \multicolumn{1}{r@{}}{(0.005)} &
  \multicolumn{1}{@{}l}{} \\
\multicolumn{1}{l}{Difference of earned income (normalised by UB II needs)} &
  \multicolumn{1}{r@{}}{} &
  \multicolumn{1}{@{}l}{} &
  \multicolumn{1}{r@{}}{} &
  \multicolumn{1}{@{}l}{} &
  \multicolumn{1}{r@{}}{} &
  \multicolumn{1}{@{}l}{} &
  \multicolumn{1}{r@{}}{-0.044} &
  \multicolumn{1}{@{}l}{***} \\
\multicolumn{1}{l}{} &
  \multicolumn{1}{r@{}}{} &
  \multicolumn{1}{@{}l}{} &
  \multicolumn{1}{r@{}}{} &
  \multicolumn{1}{@{}l}{} &
  \multicolumn{1}{r@{}}{} &
  \multicolumn{1}{@{}l}{} &
  \multicolumn{1}{r@{}}{(0.003)} &
  \multicolumn{1}{@{}l}{} \\
\multicolumn{1}{l}{Standard deviation (within) of lagged real equiv. income} &
  \multicolumn{1}{r@{}}{} &
  \multicolumn{1}{@{}l}{} &
  \multicolumn{1}{r@{}}{} &
  \multicolumn{1}{@{}l}{} &
  \multicolumn{1}{r@{}}{} &
  \multicolumn{1}{@{}l}{} &
  \multicolumn{1}{r@{}}{-0.002} &
  \multicolumn{1}{@{}l}{} \\
\multicolumn{1}{l}{} &
  \multicolumn{1}{r@{}}{} &
  \multicolumn{1}{@{}l}{} &
  \multicolumn{1}{r@{}}{} &
  \multicolumn{1}{@{}l}{} &
  \multicolumn{1}{r@{}}{} &
  \multicolumn{1}{@{}l}{} &
  \multicolumn{1}{r@{}}{(0.008)} &
  \multicolumn{1}{@{}l}{} \\
\cline{1-9}
\multicolumn{1}{l}{Number of observations} &
  \multicolumn{1}{r@{}}{27042} &
  \multicolumn{1}{@{}l}{} &
  \multicolumn{1}{r@{}}{25263} &
  \multicolumn{1}{@{}l}{} &
  \multicolumn{1}{r@{}}{24516} &
  \multicolumn{1}{@{}l}{} &
  \multicolumn{1}{r@{}}{24486} &
  \multicolumn{1}{@{}l}{} \\
\multicolumn{1}{l}{Log-likelihood} &
  \multicolumn{1}{r@{}}{-7.4464e+03} &
  \multicolumn{1}{@{}l}{} &
  \multicolumn{1}{r@{}}{-4.9415e+03} &
  \multicolumn{1}{@{}l}{} &
  \multicolumn{1}{r@{}}{-4.5948e+03} &
  \multicolumn{1}{@{}l}{} &
  \multicolumn{1}{r@{}}{-4.4147e+03} &
  \multicolumn{1}{@{}l}{} \\
\multicolumn{1}{l}{Panel-\(\rho\)} &
  \multicolumn{1}{r@{}}{0.712} &
  \multicolumn{1}{@{}l}{***} &
  \multicolumn{1}{r@{}}{0.569} &
  \multicolumn{1}{@{}l}{***} &
  \multicolumn{1}{r@{}}{0.550} &
  \multicolumn{1}{@{}l}{***} &
  \multicolumn{1}{r@{}}{0.576} &
  \multicolumn{1}{@{}l}{***} \\
\cline{1-9}
\end{longtable}

\end{ThreePartTable}
 \end{landscape}

\section{Robustness check: Alternative measures of income volatility\label{sec:AppRobustness}}

\setcounter{table}{0}\setcounter{figure}{0}
\renewcommand\thetable{\thesection\arabic{table}} 
\renewcommand\thefigure{\thesection\arabic{figure}} 

\revisedfloat{\begin{table}[!ht]
\footnotesize
\centering
\begin{threeparttable}
\caption{Income volatility in the take-up model -- alternative volatility measures}
\phantomsection\label{tab:VolatilityRobustness}
\renewcommand{\arraystretch}{1.2}

\begin{tabular}{lrr}
\hline\hline
 & (1) Level SD & (2) asinh SD \\
 & (M3 in main text) & (robustness check) \\
\hline
Income volatility (AME) & $-0.002$ & $-0.006$ \\
\quad (std.\ err.)        & $(0.008)$ & $(0.011)$ \\
\quad $p$-value           & $0.785$  & $0.596$  \\[2pt]
Income shock (AME)      & $-0.044$ & $-0.042$ \\
\quad (std.\ err.)        & $(0.003)$ & $(0.003)$ \\
\quad $p$-value           & $<0.001$ & $<0.001$ \\[2pt]
Observations              & 24,486 & 24,486 \\
\hline
\end{tabular}

\begin{tablenotes}
\item \scriptsize \textsc{Note. ---} Average marginal effects (AME) on the probability of take-up for the income-volatility and income-shock terms in the full take-up specification (M3), which also contains the baseline covariates, the benefit-history and income-potential blocks, time fixed effects, and a volatility$\times$shock interaction. Marginal effects are averaged over the estimation sample (random-effects probit). Column (1) measures volatility as the within-household standard deviation of real equivalised quarterly earned income in levels (the measure reported in the main text). Column (2) replaces it with the standard deviation of the inverse-hyperbolic-sine (asinh) transform of the same quarterly earnings, which behaves like the logarithm for larger values but is defined at zero earnings. The volatility$\times$income-shock interaction itself is a raw probit coefficient rather than an AME and so is not shown as a row above; it is likewise statistically insignificant in both columns ($p=0.455$ in column (1), $p=0.184$ in column (2)). Standard errors are cluster-robust at the household level. Source: GETTSIM v0.7.0, PASS 0620 v1, PASS-ADIAB 7520 v1, own calculations.
\end{tablenotes}

\end{threeparttable}
\end{table}
 }

\newpage
\section{Marginal effects take-up models: Weighted vs. unweighted estimations\label{sec:AppWeightedVsUnweighted}}

\setcounter{table}{0}\setcounter{figure}{0}
\renewcommand\thetable{\thesection\arabic{table}} 
\renewcommand\thefigure{\thesection\arabic{figure}}

\begin{figure}[!htb]
\caption{Take-up regression, baseline specification, weighted vs. unweighted estimation\label{fig:est_base_weighted_vs_unweighted}}
\includegraphics[width=1\textwidth]{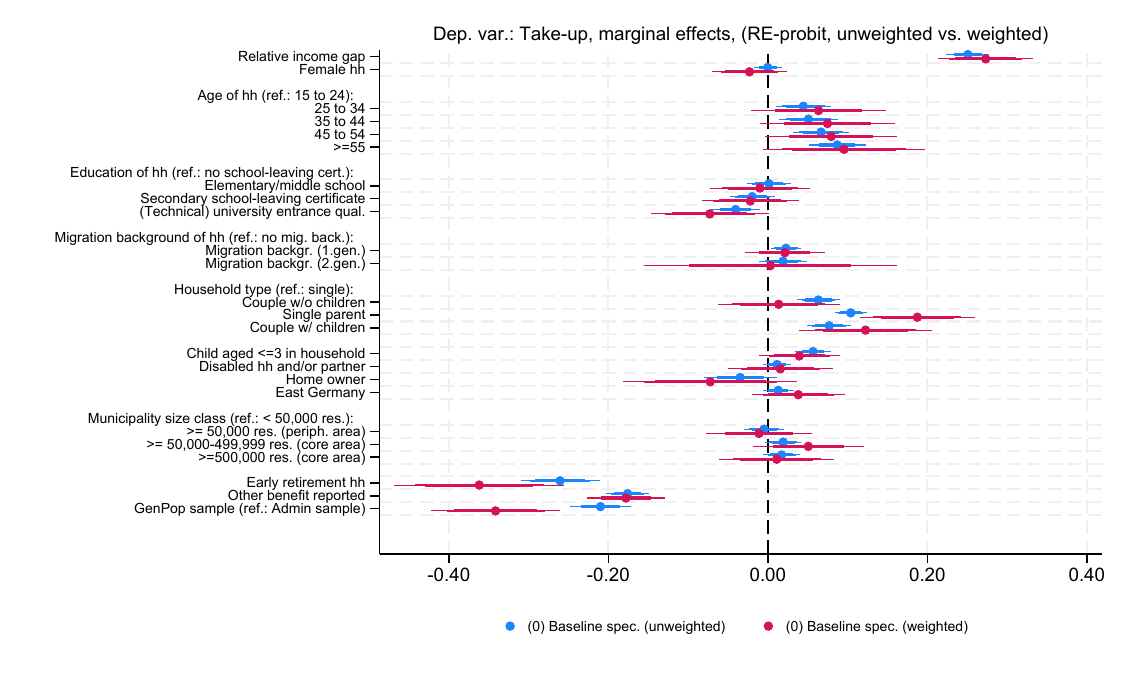}
\vspace{-1cm}\floatfoot{\scriptsize \textsc{Note. ---} Random effects probit model with binary indicator for UB~II take-up (take-up = 1) as dependent variable. Dots mark the point estimate of the respective marginal effect, while the corresponding horizontal lines represent the confidence intervals at the 99/95/90~percent level (thin/medium/thick line). The model contains time fixed effects. hh=head of household. Unweighted (weighted) estimation shown in blue (red). Source: Own calculations, GETTSIM v0.7.0, PASS 0620 v1, PASS-ADIAB 7520 v1.}   
\end{figure}

\begin{figure}[!htb]
\caption{Take-up regression, M1, weighted vs. unweighted estimation\label{fig:est_M1_weighted_vs_unweighted}}
\includegraphics[width=1\textwidth]{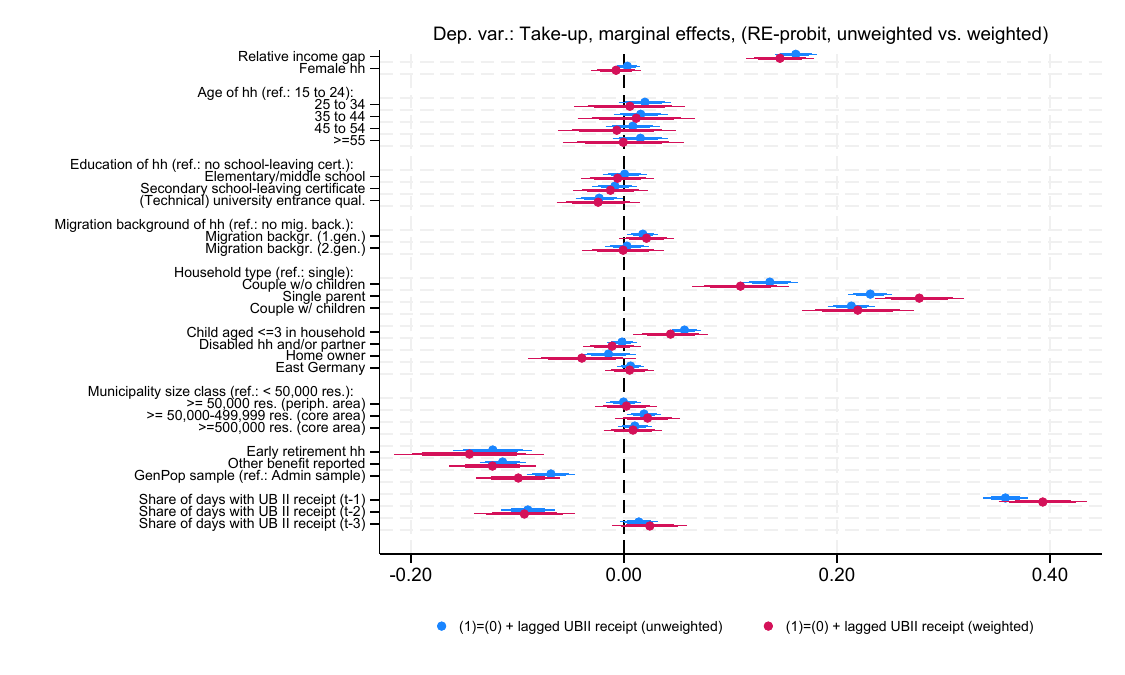}
\vspace{-1cm}\floatfoot{\scriptsize \textsc{Note. ---} Random effects probit model with binary indicator for UB~II take-up (take-up = 1) as dependent variable. Dots mark the point estimate of the respective marginal effect, while the corresponding horizontal lines represent the confidence intervals at the 99/95/90~percent level (thin/medium/thick line). The model contains time fixed effects. hh=head of household. Unweighted (weighted) estimation shown in blue (red). Source: Own calculations, GETTSIM v0.7.0, PASS 0620 v1, PASS-ADIAB 7520 v1.}    
\end{figure}

\begin{figure}[!htb]
\caption{Take-up regression, M2, weighted vs. unweighted estimation\label{fig:est_M2_weighted_vs_unweighted}}
\includegraphics[width=1\textwidth]{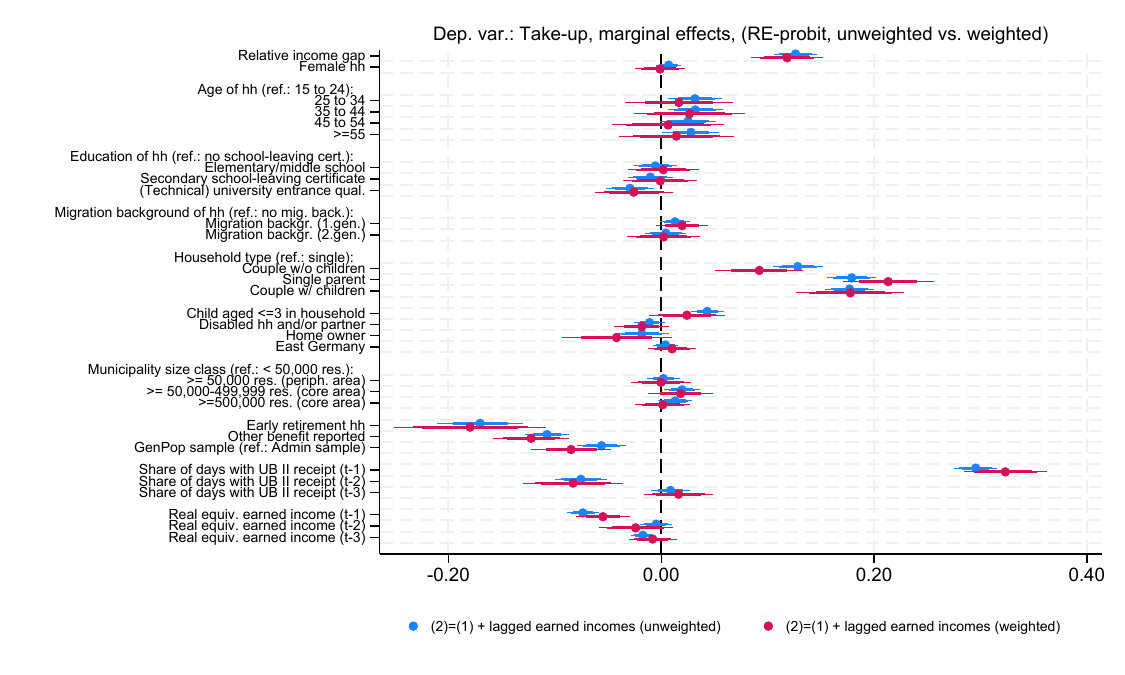}
\vspace{-1cm}\floatfoot{\scriptsize \textsc{Note. ---} Random effects probit model with binary indicator for UB~II take-up (take-up = 1) as dependent variable. Dots mark the point estimate of the respective marginal effect, while the corresponding horizontal lines represent the confidence intervals at the 99/95/90~percent level (thin/medium/thick line). Real equivalised earned income is measured in 1,000 euros/month in 2020 prices using the new OECD scale. The model contains time fixed effects. hh=head of household. Unweighted (weighted) estimation shown in blue (red). Source: Own calculations, GETTSIM v0.7.0, PASS 0620 v1, PASS-ADIAB 7520 v1.}    
\end{figure}

\begin{figure}[!htb]
\caption{Take-up regression, M3, weighted vs. unweighted estimation\label{fig:est_M3_weighted_vs_unweighted}}
\includegraphics[width=1\textwidth]{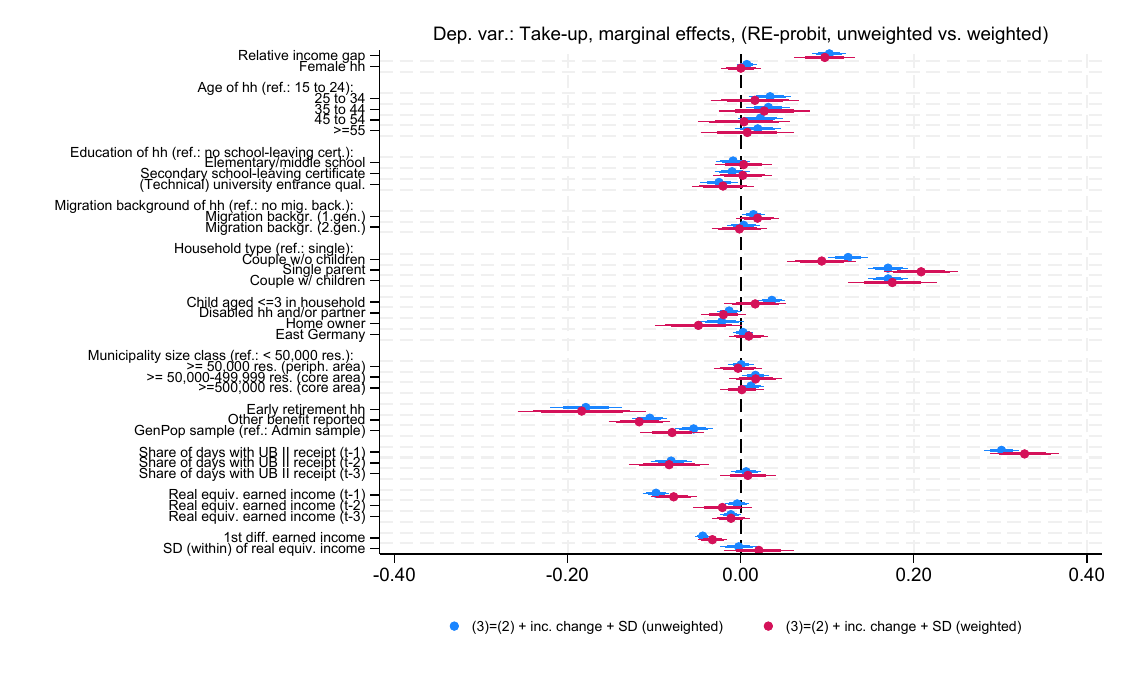}
\vspace{-1cm}\floatfoot{\scriptsize \textsc{Note. ---} Random effects probit model with binary indicator for UB~II take-up (take-up = 1) as dependent variable. Dots mark the point estimate of the respective marginal effect, while the corresponding horizontal lines represent the confidence intervals at the 99/95/90~percent level (thin/medium/thick line). Real equivalised earned income is measured in 1,000 euros/month in 2020 prices using the new OECD scale. 1st diff. earned income=year-over-year change in income preceding the quarter of the PASS interview. SD (within) of real equiv. income=within-household standard deviation of real equivalised earned income in the twelve quarters preceding the current interview. The model contains time fixed effects. hh=head of household. Unweighted (weighted) estimation shown in blue (red). Source: Own calculations, GETTSIM v0.7.0, PASS 0620 v1, PASS-ADIAB 7520 v1.}    
\end{figure}

\end{appendix}

\end{document}